\documentclass[longauth,traditabstract]{aa} % for the abstract without structuration

\def\setsymbol#1#2{\expandafter\def\csname #1\endcsname{#2}}
\def\getsymbol#1{\csname #1\endcsname}

\def\Planck{\textit{Planck}}

\def\all2013resultspapers{\nocite{planck2013-p01, planck2013-p02, planck2013-p02a, planck2013-p02d, planck2013-p02b, planck2013-p03, planck2013-p03c, planck2013-p03f, planck2013-p03d, planck2013-p03e, planck2013-p01a, planck2013-p06, planck2013-p03a, planck2013-pip88, planck2013-p08, planck2013-p11, planck2013-p12, planck2013-p13, planck2013-p14, planck2013-p15, planck2013-p05b, planck2013-p17, planck2013-p09, planck2013-p09a, planck2013-p20, planck2013-p19, planck2013-pipaberration, planck2013-p05, planck2013-p05a, planck2013-pip56, planck2013-p06b}}

\newbox\tablebox    \newdimen\tablewidth
\def\leaderfil{\leaders\hbox to 5pt{\hss.\hss}\hfil}
\def\endPlancktablewide{\tablewidth=\textwidth 
    $$\hss\copy\tablebox\hss$$
    \vskip-\lastskip\vskip -2pt}
\def\tablenote#1 #2\par{\begingroup \parindent=0.8em
    \abovedisplayshortskip=0pt\belowdisplayshortskip=0pt
    \noindent
    $$\hss\vbox{\hsize\tablewidth \hangindent=\parindent \hangafter=1 \noindent
    \hbox to \parindent{$^#1$\hss}\strut#2\strut\par}\hss$$
    \endgroup}
\def\doubleline{\vskip 3pt\hrule \vskip 1.5pt \hrule \vskip 5pt}

\def\L2{\ifmmode L_2\else $L_2$\fi}

\def\DeltaT{\ifmmode \Delta T\else $\Delta T$\fi}
\def\deltat{\ifmmode \Delta t\else $\Delta t$\fi}
\def\fknee{\ifmmode f_{\rm knee}\else $f_{\rm knee}$\fi}
\def\Fmax{\ifmmode F_{\rm max}\else $F_{\rm max}$\fi}
\def\solar{\ifmmode{\rm M}_{\mathord\odot}\else${\rm M}_{\mathord\odot}$\fi}
\def\Msolar{\ifmmode{\rm M}_{\mathord\odot}\else${\rm M}_{\mathord\odot}$\fi}
\def\Lsolar{\ifmmode{\rm L}_{\mathord\odot}\else${\rm L}_{\mathord\odot}$\fi}

\def\inv{\ifmmode^{-1}\else$^{-1}$\fi}
\def\mo{\ifmmode^{-1}\else$^{-1}$\fi}
\def\sup#1{\ifmmode ^{\rm #1}\else $^{\rm #1}$\fi}
\def\expo#1{\ifmmode \times 10^{#1}\else $\times 10^{#1}$\fi}
\def\,{\thinspace}
\def\lsim{\mathrel{\raise .4ex\hbox{\rlap{$<$}\lower 1.2ex\hbox{$\sim$}}}}
\def\gsim{\mathrel{\raise .4ex\hbox{\rlap{$>$}\lower 1.2ex\hbox{$\sim$}}}}

\def\simprop{\mathrel{\raise .4ex\hbox{\rlap{$\propto$}\lower 1.2ex\hbox{$\sim$}}}}
\def\deg{\ifmmode^\circ\else$^\circ$\fi}
\def\pdeg{\ifmmode $\setbox0=\hbox{$^{\circ}$}\rlap{\hskip.11\wd0 .}$^{\circ}
          \else \setbox0=\hbox{$^{\circ}$}\rlap{\hskip.11\wd0 .}$^{\circ}$\fi}
\def\arcs{\ifmmode {^{\scriptstyle\prime\prime}}
          \else $^{\scriptstyle\prime\prime}$\fi}
\def\arcm{\ifmmode {^{\scriptstyle\prime}}
          \else $^{\scriptstyle\prime}$\fi}
\newdimen\sa  \newdimen\sb
\def\parcs{\sa=.07em \sb=.03em
     \ifmmode \hbox{\rlap{.}}^{\scriptstyle\prime\kern -\sb\prime}\hbox{\kern -\sa}
     \else \rlap{.}$^{\scriptstyle\prime\kern -\sb\prime}$\kern -\sa\fi}
\def\parcm{\sa=.08em \sb=.03em
     \ifmmode \hbox{\rlap{.}\kern\sa}^{\scriptstyle\prime}\hbox{\kern-\sb}
     \else \rlap{.}\kern\sa$^{\scriptstyle\prime}$\kern-\sb\fi}
\def\ra[#1 #2 #3.#4]{#1\sup{h}#2\sup{m}#3\sup{s}\llap.#4}
\def\dec[#1 #2 #3.#4]{#1\deg#2\arcm#3\arcs\llap.#4}
\def\deco[#1 #2 #3]{#1\deg#2\arcm#3\arcs}
\def\rra[#1 #2]{#1\sup{h}#2\sup{m}}

\def\dots{\relax\ifmmode \ldots\else $\ldots$\fi}
\def\WHzsr{\ifmmode $W\,Hz\mo\,sr\mo$\else W\,Hz\mo\,sr\mo\fi}
\def\mHz{\ifmmode $\,mHz$\else \,mHz\fi}
\def\GHz{\ifmmode $\,GHz$\else \,GHz\fi}
\def\mKs{\ifmmode $\,mK\,s$^{1/2}\else \,mK\,s$^{1/2}$\fi}
\def\muKs{\ifmmode \,\mu$K\,s$^{1/2}\else \,$\mu$K\,s$^{1/2}$\fi}
\def\muKRJs{\ifmmode \,\mu$K$_{\rm RJ}$\,s$^{1/2}\else \,$\mu$K$_{\rm RJ}$\,s$^{1/2}$\fi}
\def\muKHz{\ifmmode \,\mu$K\,Hz$^{-1/2}\else \,$\mu$K\,Hz$^{-1/2}$\fi}
\def\MJysr{\ifmmode \,$MJy\,sr\mo$\else \,MJy\,sr\mo\fi}
\def\MJysrmK{\ifmmode \,$MJy\,sr\mo$\,mK$_{\rm CMB}\mo\else \,MJy\,sr\mo\,mK$_{\rm CMB}\mo$\fi}
\def\microns{\ifmmode \,\mu$m$\else \,$\mu$m\fi}

\def\muK{\ifmmode \,\mu$K$\else \,$\mu$\hbox{K}\fi}
\def\microK{\ifmmode \,\mu$K$\else \,$\mu$\hbox{K}\fi}
\def\muW{\ifmmode \,\mu$W$\else \,$\mu$\hbox{W}\fi}
\def\kms{\ifmmode $\,km\,s$^{-1}\else \,km\,s$^{-1}$\fi}
\def\kmsMpc{\ifmmode $\,\kms\,Mpc\mo$\else \,\kms\,Mpc\mo\fi}
\setsymbol{LFI:center:frequency:70GHz:units}{70.3\,GHz}
\setsymbol{LFI:center:frequency:44GHz:units}{44.1\,GHz}
\setsymbol{LFI:center:frequency:30GHz:units}{28.5\,GHz}

\setsymbol{LFI:center:frequency:70GHz}{70.3}
\setsymbol{LFI:center:frequency:44GHz}{44.1}
\setsymbol{LFI:center:frequency:30GHz}{28.5}

\setsymbol{LFI:center:frequency:LFI18:Rad:M:units}{71.7\GHz}
\setsymbol{LFI:center:frequency:LFI19:Rad:M:units}{67.5\GHz}
\setsymbol{LFI:center:frequency:LFI20:Rad:M:units}{69.2\GHz}
\setsymbol{LFI:center:frequency:LFI21:Rad:M:units}{70.4\GHz}
\setsymbol{LFI:center:frequency:LFI22:Rad:M:units}{71.5\GHz}
\setsymbol{LFI:center:frequency:LFI23:Rad:M:units}{70.8\GHz}
\setsymbol{LFI:center:frequency:LFI24:Rad:M:units}{44.4\GHz}
\setsymbol{LFI:center:frequency:LFI25:Rad:M:units}{44.0\GHz}
\setsymbol{LFI:center:frequency:LFI26:Rad:M:units}{43.9\GHz}
\setsymbol{LFI:center:frequency:LFI27:Rad:M:units}{28.3\GHz}
\setsymbol{LFI:center:frequency:LFI28:Rad:M:units}{28.8\GHz}
\setsymbol{LFI:center:frequency:LFI18:Rad:S:units}{70.1\GHz}
\setsymbol{LFI:center:frequency:LFI19:Rad:S:units}{69.6\GHz}
\setsymbol{LFI:center:frequency:LFI20:Rad:S:units}{69.5\GHz}
\setsymbol{LFI:center:frequency:LFI21:Rad:S:units}{69.5\GHz}
\setsymbol{LFI:center:frequency:LFI22:Rad:S:units}{72.8\GHz}
\setsymbol{LFI:center:frequency:LFI23:Rad:S:units}{71.3\GHz}
\setsymbol{LFI:center:frequency:LFI24:Rad:S:units}{44.1\GHz}
\setsymbol{LFI:center:frequency:LFI25:Rad:S:units}{44.1\GHz}
\setsymbol{LFI:center:frequency:LFI26:Rad:S:units}{44.1\GHz}
\setsymbol{LFI:center:frequency:LFI27:Rad:S:units}{28.5\GHz}
\setsymbol{LFI:center:frequency:LFI28:Rad:S:units}{28.2\GHz}

\setsymbol{LFI:center:frequency:LFI18:Rad:M}{71.7}
\setsymbol{LFI:center:frequency:LFI19:Rad:M}{67.5}
\setsymbol{LFI:center:frequency:LFI20:Rad:M}{69.2}
\setsymbol{LFI:center:frequency:LFI21:Rad:M}{70.4}
\setsymbol{LFI:center:frequency:LFI22:Rad:M}{71.5}
\setsymbol{LFI:center:frequency:LFI23:Rad:M}{70.8}
\setsymbol{LFI:center:frequency:LFI24:Rad:M}{44.4}
\setsymbol{LFI:center:frequency:LFI25:Rad:M}{44.0}
\setsymbol{LFI:center:frequency:LFI26:Rad:M}{43.9}
\setsymbol{LFI:center:frequency:LFI27:Rad:M}{28.3}
\setsymbol{LFI:center:frequency:LFI28:Rad:M}{28.8}
\setsymbol{LFI:center:frequency:LFI18:Rad:S}{70.1}
\setsymbol{LFI:center:frequency:LFI19:Rad:S}{69.6}
\setsymbol{LFI:center:frequency:LFI20:Rad:S}{69.5}
\setsymbol{LFI:center:frequency:LFI21:Rad:S}{69.5}
\setsymbol{LFI:center:frequency:LFI22:Rad:S}{72.8}
\setsymbol{LFI:center:frequency:LFI23:Rad:S}{71.3}
\setsymbol{LFI:center:frequency:LFI24:Rad:S}{44.1}
\setsymbol{LFI:center:frequency:LFI25:Rad:S}{44.1}
\setsymbol{LFI:center:frequency:LFI26:Rad:S}{44.1}
\setsymbol{LFI:center:frequency:LFI27:Rad:S}{28.5}
\setsymbol{LFI:center:frequency:LFI28:Rad:S}{28.2}

\setsymbol{LFI:white:noise:sensitivity:70GHz:units}{134.7\muKs}
\setsymbol{LFI:white:noise:sensitivity:44GHz:units}{164.7\muKs}
\setsymbol{LFI:white:noise:sensitivity:30GHz:units}{143.4\muKs}

\setsymbol{LFI:white:noise:sensitivity:70GHz}{134.7}
\setsymbol{LFI:white:noise:sensitivity:44GHz}{164.7}
\setsymbol{LFI:white:noise:sensitivity:30GHz}{143.4}

\setsymbol{LFI:white:noise:sensitivity:LFI18:Rad:M:units}{512.0\muKs}
\setsymbol{LFI:white:noise:sensitivity:LFI19:Rad:M:units}{581.4\muKs}
\setsymbol{LFI:white:noise:sensitivity:LFI20:Rad:M:units}{590.8\muKs}
\setsymbol{LFI:white:noise:sensitivity:LFI21:Rad:M:units}{455.2\muKs}
\setsymbol{LFI:white:noise:sensitivity:LFI22:Rad:M:units}{492.0\muKs}
\setsymbol{LFI:white:noise:sensitivity:LFI23:Rad:M:units}{507.7\muKs}
\setsymbol{LFI:white:noise:sensitivity:LFI24:Rad:M:units}{462.2\muKs}
\setsymbol{LFI:white:noise:sensitivity:LFI25:Rad:M:units}{413.6\muKs}
\setsymbol{LFI:white:noise:sensitivity:LFI26:Rad:M:units}{478.6\muKs}
\setsymbol{LFI:white:noise:sensitivity:LFI27:Rad:M:units}{277.7\muKs}
\setsymbol{LFI:white:noise:sensitivity:LFI28:Rad:M:units}{312.3\muKs}
\setsymbol{LFI:white:noise:sensitivity:LFI18:Rad:S:units}{465.7\muKs}
\setsymbol{LFI:white:noise:sensitivity:LFI19:Rad:S:units}{555.6\muKs}
\setsymbol{LFI:white:noise:sensitivity:LFI20:Rad:S:units}{623.2\muKs}
\setsymbol{LFI:white:noise:sensitivity:LFI21:Rad:S:units}{564.1\muKs}
\setsymbol{LFI:white:noise:sensitivity:LFI22:Rad:S:units}{534.4\muKs}
\setsymbol{LFI:white:noise:sensitivity:LFI23:Rad:S:units}{542.4\muKs}
\setsymbol{LFI:white:noise:sensitivity:LFI24:Rad:S:units}{399.2\muKs}
\setsymbol{LFI:white:noise:sensitivity:LFI25:Rad:S:units}{392.6\muKs}
\setsymbol{LFI:white:noise:sensitivity:LFI26:Rad:S:units}{418.6\muKs}
\setsymbol{LFI:white:noise:sensitivity:LFI27:Rad:S:units}{302.9\muKs}
\setsymbol{LFI:white:noise:sensitivity:LFI28:Rad:S:units}{285.3\muKs}

\setsymbol{LFI:white:noise:sensitivity:LFI18:Rad:M}{512.0}
\setsymbol{LFI:white:noise:sensitivity:LFI19:Rad:M}{581.4}
\setsymbol{LFI:white:noise:sensitivity:LFI20:Rad:M}{590.8}
\setsymbol{LFI:white:noise:sensitivity:LFI21:Rad:M}{455.2}
\setsymbol{LFI:white:noise:sensitivity:LFI22:Rad:M}{492.0}
\setsymbol{LFI:white:noise:sensitivity:LFI23:Rad:M}{507.7}
\setsymbol{LFI:white:noise:sensitivity:LFI24:Rad:M}{462.2}
\setsymbol{LFI:white:noise:sensitivity:LFI25:Rad:M}{413.6}
\setsymbol{LFI:white:noise:sensitivity:LFI26:Rad:M}{478.6}
\setsymbol{LFI:white:noise:sensitivity:LFI27:Rad:M}{277.7}
\setsymbol{LFI:white:noise:sensitivity:LFI28:Rad:M}{312.3}
\setsymbol{LFI:white:noise:sensitivity:LFI18:Rad:S}{465.7}
\setsymbol{LFI:white:noise:sensitivity:LFI19:Rad:S}{555.6}
\setsymbol{LFI:white:noise:sensitivity:LFI20:Rad:S}{623.2}
\setsymbol{LFI:white:noise:sensitivity:LFI21:Rad:S}{564.1}
\setsymbol{LFI:white:noise:sensitivity:LFI22:Rad:S}{534.4}
\setsymbol{LFI:white:noise:sensitivity:LFI23:Rad:S}{542.4}
\setsymbol{LFI:white:noise:sensitivity:LFI24:Rad:S}{399.2}
\setsymbol{LFI:white:noise:sensitivity:LFI25:Rad:S}{392.6}
\setsymbol{LFI:white:noise:sensitivity:LFI26:Rad:S}{418.6}
\setsymbol{LFI:white:noise:sensitivity:LFI27:Rad:S}{302.9}
\setsymbol{LFI:white:noise:sensitivity:LFI28:Rad:S}{285.3}

\setsymbol{LFI:knee:frequency:70GHz:units}{29.5\mHz}
\setsymbol{LFI:knee:frequency:44GHz:units}{56.2\mHz}
\setsymbol{LFI:knee:frequency:30GHz:units}{113.7\mHz}

\setsymbol{LFI:knee:frequency:70GHz}{29.5}
\setsymbol{LFI:knee:frequency:44GHz}{56.2}
\setsymbol{LFI:knee:frequency:30GHz}{113.7}

\setsymbol{LFI:knee:frequency:LFI18:Rad:M:units}{16.3\mHz}
\setsymbol{LFI:knee:frequency:LFI19:Rad:M:units}{15.1\mHz}
\setsymbol{LFI:knee:frequency:LFI20:Rad:M:units}{18.7\mHz}
\setsymbol{LFI:knee:frequency:LFI21:Rad:M:units}{37.2\mHz}
\setsymbol{LFI:knee:frequency:LFI22:Rad:M:units}{12.7\mHz}
\setsymbol{LFI:knee:frequency:LFI23:Rad:M:units}{34.6\mHz}
\setsymbol{LFI:knee:frequency:LFI24:Rad:M:units}{46.2\mHz}
\setsymbol{LFI:knee:frequency:LFI25:Rad:M:units}{24.9\mHz}
\setsymbol{LFI:knee:frequency:LFI26:Rad:M:units}{67.6\mHz}
\setsymbol{LFI:knee:frequency:LFI27:Rad:M:units}{187.4\mHz}
\setsymbol{LFI:knee:frequency:LFI28:Rad:M:units}{122.2\mHz}
\setsymbol{LFI:knee:frequency:LFI18:Rad:S:units}{17.7\mHz}
\setsymbol{LFI:knee:frequency:LFI19:Rad:S:units}{22.0\mHz}
\setsymbol{LFI:knee:frequency:LFI20:Rad:S:units}{8.7\mHz}
\setsymbol{LFI:knee:frequency:LFI21:Rad:S:units}{25.9\mHz}
\setsymbol{LFI:knee:frequency:LFI22:Rad:S:units}{15.8\mHz}
\setsymbol{LFI:knee:frequency:LFI23:Rad:S:units}{129.8\mHz}
\setsymbol{LFI:knee:frequency:LFI24:Rad:S:units}{100.9\mHz}
\setsymbol{LFI:knee:frequency:LFI25:Rad:S:units}{38.9\mHz}
\setsymbol{LFI:knee:frequency:LFI26:Rad:S:units}{58.9\mHz}
\setsymbol{LFI:knee:frequency:LFI27:Rad:S:units}{104.4\mHz}
\setsymbol{LFI:knee:frequency:LFI28:Rad:S:units}{40.7\mHz}

\setsymbol{LFI:knee:frequency:LFI18:Rad:M}{16.3}
\setsymbol{LFI:knee:frequency:LFI19:Rad:M}{15.1}
\setsymbol{LFI:knee:frequency:LFI20:Rad:M}{18.7}
\setsymbol{LFI:knee:frequency:LFI21:Rad:M}{37.2}
\setsymbol{LFI:knee:frequency:LFI22:Rad:M}{12.7}
\setsymbol{LFI:knee:frequency:LFI23:Rad:M}{34.6}
\setsymbol{LFI:knee:frequency:LFI24:Rad:M}{46.2}
\setsymbol{LFI:knee:frequency:LFI25:Rad:M}{24.9}
\setsymbol{LFI:knee:frequency:LFI26:Rad:M}{67.6}
\setsymbol{LFI:knee:frequency:LFI27:Rad:M}{187.4}
\setsymbol{LFI:knee:frequency:LFI28:Rad:M}{122.2}
\setsymbol{LFI:knee:frequency:LFI18:Rad:S}{17.7}
\setsymbol{LFI:knee:frequency:LFI19:Rad:S}{22.0}
\setsymbol{LFI:knee:frequency:LFI20:Rad:S}{8.7}
\setsymbol{LFI:knee:frequency:LFI21:Rad:S}{25.9}
\setsymbol{LFI:knee:frequency:LFI22:Rad:S}{15.8}
\setsymbol{LFI:knee:frequency:LFI23:Rad:S}{129.8}
\setsymbol{LFI:knee:frequency:LFI24:Rad:S}{100.9}
\setsymbol{LFI:knee:frequency:LFI25:Rad:S}{38.9}
\setsymbol{LFI:knee:frequency:LFI26:Rad:S}{58.9}
\setsymbol{LFI:knee:frequency:LFI27:Rad:S}{104.4}
\setsymbol{LFI:knee:frequency:LFI28:Rad:S}{40.7}

\setsymbol{LFI:slope:70GHz:units}{$-1.03$\mHz}
\setsymbol{LFI:slope:44GHz:units}{$-0.89$\mHz}
\setsymbol{LFI:slope:30GHz:units}{$-0.87$\mHz}

\setsymbol{LFI:slope:70GHz}{$-1.03$}
\setsymbol{LFI:slope:44GHz}{$-0.89$}
\setsymbol{LFI:slope:30GHz}{$-0.87$}

\setsymbol{LFI:slope:LFI18:Rad:M:units}{$-1.04$\mHz}
\setsymbol{LFI:slope:LFI19:Rad:M:units}{$-1.09$\mHz}
\setsymbol{LFI:slope:LFI20:Rad:M:units}{$-0.69$\mHz}
\setsymbol{LFI:slope:LFI21:Rad:M:units}{$-1.56$\mHz}
\setsymbol{LFI:slope:LFI22:Rad:M:units}{$-1.01$\mHz}
\setsymbol{LFI:slope:LFI23:Rad:M:units}{$-0.96$\mHz}
\setsymbol{LFI:slope:LFI24:Rad:M:units}{$-0.83$\mHz}
\setsymbol{LFI:slope:LFI25:Rad:M:units}{$-0.91$\mHz}
\setsymbol{LFI:slope:LFI26:Rad:M:units}{$-0.95$\mHz}
\setsymbol{LFI:slope:LFI27:Rad:M:units}{$-0.87$\mHz}
\setsymbol{LFI:slope:LFI28:Rad:M:units}{$-0.88$\mHz}
\setsymbol{LFI:slope:LFI18:Rad:S:units}{$-1.15$\mHz}
\setsymbol{LFI:slope:LFI19:Rad:S:units}{$-1.00$\mHz}
\setsymbol{LFI:slope:LFI20:Rad:S:units}{$-0.95$\mHz}
\setsymbol{LFI:slope:LFI21:Rad:S:units}{$-0.92$\mHz}
\setsymbol{LFI:slope:LFI22:Rad:S:units}{$-1.01$\mHz}
\setsymbol{LFI:slope:LFI23:Rad:S:units}{$-0.95$\mHz}
\setsymbol{LFI:slope:LFI24:Rad:S:units}{$-0.73$\mHz}
\setsymbol{LFI:slope:LFI25:Rad:S:units}{$-1.16$\mHz}
\setsymbol{LFI:slope:LFI26:Rad:S:units}{$-0.79$\mHz}
\setsymbol{LFI:slope:LFI27:Rad:S:units}{$-0.82$\mHz}
\setsymbol{LFI:slope:LFI28:Rad:S:units}{$-0.91$\mHz}

\setsymbol{LFI:slope:LFI18:Rad:M}{$-1.04$}
\setsymbol{LFI:slope:LFI19:Rad:M}{$-1.09$}
\setsymbol{LFI:slope:LFI20:Rad:M}{$-0.69$}
\setsymbol{LFI:slope:LFI21:Rad:M}{$-1.56$}
\setsymbol{LFI:slope:LFI22:Rad:M}{$-1.01$}
\setsymbol{LFI:slope:LFI23:Rad:M}{$-0.96$}
\setsymbol{LFI:slope:LFI24:Rad:M}{$-0.83$}
\setsymbol{LFI:slope:LFI25:Rad:M}{$-0.91$}
\setsymbol{LFI:slope:LFI26:Rad:M}{$-0.95$}
\setsymbol{LFI:slope:LFI27:Rad:M}{$-0.87$}
\setsymbol{LFI:slope:LFI28:Rad:M}{$-0.88$}
\setsymbol{LFI:slope:LFI18:Rad:S}{$-1.15$}
\setsymbol{LFI:slope:LFI19:Rad:S}{$-1.00$}
\setsymbol{LFI:slope:LFI20:Rad:S}{$-0.95$}
\setsymbol{LFI:slope:LFI21:Rad:S}{$-0.92$}
\setsymbol{LFI:slope:LFI22:Rad:S}{$-1.01$}
\setsymbol{LFI:slope:LFI23:Rad:S}{$-0.95$}
\setsymbol{LFI:slope:LFI24:Rad:S}{$-0.73$}
\setsymbol{LFI:slope:LFI25:Rad:S}{$-1.16$}
\setsymbol{LFI:slope:LFI26:Rad:S}{$-0.79$}
\setsymbol{LFI:slope:LFI27:Rad:S}{$-0.82$}
\setsymbol{LFI:slope:LFI28:Rad:S}{$-0.91$}

\setsymbol{LFI:FWHM:70GHz:units}{13\parcm01}
\setsymbol{LFI:FWHM:44GHz:units}{27\parcm92}
\setsymbol{LFI:FWHM:30GHz:units}{32\parcm65}

\setsymbol{LFI:FWHM:70GHz}{13.01}
\setsymbol{LFI:FWHM:44GHz}{27.92}
\setsymbol{LFI:FWHM:30GHz}{32.65}

\setsymbol{LFI:FWHM:LFI18:units}{13\parcm39}
\setsymbol{LFI:FWHM:LFI19:units}{13\parcm01}
\setsymbol{LFI:FWHM:LFI20:units}{12\parcm75}
\setsymbol{LFI:FWHM:LFI21:units}{12\parcm74}
\setsymbol{LFI:FWHM:LFI22:units}{12\parcm87}
\setsymbol{LFI:FWHM:LFI23:units}{13\parcm27}
\setsymbol{LFI:FWHM:LFI24:units}{22\parcm98}
\setsymbol{LFI:FWHM:LFI25:units}{30\parcm46}
\setsymbol{LFI:FWHM:LFI26:units}{30\parcm31}
\setsymbol{LFI:FWHM:LFI27:units}{32\parcm65}
\setsymbol{LFI:FWHM:LFI28:units}{32\parcm66}

\setsymbol{LFI:FWHM:LFI18}{13.39}
\setsymbol{LFI:FWHM:LFI19}{13.01}
\setsymbol{LFI:FWHM:LFI20}{12.75}
\setsymbol{LFI:FWHM:LFI21}{12.74}
\setsymbol{LFI:FWHM:LFI22}{12.87}
\setsymbol{LFI:FWHM:LFI23}{13.27}
\setsymbol{LFI:FWHM:LFI24}{22.98}
\setsymbol{LFI:FWHM:LFI25}{30.46}
\setsymbol{LFI:FWHM:LFI26}{30.31}
\setsymbol{LFI:FWHM:LFI27}{32.65}
\setsymbol{LFI:FWHM:LFI28}{32.66}

\setsymbol{LFI:FWHM:uncertainty:LFI18:units}{0.170\arcm}
\setsymbol{LFI:FWHM:uncertainty:LFI19:units}{0.174\arcm}
\setsymbol{LFI:FWHM:uncertainty:LFI20:units}{0.170\arcm}
\setsymbol{LFI:FWHM:uncertainty:LFI21:units}{0.156\arcm}
\setsymbol{LFI:FWHM:uncertainty:LFI22:units}{0.164\arcm}
\setsymbol{LFI:FWHM:uncertainty:LFI23:units}{0.171\arcm}
\setsymbol{LFI:FWHM:uncertainty:LFI24:units}{0.652\arcm}
\setsymbol{LFI:FWHM:uncertainty:LFI25:units}{1.075\arcm}
\setsymbol{LFI:FWHM:uncertainty:LFI26:units}{1.131\arcm}
\setsymbol{LFI:FWHM:uncertainty:LFI27:units}{1.266\arcm}
\setsymbol{LFI:FWHM:uncertainty:LFI28:units}{1.287\arcm}

\setsymbol{LFI:FWHM:uncertainty:LFI18}{0.170}
\setsymbol{LFI:FWHM:uncertainty:LFI19}{0.174}
\setsymbol{LFI:FWHM:uncertainty:LFI20}{0.170}
\setsymbol{LFI:FWHM:uncertainty:LFI21}{0.156}
\setsymbol{LFI:FWHM:uncertainty:LFI22}{0.164}
\setsymbol{LFI:FWHM:uncertainty:LFI23}{0.171}
\setsymbol{LFI:FWHM:uncertainty:LFI24}{0.652}
\setsymbol{LFI:FWHM:uncertainty:LFI25}{1.075}
\setsymbol{LFI:FWHM:uncertainty:LFI26}{1.131}
\setsymbol{LFI:FWHM:uncertainty:LFI27}{1.266}
\setsymbol{LFI:FWHM:uncertainty:LFI28}{1.287}

\setsymbol{HFI:center:frequency:100GHz:units}{100\,GHz}
\setsymbol{HFI:center:frequency:143GHz:units}{143\,GHz}
\setsymbol{HFI:center:frequency:217GHz:units}{217\,GHz}
\setsymbol{HFI:center:frequency:353GHz:units}{353\,GHz}
\setsymbol{HFI:center:frequency:545GHz:units}{545\,GHz}
\setsymbol{HFI:center:frequency:857GHz:units}{857\,GHz}

\setsymbol{HFI:center:frequency:100GHz}{100}
\setsymbol{HFI:center:frequency:143GHz}{143}
\setsymbol{HFI:center:frequency:217GHz}{217}
\setsymbol{HFI:center:frequency:353GHz}{353}
\setsymbol{HFI:center:frequency:545GHz}{545}
\setsymbol{HFI:center:frequency:857GHz}{857}

\setsymbol{HFI:Ndetectors:100GHz}{8}
\setsymbol{HFI:Ndetectors:143GHz}{11}
\setsymbol{HFI:Ndetectors:217GHz}{12}
\setsymbol{HFI:Ndetectors:353GHz}{12}
\setsymbol{HFI:Ndetectors:545GHz}{3}
\setsymbol{HFI:Ndetectors:857GHz}{4}

\setsymbol{HFI:FWHM:Maps:100GHz:units}{9\parcm88}
\setsymbol{HFI:FWHM:Maps:143GHz:units}{7\parcm18}
\setsymbol{HFI:FWHM:Maps:217GHz:units}{4\parcm87}
\setsymbol{HFI:FWHM:Maps:353GHz:units}{4\parcm65}
\setsymbol{HFI:FWHM:Maps:545GHz:units}{4\parcm72}
\setsymbol{HFI:FWHM:Maps:857GHz:units}{4\parcm39}
\setsymbol{HFI:FWHM:Maps:100GHz}{9.88}
\setsymbol{HFI:FWHM:Maps:143GHz}{7.18}
\setsymbol{HFI:FWHM:Maps:217GHz}{4.87}
\setsymbol{HFI:FWHM:Maps:353GHz}{4.65}
\setsymbol{HFI:FWHM:Maps:545GHz}{4.72}
\setsymbol{HFI:FWHM:Maps:857GHz}{4.39}

\setsymbol{HFI:beam:ellipticity:Maps:100GHz}{1.15}
\setsymbol{HFI:beam:ellipticity:Maps:143GHz}{1.01}
\setsymbol{HFI:beam:ellipticity:Maps:217GHz}{1.06}
\setsymbol{HFI:beam:ellipticity:Maps:353GHz}{1.05}
\setsymbol{HFI:beam:ellipticity:Maps:545GHz}{1.14}
\setsymbol{HFI:beam:ellipticity:Maps:857GHz}{1.19}

\setsymbol{HFI:FWHM:Mars:100GHz:units}{9\parcm37}
\setsymbol{HFI:FWHM:Mars:143GHz:units}{7\parcm04}
\setsymbol{HFI:FWHM:Mars:217GHz:units}{4\parcm68}
\setsymbol{HFI:FWHM:Mars:353GHz:units}{4\parcm43}
\setsymbol{HFI:FWHM:Mars:545GHz:units}{3\parcm80}
\setsymbol{HFI:FWHM:Mars:857GHz:units}{3\parcm67}

\setsymbol{HFI:FWHM:Mars:100GHz}{9.37}
\setsymbol{HFI:FWHM:Mars:143GHz}{7.04}
\setsymbol{HFI:FWHM:Mars:217GHz}{4.68}
\setsymbol{HFI:FWHM:Mars:353GHz}{4.43}
\setsymbol{HFI:FWHM:Mars:545GHz}{3.80}
\setsymbol{HFI:FWHM:Mars:857GHz}{3.67}

\setsymbol{HFI:beam:ellipticity:Mars:100GHz}{1.18}
\setsymbol{HFI:beam:ellipticity:Mars:143GHz}{1.03}
\setsymbol{HFI:beam:ellipticity:Mars:217GHz}{1.14}
\setsymbol{HFI:beam:ellipticity:Mars:353GHz}{1.09}
\setsymbol{HFI:beam:ellipticity:Mars:545GHz}{1.25}
\setsymbol{HFI:beam:ellipticity:Mars:857GHz}{1.03}

\setsymbol{HFI:CMB:relative:calibration:100GHz}{$\lsim 1\%$}
\setsymbol{HFI:CMB:relative:calibration:143GHz}{$\lsim 1\%$}
\setsymbol{HFI:CMB:relative:calibration:217GHz}{$\lsim 1\%$}
\setsymbol{HFI:CMB:relative:calibration:353GHz}{$\lsim 1\%$}
\setsymbol{HFI:CMB:relative:calibration:545GHz}{}
\setsymbol{HFI:CMB:relative:calibration:857GHz}{}

\setsymbol{HFI:CMB:absolute:calibration:100GHz}{$\lsim 2\%$}
\setsymbol{HFI:CMB:absolute:calibration:143GHz}{$\lsim 2\%$}
\setsymbol{HFI:CMB:absolute:calibration:217GHz}{$\lsim 2\%$}
\setsymbol{HFI:CMB:absolute:calibration:353GHz}{$\lsim 2\%$}
\setsymbol{HFI:CMB:absolute:calibration:545GHz}{}
\setsymbol{HFI:CMB:absolute:calibration:857GHz}{}

\setsymbol{HFI:FIRAS:gain:calibration:accuracy:statistical:100GHz}{}
\setsymbol{HFI:FIRAS:gain:calibration:accuracy:statistical:143GHz}{}
\setsymbol{HFI:FIRAS:gain:calibration:accuracy:statistical:217GHz}{}
\setsymbol{HFI:FIRAS:gain:calibration:accuracy:statistical:353GHz}{2.5\%}
\setsymbol{HFI:FIRAS:gain:calibration:accuracy:statistical:545GHz}{1\%}
\setsymbol{HFI:FIRAS:gain:calibration:accuracy:statistical:857GHz}{0.5\%}

\setsymbol{HFI:FIRAS:gain:calibration:accuracy:systematic:100GHz}{}
\setsymbol{HFI:FIRAS:gain:calibration:accuracy:systematic:143GHz}{}
\setsymbol{HFI:FIRAS:gain:calibration:accuracy:systematic:217GHz}{}
\setsymbol{HFI:FIRAS:gain:calibration:accuracy:systematic:353GHz}{}
\setsymbol{HFI:FIRAS:gain:calibration:accuracy:systematic:545GHz}{7\%}
\setsymbol{HFI:FIRAS:gain:calibration:accuracy:systematic:857GHz}{7\%}

\setsymbol{HFI:FIRAS:zero:point:accuracy:100GHz:units}{0.8\MJysr}
\setsymbol{HFI:FIRAS:zero:point:accuracy:143GHz:units}{}
\setsymbol{HFI:FIRAS:zero:point:accuracy:217GHz:units}{}
\setsymbol{HFI:FIRAS:zero:point:accuracy:353GHz:units}{1.4\MJysr}
\setsymbol{HFI:FIRAS:zero:point:accuracy:545GHz:units}{2.2\MJysr}
\setsymbol{HFI:FIRAS:zero:point:accuracy:857GHz:units}{1.7\MJysr}

\setsymbol{HFI:FIRAS:zero:point:accuracy:100GHz}{0.8}
\setsymbol{HFI:FIRAS:zero:point:accuracy:143GHz}{}
\setsymbol{HFI:FIRAS:zero:point:accuracy:217GHz}{}
\setsymbol{HFI:FIRAS:zero:point:accuracy:353GHz}{1.4}
\setsymbol{HFI:FIRAS:zero:point:accuracy:545GHz}{2.2}
\setsymbol{HFI:FIRAS:zero:point:accuracy:857GHz}{1.7}

\setsymbol{HFI:unit:conversion:100GHz:units}{0.2415\MJysrmK}
\setsymbol{HFI:unit:conversion:143GHz:units}{0.3694\MJysrmK}
\setsymbol{HFI:unit:conversion:217GHz:units}{0.4811\MJysrmK}
\setsymbol{HFI:unit:conversion:353GHz:units}{0.2883\MJysrmK}
\setsymbol{HFI:unit:conversion:545GHz:units}{0.05826\MJysrmK}
\setsymbol{HFI:unit:conversion:857GHz:units}{0.002238\MJysrmK}

\setsymbol{HFI:unit:conversion:100GHz}{0.2415}
\setsymbol{HFI:unit:conversion:143GHz}{0.3694}
\setsymbol{HFI:unit:conversion:217GHz}{0.4811}
\setsymbol{HFI:unit:conversion:353GHz}{0.2883}
\setsymbol{HFI:unit:conversion:545GHz}{0.05826}
\setsymbol{HFI:unit:conversion:857GHz}{0.002238}

\setsymbol{HFI:colour:correction:alpha=-2:V1.01:100GHz}{0.9893}
\setsymbol{HFI:colour:correction:alpha=-2:V1.01:143GHz}{0.9759}
\setsymbol{HFI:colour:correction:alpha=-2:V1.01:217GHz}{1.0007}
\setsymbol{HFI:colour:correction:alpha=-2:V1.01:353GHz}{1.0028}
\setsymbol{HFI:colour:correction:alpha=-2:V1.01:545GHz}{1.0019}
\setsymbol{HFI:colour:correction:alpha=-2:V1.01:857GHz}{0.9889}

\setsymbol{HFI:colour:correction:alpha=0:V1.01:100GHz}{1.0008}
\setsymbol{HFI:colour:correction:alpha=0:V1.01:143GHz}{1.0148}
\setsymbol{HFI:colour:correction:alpha=0:V1.01:217GHz}{0.9909}
\setsymbol{HFI:colour:correction:alpha=0:V1.01:353GHz}{0.9888}
\setsymbol{HFI:colour:correction:alpha=0:V1.01:545GHz}{0.9878}
\setsymbol{HFI:colour:correction:alpha=0:V1.01:857GHz}{1.0014}

\providecommand{\sorthelp}[1]{}

\usepackage{graphicx,amsmath}
\usepackage{epstopdf}
\usepackage{epsf,color}
\usepackage{txfonts}
\usepackage[colorlinks=true,linkcolor=red,citecolor=blue]{hyperref}
\usepackage{natbib}
\usepackage{multirow}
\usepackage{url}
\usepackage{fixltx2e}
\usepackage{float}
\bibpunct{(}{)}{;}{a}{}{,} % to follow the A&A style

\newcommand{\hi}{\ion{H}{i}}
\newcommand{\hii}{\ion{H}{ii}}
\newcommand{\healpix}{HEALPix}
\newcommand{\planck}{\textit{Planck}}  
\newcommand{\DIRBE}{{\it DIRBE}}

\newcommand{\wmap}{\textit{WMAP}}  
\newcommand{\COBE}{{\it COBE}}
\def\NH{\ifmmode N_{\rm HI}\else $N_{\rm HI}$\fi}
\newcommand{\vtHI}{{\hat I_{\rm HI}}}
\newcommand{\vnT}{{\hat T_{\nu}}}

\def\ben{\begin{enumerate}}
\def\een{\end{enumerate}}
\def\bi{\begin{itemize}}
\def\ei{\end{itemize}}
\def\be{\begin{equation}}
\def\ee{\end{equation}}
\def\bea{\begin{eqnarray}}
\def\eea{\end{eqnarray}}

\def\MJH{\ifmmode {\rm MJy}\,{\rm sr}\mo\,(10^{20}\,{\rm H}\,{\rm cm}^{-2})\mo
         \else MJy\,sr\mo\,$(10^{20}\,{\rm H}\,{\rm cm}^{-2})\mo$\fi}
\def\microKH{\ifmmode {\rm \muK_{\rm RJ}}\,(10^{20}\,{\rm H}\,{\rm cm}^{-2})\mo
         \else $\mu{\rm K_{\rm RJ}}\,(10^{20}\,{\rm H}\,{\rm cm}^{-2})\mo$\fi}
\begin{document}

%==============================
\title{Planck intermediate results. XVI. Emission of dust in the diffuse interstellar medium from the far-infrared to microwave frequencies}
%==============================
%\thanks{Corresponding author:  francois.boulanger@ias.u-psud.fr}

%This author list corresponds to \title{Author list for SVN PIP\_82\_Proj\_7\_12\_Boulanger: Emission of dust in the diffuse interstellar medium from the far-infrared to microwave frequencies}
%Prepared by R. Leonardi (rleonardi@sciops.esa.int), ESAC/ESA
%This version is from Thu Nov 14 15:10:47 2013 CET
%\subtitle{There are 196 co-authors in this list}
\author{\small
Planck Collaboration:
A.~Abergel\inst{56}
\and
P.~A.~R.~Ade\inst{79}
\and
N.~Aghanim\inst{56}
%\and
%D.~Alina\inst{11, 81}
\and
M.~I.~R.~Alves\inst{56}
\and
G.~Aniano\inst{56}
\and
M.~Arnaud\inst{68}
\and
M.~Ashdown\inst{65, 7}
\and
J.~Aumont\inst{56}
\and
C.~Baccigalupi\inst{78}
\and
A.~J.~Banday\inst{81, 11}
% ...
\thanks{Corresponding author:  francois.boulanger@ias.u-psud.fr}
}
\institute{\small
APC, AstroParticule et Cosmologie, Universit\'{e} Paris Diderot, CNRS/IN2P3, CEA/lrfu, Observatoire de Paris, Sorbonne Paris Cit\'{e}, 10, rue Alice Domon et L\'{e}onie Duquet, 75205 Paris Cedex 13, France\\
\and
Aalto University Mets\"{a}hovi Radio Observatory and Dept of Radio Science and Engineering, P.O. Box 13000, FI-00076 AALTO, Finland\\
\and
African Institute for Mathematical Sciences, 6-8 Melrose Road, Muizenberg, Cape Town, South Africa\\
\and
Agenzia Spaziale Italiana Science Data Center, Via del Politecnico snc, 00133, Roma, Italy\\
\and
Agenzia Spaziale Italiana, Viale Liegi 26, Roma, Italy\\
\and
Argelander-Institut f\"{u}r Astronomie, Universit\"{a}t Bonn, Auf dem H\"{u}gel 71, D-53121 Bonn, Germany\\
\and
Astrophysics Group, Cavendish Laboratory, University of Cambridge, J J Thomson Avenue, Cambridge CB3 0HE, U.K.\\
\and
Astrophysics \& Cosmology Research Unit, School of Mathematics, Statistics \& Computer Science, University of KwaZulu-Natal, Westville Campus, Private Bag X54001, Durban 4000, South Africa\\
\and
Atacama Large Millimeter/submillimeter Array, ALMA Santiago Central Offices, Alonso de Cordova 3107, Vitacura, Casilla 763 0355, Santiago, Chile\\
\and
CITA, University of Toronto, 60 St. George St., Toronto, ON M5S 3H8, Canada\\
\and
CNRS, IRAP, 9 Av. colonel Roche, BP 44346, F-31028 Toulouse cedex 4, France\\
\and
California Institute of Technology, Pasadena, California, U.S.A.\\
\and
Centro de Estudios de F\'{i}sica del Cosmos de Arag\'{o}n (CEFCA), Plaza San Juan, 1, planta 2, E-44001, Teruel, Spain\\
\and
Computational Cosmology Center, Lawrence Berkeley National Laboratory, Berkeley, California, U.S.A.\\
\and
Consejo Superior de Investigaciones Cient\'{\i}ficas (CSIC), Madrid, Spain\\
\and
DSM/Irfu/SPP, CEA-Saclay, F-91191 Gif-sur-Yvette Cedex, France\\
\and
DTU Space, National Space Institute, Technical University of Denmark, Elektrovej 327, DK-2800 Kgs. Lyngby, Denmark\\
\and
D\'{e}partement de Physique Th\'{e}orique, Universit\'{e} de Gen\`{e}ve, 24, Quai E. Ansermet,1211 Gen\`{e}ve 4, Switzerland\\
\and
D\'{e}partement de physique, de g\'{e}nie physique et d'optique, Universit\'{e} Laval, Qu\'{e}bec, Canada\\
\and
Departamento de F\'{\i}sica, Universidad de Oviedo, Avda. Calvo Sotelo s/n, Oviedo, Spain\\
\and
Department of Astrophysics/IMAPP, Radboud University Nijmegen, P.O. Box 9010, 6500 GL Nijmegen, The Netherlands\\
\and
Department of Electrical Engineering and Computer Sciences, University of California, Berkeley, California, U.S.A.\\
\and
Department of Physics and Astronomy, Dana and David Dornsife College of Letter, Arts and Sciences, University of Southern California, Los Angeles, CA 90089, U.S.A.\\
\and
Department of Physics and Astronomy, University College London, London WC1E 6BT, U.K.\\
\and
Department of Physics, Florida State University, Keen Physics Building, 77 Chieftan Way, Tallahassee, Florida, U.S.A.\\
\and
Department of Physics, Gustaf H\"{a}llstr\"{o}min katu 2a, University of Helsinki, Helsinki, Finland\\
\and
Department of Physics, Princeton University, Princeton, New Jersey, U.S.A.\\
\and
Department of Physics, University of California, Santa Barbara, California, U.S.A.\\
\and
Department of Physics, University of Illinois at Urbana-Champaign, 1110 West Green Street, Urbana, Illinois, U.S.A.\\
\and
Dipartimento di Fisica e Astronomia G. Galilei, Universit\`{a} degli Studi di Padova, via Marzolo 8, 35131 Padova, Italy\\
\and
Dipartimento di Fisica e Scienze della Terra, Universit\`{a} di Ferrara, Via Saragat 1, 44122 Ferrara, Italy\\
\and
Dipartimento di Fisica, Universit\`{a} La Sapienza, P. le A. Moro 2, Roma, Italy\\
\and
Dipartimento di Fisica, Universit\`{a} degli Studi di Milano, Via Celoria, 16, Milano, Italy\\
\and
Dipartimento di Fisica, Universit\`{a} degli Studi di Trieste, via A. Valerio 2, Trieste, Italy\\
\and
Dipartimento di Fisica, Universit\`{a} di Roma Tor Vergata, Via della Ricerca Scientifica, 1, Roma, Italy\\
\and
Discovery Center, Niels Bohr Institute, Blegdamsvej 17, Copenhagen, Denmark\\
\and
Dpto. Astrof\'{i}sica, Universidad de La Laguna (ULL), E-38206 La Laguna, Tenerife, Spain\\
\and
European Southern Observatory, ESO Vitacura, Alonso de Cordova 3107, Vitacura, Casilla 19001, Santiago, Chile\\
\and
European Space Agency, ESAC, Planck Science Office, Camino bajo del Castillo, s/n, Urbanizaci\'{o}n Villafranca del Castillo, Villanueva de la Ca\~{n}ada, Madrid, Spain\\
\and
European Space Agency, ESTEC, Keplerlaan 1, 2201 AZ Noordwijk, The Netherlands\\
\and
Helsinki Institute of Physics, Gustaf H\"{a}llstr\"{o}min katu 2, University of Helsinki, Helsinki, Finland\\
\and
INAF - Osservatorio Astrofisico di Catania, Via S. Sofia 78, Catania, Italy\\
\and
INAF - Osservatorio Astronomico di Padova, Vicolo dell'Osservatorio 5, Padova, Italy\\
\and
INAF - Osservatorio Astronomico di Roma, via di Frascati 33, Monte Porzio Catone, Italy\\
\and
INAF - Osservatorio Astronomico di Trieste, Via G.B. Tiepolo 11, Trieste, Italy\\
\and
INAF Istituto di Radioastronomia, Via P. Gobetti 101, 40129 Bologna, Italy\\
\and
INAF/IASF Bologna, Via Gobetti 101, Bologna, Italy\\
\and
INAF/IASF Milano, Via E. Bassini 15, Milano, Italy\\
\and
INFN, Sezione di Bologna, Via Irnerio 46, I-40126, Bologna, Italy\\
\and
INFN, Sezione di Roma 1, Universit\`{a} di Roma Sapienza, Piazzale Aldo Moro 2, 00185, Roma, Italy\\
\and
IPAG: Institut de Plan\'{e}tologie et d'Astrophysique de Grenoble, Universit\'{e} Joseph Fourier, Grenoble 1 / CNRS-INSU, UMR 5274, Grenoble, F-38041, France\\
\and
IUCAA, Post Bag 4, Ganeshkhind, Pune University Campus, Pune 411 007, India\\
\and
Imperial College London, Astrophysics group, Blackett Laboratory, Prince Consort Road, London, SW7 2AZ, U.K.\\
\and
Infrared Processing and Analysis Center, California Institute of Technology, Pasadena, CA 91125, U.S.A.\\
\and
Institut Universitaire de France, 103, bd Saint-Michel, 75005, Paris, France\\
\and
Institut d'Astrophysique Spatiale, CNRS (UMR8617) Universit\'{e} Paris-Sud 11, B\^{a}timent 121, Orsay, France\\
\and
Institut d'Astrophysique de Paris, CNRS (UMR7095), 98 bis Boulevard Arago, F-75014, Paris, France\\
\and
Institute for Space Sciences, Bucharest-Magurale, Romania\\
\and
Institute of Astronomy, University of Cambridge, Madingley Road, Cambridge CB3 0HA, U.K.\\
\and
Institute of Theoretical Astrophysics, University of Oslo, Blindern, Oslo, Norway\\
\and
Instituto de Astrof\'{\i}sica de Canarias, C/V\'{\i}a L\'{a}ctea s/n, La Laguna, Tenerife, Spain\\
\and
Instituto de F\'{\i}sica de Cantabria (CSIC-Universidad de Cantabria), Avda. de los Castros s/n, Santander, Spain\\
\and
Jet Propulsion Laboratory, California Institute of Technology, 4800 Oak Grove Drive, Pasadena, California, U.S.A.\\
\and
Jodrell Bank Centre for Astrophysics, Alan Turing Building, School of Physics and Astronomy, The University of Manchester, Oxford Road, Manchester, M13 9PL, U.K.\\
\and
Kavli Institute for Cosmology Cambridge, Madingley Road, Cambridge, CB3 0HA, U.K.\\
\and
LAL, Universit\'{e} Paris-Sud, CNRS/IN2P3, Orsay, France\\
\and
LERMA, CNRS, Observatoire de Paris, 61 Avenue de l'Observatoire, Paris, France\\
\and
Laboratoire AIM, IRFU/Service d'Astrophysique - CEA/DSM - CNRS - Universit\'{e} Paris Diderot, B\^{a}t. 709, CEA-Saclay, F-91191 Gif-sur-Yvette Cedex, France\\
\and
Laboratoire Traitement et Communication de l'Information, CNRS (UMR 5141) and T\'{e}l\'{e}com ParisTech, 46 rue Barrault F-75634 Paris Cedex 13, France\\
\and
Laboratoire de Physique Subatomique et de Cosmologie, Universit\'{e} Joseph Fourier Grenoble I, CNRS/IN2P3, Institut National Polytechnique de Grenoble, 53 rue des Martyrs, 38026 Grenoble cedex, France\\
\and
Laboratoire de Physique Th\'{e}orique, Universit\'{e} Paris-Sud 11 \& CNRS, B\^{a}timent 210, 91405 Orsay, France\\
\and
Lawrence Berkeley National Laboratory, Berkeley, California, U.S.A.\\
\and
Max-Planck-Institut f\"{u}r Astrophysik, Karl-Schwarzschild-Str. 1, 85741 Garching, Germany\\
\and
National University of Ireland, Department of Experimental Physics, Maynooth, Co. Kildare, Ireland\\
\and
Niels Bohr Institute, Blegdamsvej 17, Copenhagen, Denmark\\
\and
Observational Cosmology, Mail Stop 367-17, California Institute of Technology, Pasadena, CA, 91125, U.S.A.\\
\and
Optical Science Laboratory, University College London, Gower Street, London, U.K.\\
\and
SISSA, Astrophysics Sector, via Bonomea 265, 34136, Trieste, Italy\\
\and
School of Physics and Astronomy, Cardiff University, Queens Buildings, The Parade, Cardiff, CF24 3AA, U.K.\\
\and
UPMC Univ Paris 06, UMR7095, 98 bis Boulevard Arago, F-75014, Paris, France\\
\and
Universit\'{e} de Toulouse, UPS-OMP, IRAP, F-31028 Toulouse cedex 4, France\\
% ...
}

\date{\today}

% \abstract{}{}{}{}{} 
 
  \abstract{
  %{\it Context:}
The dust-\hi\ correlation is used to characterize the emission
properties of dust in the diffuse interstellar medium (ISM) from far infrared
wavelengths to microwave frequencies.
 The field of this investigation encompasses the part of the southern sky best suited to study the cosmic infrared and microwave backgrounds. %, and the trailing section of the Magellanic Stream. 
%{\it Method:}
We cross-correlate sky maps from \planck, the Wilkinson microwave anisotropy probe (\wmap), and the diffuse infrared background experiment (\DIRBE), at 17
frequencies from 23 to $3000\,$GHz, with the Parkes survey of the $21\,$cm
line emission of neutral atomic hydrogen, over a contiguous area of 7500\,deg$^2$ centred on the southern Galactic pole. 
We present a general methodology to study the dust-\hi\ correlation
over the sky, including simulations to quantify uncertainties.
%{\it Results:}
Our analysis yields four specific results. 
\quad(1) We map the temperature, submillimetre emissivity, and opacity of the dust per H-atom.
The dust temperature is observed to be anti-correlated with the dust emissivity and opacity.  We interpret this result as evidence of dust evolution within the diffuse ISM.
The mean dust opacity is measured to be $(7.1 \pm 0.6) \times 10^{-27} \,{\rm cm^2\,H}^{-1} \times\, (\nu/{\rm 353\,GHz})^{1.53\pm0.03}$ for $100 \le \nu \le 353$\,GHz.  This is a reference value to estimate hydrogen column densities from dust emission at submillimetre and millimetre wavelengths.  
\quad(2) We map the spectral index $\beta_{\rm mm}$ of dust emission at millimetre wavelengths (defined here as $\nu \le 353$\,GHz), and find it to be remarkably constant at $\beta_{\rm mm} = 1.51\pm 0.13$.  We compare it with the far infrared spectral index $\beta_{\rm FIR}$ derived from greybody fits at higher frequencies, and find a systematic difference, $\beta_{\rm mm}-\beta_{\rm FIR} = -0.15$, which suggests that the dust spectral energy distribution (SED) flattens at $\nu \le 353\,$GHz.
\quad(3) We present spectral fits of the microwave emission correlated with \hi\ from 23 to 353\,GHz, which separate dust and 
anomalous microwave emission (AME).  We show that the flattening of the dust SED can be accounted for with an additional component with a blackbody spectrum. This additional component, which accounts for $(26 \pm 6)\,\%$ of the dust emission at 100\,GHz, could represent magnetic dipole emission.  Alternatively, it could account for an increasing contribution of carbon dust, or a flattening of the emissivity of amorphous silicates, at millimetre wavelengths. These interpretations make different predictions for the dust polarization SED. 
\quad(4) We analyse the residuals of the dust-\hi\ correlation.  We identify a Galactic contribution to these residuals, which we model with variations of the dust emissivity on angular scales smaller than that  of our correlation analysis.  This model of the residuals is used to quantify uncertainties of the CIB power spectrum in a companion \Planck\ paper. 
}


\titlerunning{Dust  emission from the diffuse interstellar medium}

\authorrunning{Planck collaboration}

\date{Received 18 December  2013 / Accepted 29 January 2014}
\maketitle
   
%\tableofcontents

%

%$$$$$$$$$$$$$$$$$$$$$$$$$$$$$$$$

\section{Introduction}
%==================

Understanding interstellar dust is a major challenge in astrophysics
related to physical and chemical processes in interstellar space.
The composition of interstellar dust reflects the processes that
contribute to breaking down and rebuilding grains over timescales much
shorter than that of the injection of newly formed circumstellar or
supernova dust. While there is wide consensus on this view, the
composition of interstellar dust and the processes that drive its
evolution are still poorly understood \citep{Zhukovska08,Draine09c,Jones11}.  Observations
of dust emission are essential in constraining the nature of interstellar grains and their size
distribution. 

The \Planck\footnote{Planck (http://www.esa.int/Planck) is a project
  of the European Space Agency (ESA) with instruments provided by two
  scientific consortia funded by ESA member states (in particular the
  lead countries France and Italy), with contributions from NASA (USA)
  and telescope reflectors provided by a collaboration between ESA and
  a scientific consortium led and funded by Denmark.}  
all-sky survey has opened a new era in dust studies by extending to
submillimetre wavelengths and microwave frequencies the detailed
mapping of the interstellar dust emission provided by past infrared
space missions.  For the first time we have the sensitivity to map the
long wavelength emission of dust in the diffuse interstellar medium
(ISM).  Large dust grains (size $>$\,10\,nm) dominate the dust mass.
Far from luminous stars, the grains are cold (10--20\,K) so that a
significant fraction of their emission is over the \Planck\ frequency
range.  Dipolar emission from small, rapidly spinning, dust particles
is an additional emission component accounting for the so-called
anomalous microwave emission (AME) revealed by observations of the
cosmic microwave background (CMB)
\citep[e.g.][]{Leitch97,Banday03,Davies06,Ghosh12,planck2011-7.2}.  Magnetic dipole radiation
from thermal fluctuations in magnetic nano-particles may also be a
significant emission component over the frequency range relevant to
CMB studies \citep{Draine99,Draine13}, a possibility that has yet to
be tested.

The separation of the dust emission from anisotropies of the cosmic
infrared background (CIB) and the CMB is a difficulty for both dust
and background studies.  The dust-gas correlation provides a means to
separate these emission components from an astrophysics perspective,
complementary to mathematical component separation methods
\citep{planck2013-p06}. At high Galactic latitudes, the dust emission
is known to be correlated with the $21\,$cm line emission from neutral
atomic hydrogen \citep{Boulanger88}. This correlation has been used to separate the dust
emission from CIB anisotropies and characterize the
emission properties of dust in the diffuse ISM using data from the cosmic background explorer \citep[\COBE,][]{Boulanger96,Dwek97,Arendt98}, 
the Wilkinson microwave anisotropy probe \citep[\wmap][]{Lagache:2003a}, and \Planck\ \citep{planck2011-7.12}.  The residual
maps obtained after subtraction of the dust emission correlated with
\hi\ have been used successfully to study CIB anisotropies
\citep{Puget96,Fixsen98,Hauser98,planck2011-6.6}.  The correlation analysis
also yields the spectral energy distribution (SED) of the dust
emission normalized per unit hydrogen column density, which is an essential
input to dust models, and a prerequisite for determining the dust
temperature and opacity (i.e. the optical depth per hydrogen atom).

The \COBE\ satellite provided the first data on the thermal emission from large dust
grains at long wavelengths.  These data were used to define the dust
models of \citet{DraineLi07}, \citet{Compiegne11} and \citet{Siebenmorgen13}, and the
analytical fit proposed by \cite{Finkbeiner99}, 
which has been widely used by
the CMB community to extrapolate the IRAS all-sky survey to microwave frequencies.  
Today the \Planck\ data allow us to characterize
the dust emission at millimetre wavelengths directly from observations.
A first analysis of the correlation between Planck and \hi\
observations was presented in \citet{planck2011-7.12}. In that study, the
IRAS $100 \,\mu$m and the 857, 545, and 353\,GHz Planck maps were
correlated with \hi\ observations made with the Green Bank Telescope
(GBT) for a set of fields sampling a range of \hi\ column densities.
We extend this early work  to microwave
frequencies, and to a total sky area more than an order of magnitude
higher.

The goal of this paper is to characterize the emission
properties of dust in the diffuse ISM, from far infrared to microwave
frequencies, for dust, CIB, and CMB studies. We achieve this by
cross-correlating the \planck\ data with atomic hydrogen
emission surveyed over the southern sky with the Parkes telescope
(the Galactic All Sky Survey, hereafter GASS; \citealt{McClure-Griffiths09,Kalberla10}).  
We focus on the southern Galactic polar cap ($b < -25^\circ$) where 
the dust-gas correlation is most easily characterized using \hi\ data
because the fraction of the sky with significant H$_2$ column density is low \citep{Gillmon06}.  
This is also the cleanest part of the southern sky for CIB and CMB studies.

The paper is organized as follows.  We start by presenting the
\planck\ and the ancillary data from the COBE diffuse infrared background experiment (\DIRBE) and \wmap\ that we are
correlating with the \hi\ GASS survey
(Sect.~\ref{sec:observations}). 
The methodology we follow to quantify the dust-gas correlation is
described in Sect.~\ref{sec:gas-dust}.
We use the results from the correlation analysis to characterize the
variations of the dust emission properties across the southern
Galactic polar cap in Sect.~\ref{sec:southern_Cap} 
and determine the spectral index of the thermal dust emission from submm to millimetre wavelengths
in Sect.~\ref{sec:microwave_index}.
In Sect.~\ref{sec:Dust_SED}, we present the mean SED of dust from
far infrared to millimetre wavelengths, and a comparison with models of the thermal
dust emission. 
Section~\ref{sec:microwave} focuses on the SED of the \hi-correlated
emission at microwave frequencies, which we quantify and model over
the full spectral range relevant to CMB studies from 23 to $353\,$GHz.
The main results of the paper are summarized in
Sect.~\ref{sec:summary}.  
The paper contains four appendices where we detail specific aspects of
the data analysis.  In Appendix~\ref{appendix:model}, we describe how
maps of dust emission are built from the results of the \hi\
correlation analysis.
We explain how we separate dust and CMB emission at microwave
frequencies in Appendix~\ref{appendix:cmb}.
We detail how we quantify the uncertainties of the results of the
dust-\hi\ correlation in Appendix~\ref{appendix:uncertainties}.
Appendix~\ref{appendix:simus_GalRes} presents simulations of the dust
emission that we use to quantify uncertainties.

\section{Data sets}
\label{sec:observations}

In this section, we introduce the \Planck, \hi, and ancillary sky maps
we use in the paper.

\subsection{Planck data}
\label{subsec:planck_maps}

\Planck\ is the third generation space mission to characterize the
anisotropies of the CMB.  It observed the sky in nine frequency bands
from 30 to 857\,GHz with an angular resolution from 33\arcm\ to 5\arcm\ 
\citep{planck2013-p01}. The Low Frequency Instrument
\citep[LFI,][]{Mandolesi:2010,Bersanelli:2010,Mennella:2010} observed
the 30, 44, and $70\,$GHz bands with amplifiers cooled to $20\,$K. The
High Frequency Instrument \citep[HFI,][]{Lamarre:2010} observed the 100, 143, 217, 353, 545, and 857\,GHz bands with bolometers cooled to 0.1\,K.  In this paper, we use the nine \Planck\ frequency maps made from the first 15.5\,months of the mission \citep{planck2013-p01} in \healpix\ format\footnote{\citet{Gorski05}, http://healpix.sf.net}.  Maps at 70\,GHz and below are at $N_{\rm side} = 1024$ (pixel size 3\parcm4); those at 100\,GHz and above are at $N_{\rm side} = 2048$ (1\parcm7). We
refer to previous \planck\ publications for the data processing,
map-making, photometric calibration, and photometric uncertainties
\citep{planck2013-p02,planck2013-p03,planck2013-p02b,planck2013-p03f}.
At HFI frequencies, we analyse maps produced both with and without subtraction of the zodiacal
emission \citep{planck2013-pip88}.  To quantify uncertainties associated
with noise, we use maps made from the first and second
halves of each stable pointing period \citep{planck2013-p03}.

\begin{figure*}[h!]
\centering
\includegraphics[width=0.46\textwidth]{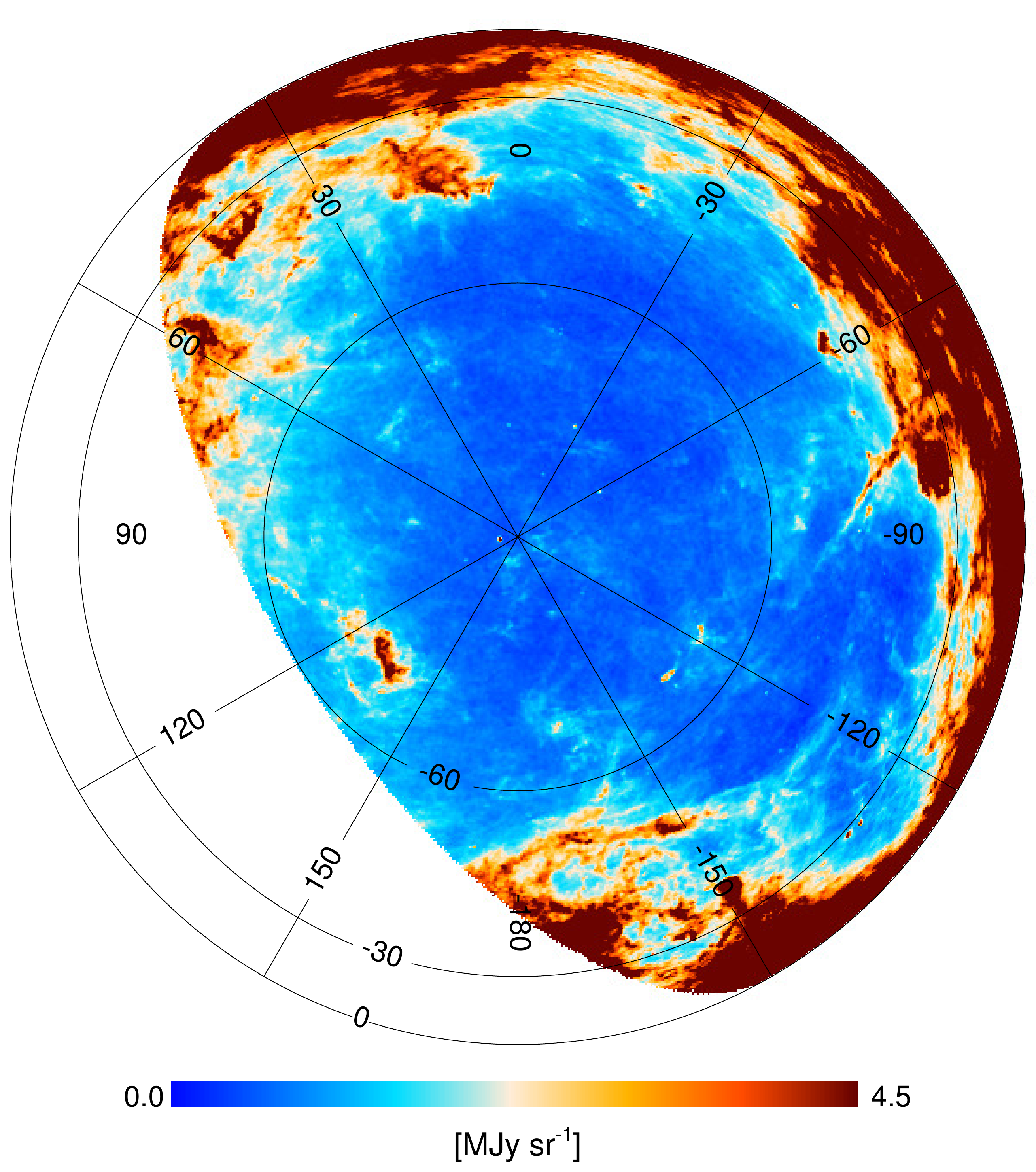}
\includegraphics[width=0.46\textwidth]{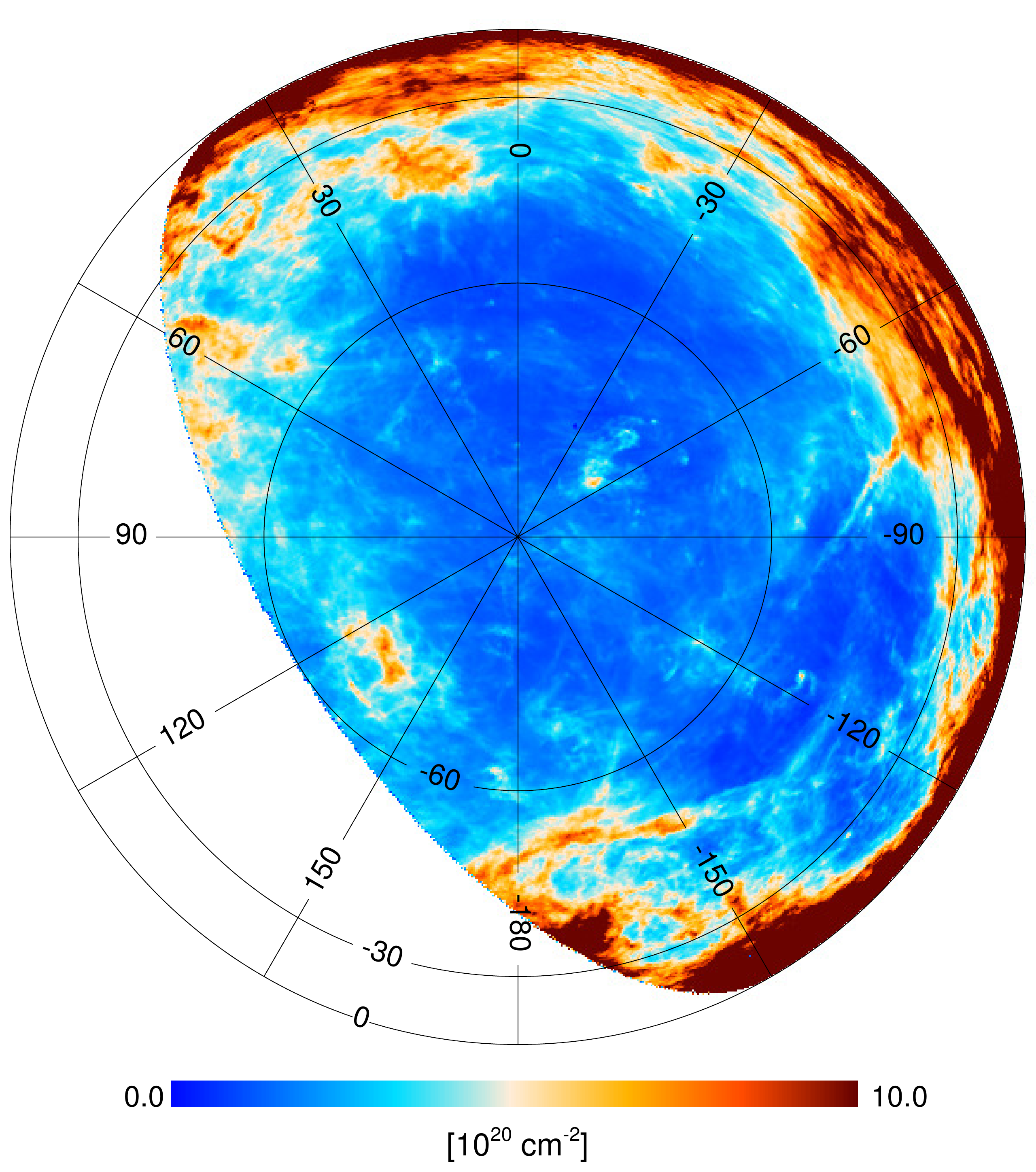}
\caption{{\it Left\/}: \planck\ map at 857\,GHz over the area where we
  have \hi\ data from the GASS survey. The center of the orthographic projection is
  the southern Galactic pole. Galactic longitudes and latitudes are
  marked by lines and circles, respectively. The \planck\ image has been smoothed to
  the $16\arcmin$ resolution of the GASS \NH\ map.   
  {\it Right\/}: GASS \NH\ map of Galactic disk emission, obtained by
  integrating over the velocity range defined by Galactic rotation
  (Sect.~\ref{subsec:GASS_MS}). }
\label{fig:HFI857}
\end{figure*}

As an example, Fig.~\ref{fig:HFI857} shows the 857\,GHz map for the area of the \hi\ GASS survey. 
%and smoothed its $16\arcmin$ resolution.

\subsection{The GASS \hi\ survey}
\label{subsec:GASS}

In this section we explain how we produce the column density map of
Galactic \hi\ gas that we will use as a spatial template in our
dust-gas correlation analysis.

\subsubsection{\hi\ observations}

We make use of data from the GASS \hi\ survey obtained with the Parkes
telescope \citep{McClure-Griffiths09}.  The $21\,$cm line emission was
mapped over the southern sky ($\delta < 1^\circ$) with $14\parcm5$ FWHM 
angular resolution and a velocity
resolution of $1\,$km\,s$^{-1}$.  At high Galactic latitudes, the
average noise for individual
emission-free
channel maps is 50~mK ($1 \sigma$).  GASS is the most sensitive,
highest angular resolution survey of Galactic \hi\ emission over the
southern sky. We use data corrected for instrumental effects, stray
radiation, and radio-frequency interference from \citet{Kalberla10}.

Maps of \hi\ emission integrated over velocities were generated from
spectra in the 3-D data cube. To minimize uncertainties from
instrumental noise and to eliminate residual instrumental problems we do not integrate the emission over all velocities. 
The problem is that weak systematic biases over a large number of channels can add up to a significant error. 
We select the channels on a smoothed data cube to ensure that weak emission around \hi\ clouds is not affected. 
Specifically, we calculate a second data cube smoothed to angular and velocity
resolutions of $30^\prime$ and
$8\,$km\,s$^{-1}$.  Velocity channels where the emission in this smoothed data cube is below 
a $5 \sigma$ level of $30\,$mK  are not used in the integration.  
This brightness threshold is applied to each
smoothed spectrum to define the velocity ranges, not necessarily
contiguous, over which to integrate the signal in the full-resolution data cube.  
The impact on the HI column density map of the selection of channels is small and noticeable only in the regions of 
lowest column densities.  The magnitude of the difference between  maps produced with and without the $5 \sigma$ selection of the channels 
is a few $10^{18}\,{\rm H\,cm^{-2}}$. This small difference is not critical for our analysis. 
%In the next section we explain how we identify the emission from the Galaxy.

\subsubsection{Separation of \hi\ emission from the Galaxy and Magellanic Stream}
\label{subsec:GASS_MS}

The southern polar cap contains Galactic \hi\ emission with typical
column densities \NH\ from one to a few times $10^{20}\,{\rm cm^{-2}}$, plus a
significant contribution from the Magellanic Stream (MS;
\citealt{Nidever08}). We need to separate the Galactic and MS gas because
the dust-to-gas mass ratio of the low metallicity MS gas is lower than
that of the Galactic \hi.
  
The velocity information permits a separation of the Galactic and MS
emission over most of the sky \citep{Venzmer12}.  To distinguish the
two components, we use a 3-D model of the Galactic \hi\ emission
presented in \citet{Kalberla08}.  The model matches the velocity
distribution of the observed emission.
We produce a 3-D data cube with the model that we use to distinguish parts of the GASS data 
cube that have emission likely to be associated with the MS from those associated with the Galaxy.
Specifically, the emission in a given velocity channel is ascribed to
the MS where $ T_{\rm model} < 60 \,$mK, and to the Galaxy where $
T_{\rm model} \ge 60 \,$mK (see Fig.~A.1 in \citet{planck2013-pip56}). 
This defines the MS and Galactic maps used in
the paper. 
The MS and Galactic emissions are clearly separated except in a
circular area of $20^\circ$ diameter centred at Magellanic longitudes
and latitudes\footnote{Defined in \citet{Nidever08}. Magellanic
  latitude is $0^\circ$ along the MS. The trailing section of the MS
  has negative longitudes.}  $l_{\rm MS} =-50^\circ $ and $b_{\rm MS}
=0^\circ$, where the radial velocity of gas in the MS merges with
Galactic velocities \citep{Nidever10}.  We do not use this area in our
dust-gas correlation analysis.

\subsubsection{The IVC and HVC contributions to the Magellanic Stream component}
\label{subsec:GASS_IVC}

Our method to identify the emission from the local \hi\ differs from that used for the GBT fields in
\citet{planck2011-7.12}, where the low velocity gas and intermediate and
high velocity clouds (IVCs and HVCs) have been distinguished based on the
specific spectral features present in each of the fields. Such a
solution is not available across the much more extended GASS field, but
our MS map may be expressed as the sum of IVC and HVC maps.

HVCs and IVCs are distinguished from gas in the Galactic disk by their
deviation velocities $v_{\rm dev}$, defined as the difference between the
observed radial velocity and that expected in a given direction from the
Galactic rotation. Clouds with $|v_{\rm dev}| > 90\,{\rm km\,s^{-1}}$ are
usually considered as HVCs,
while IVCs correspond to the velocity range $35 < |v_{\rm dev}| < 90\,{\rm
  km\,s^{-1}}$ \citep{Wakker04}.  At high Galactic latitudes, our
threshold of $60\,$mK for the \hi\ model corresponds to about $|v_{\rm dev}|
\le 45$\,km\,s\mo; a threshold of $T_{\rm model} \ge 16$\,mK corresponds to $|v_{\rm dev}| \le 90$\,km\,s\mo. To separate the MS emission into its IVC and HVC contributions, therefore, we make a
second separation using the 16\,mK threshold.  The lower threshold allows us to identify the part of the MS emission with deviation velocities in the HVC range, and the difference between
the two MS maps produced with  60 and 16\,mK thresholds identifies the
part of the MS map with deviation velocities in the IVC range.

We note that the HVC map could contain HVC gas not associated with the
MS, but also of low dust content.  The IVC map might contain Galactic gas
with more normal dust content like in Galactic IVCs \citep{planck2011-7.12}.
In addition, the Galactic gas as defined might also contain Galactic IVCs,
which often have a depleted dust content, typically by a factor
two \citep{planck2011-7.12}.  However, anomalous lines of sight are removed by our
masking process (Sect.~\ref{subsec:masking}).

\subsubsection{Column density maps}
 
\begin{figure*}[!h]
\centering
\includegraphics[width=0.46\textwidth]{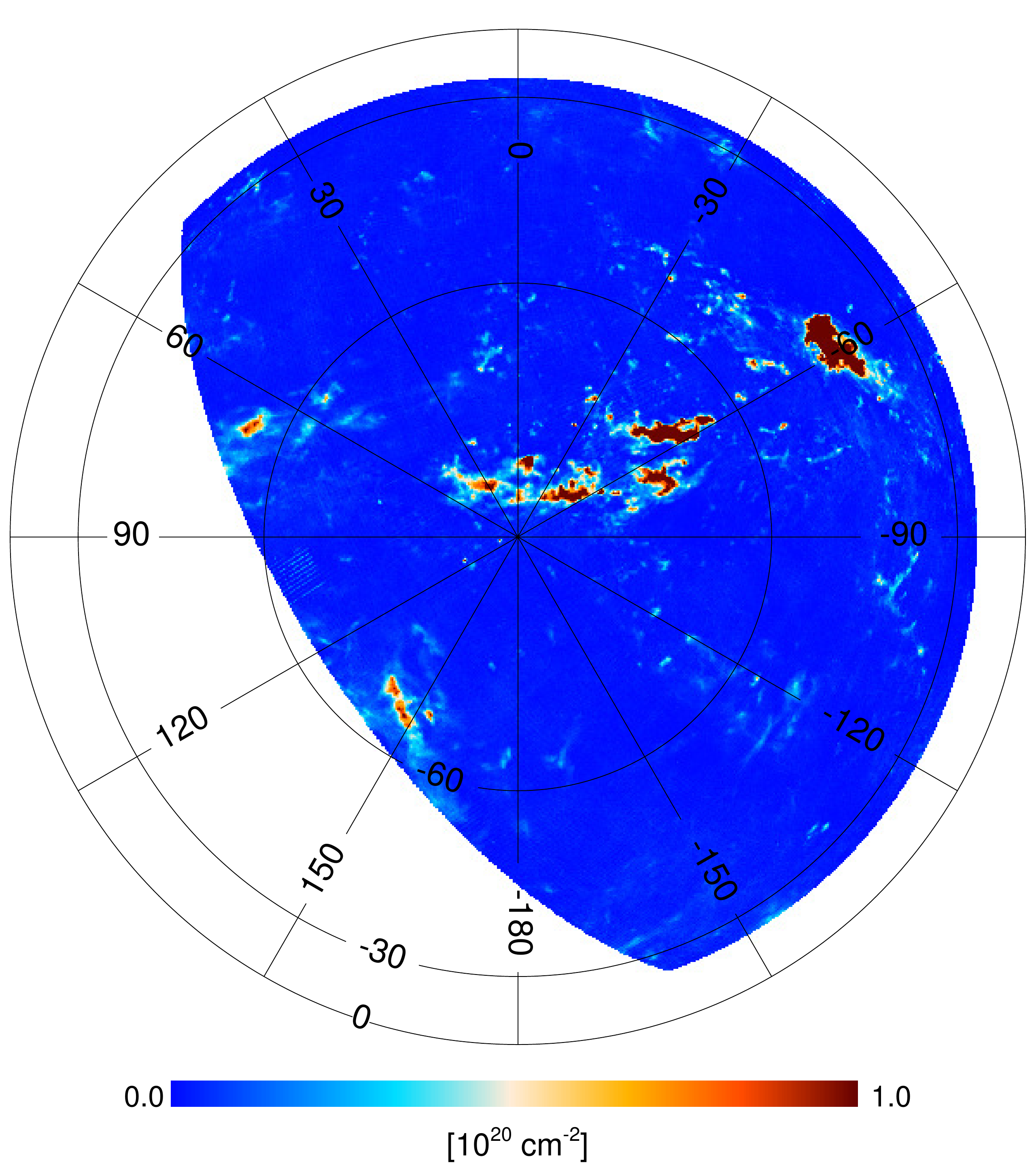}
\includegraphics[width=0.46\textwidth]{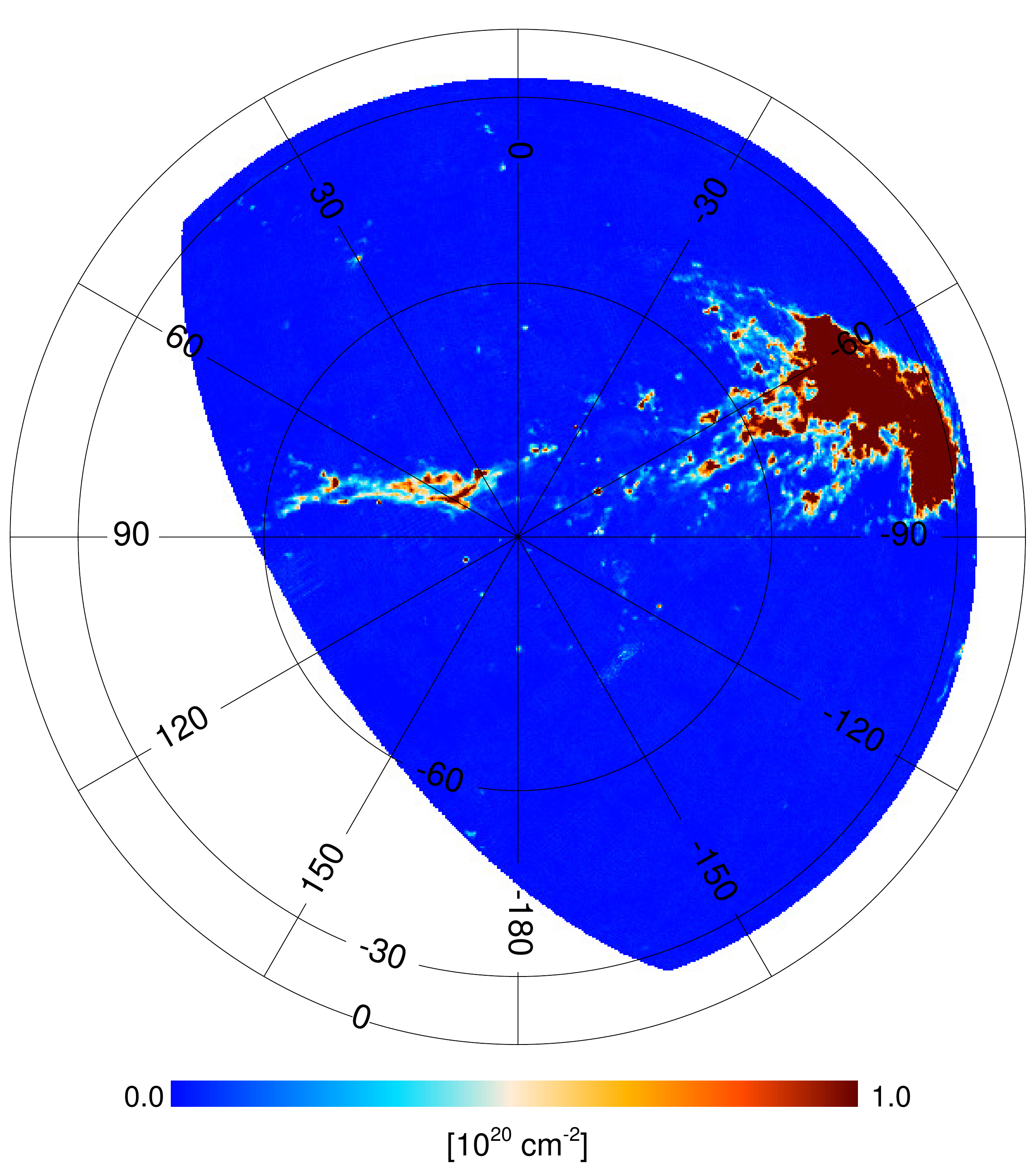}
\caption{\NH\ maps corresponding to the IVC (left) and HVC (right)
  velocity ranges as defined in Sect.~\ref{subsec:GASS_IVC}.  We show
  the data at Galactic latitudes $b < -25^\circ$ that we use in our
  correlation analysis. }
\label{fig:IVC_HVC}
\end{figure*}

The Galactic and the MS \hi\ emission maps, as well as the division of
the MS map into its IVC and HVC contributions, are projected on a
\healpix\ grid with a resolution parameter $N_{\rm side} = 1024$ using the nearest  \healpix\  pixel to each GASS 
position,  before reducing the map to $N_{\rm side} = 512$ (pixel
size $6\parcm9$) with the $\rm ud\_grade$ \healpix\  procedure.  After interpolation onto the  
\healpix\ grid, the angular resolution is $16\parcm2$.  For all maps, the \hi\ emission is
converted to \hi\ column density \NH\ assuming that the $21\,$cm line
emission is optically thin. For the column densities of one to a few $10^{20}\,$H\,cm$^{-2}$ relevant to this study, the 
opacity correction is expected to be less than 5\,\% \citep[see Fig. 4 in][]{Elvis89}. 
The Galactic \NH\ map is presented in
Fig.~Ê\ref{fig:HFI857}.  Figure~\ref{fig:IVC_HVC} shows the \NH\ maps
corresponding to the IVC and HVC velocity ranges. % and Fig.~\ref{fig:Residu857} their sum. 

We use the Galactic \NH\ map as a spatial template in our dust-gas
correlation analysis.  The IVC and HVC maps are used to quantify how
the separation of the \hi\ emission into its Galactic and MS
contributions affects the results of our
analysis.

\subsection{Ancillary sky maps}

In addition to the \planck\ maps, we use the \DIRBE\ sky maps at 100,
140, and $240 \, \mu$m \citep{Hauser98}, and the \wmap\
9-year sky maps at frequencies 23, 33, 41, 61, and 94\,GHz
\citep{Bennett12}.  The \DIRBE\ maps allow us to extend our \hi\
correlation analysis to the peak of the dust SED in the far infrared.  The
\wmap\ maps complement the LFI data, giving finer frequency sampling of the SED at microwave
frequencies.  We also use the 408\,MHz map of
\citet{Haslam82} to correct our dust-gas correlation for chance correlations of the \hi\ template with synchrotron emission.  These chance correlations are non-negligible for the lowest \planck\ and \wmap\
frequencies.

The \DIRBE, \wmap, and 408\,MHz data are available from the Legacy
Archive for Microwave Background
Data\footnote{http://lambda.gsfc.nasa.gov/}.  We use the \DIRBE\ data
corrected for zodiacal emission.  We project the data on a \healpix\
grid at $N_{\rm side} = 512$ with a Gaussian interpolation kernel that
reduces the angular resolution to 50\arcm. We compute maps of
uncertainties that take into account this slight smoothing of the
data.  
The photometric uncertainties of the DIRBE maps at 100, 140, and
240\microns\ are 13.6, 10.6, and 11.6\,\%, respectively \citep{Hauser98}.

\section{The dust-gas correlation}
\label{sec:gas-dust}

Figure~Ê\ref{fig:HFI857} illustrates the general correlation between the
dust emission and \hi\ column density over the southern Galactic
cap.  In this section we describe how we quantify this correspondence by cross
correlating locally the spatial structure in the dust and \hi\
maps.  Section~\ref{subsec:cc} describes the method that we use to cross correlate maps; Sects.~\ref{subsec:implementation} and \ref{subsec:masking} describe its implementation. 
Residuals to the dust-\hi\ correlation are discussed in Sect.~\ref{subsec:GalRes}.

\subsection{Methodology}
\label{subsec:cc}

We follow the early \Planck\ study \citep{planck2011-7.12} in cross
correlating spatially the Planck maps with the Galactic \hi\ map
(Sect.~\ref{subsec:GASS}).  
For a set of sky positions, we perform a linear fit  between the data  and the \hi\ 
template. We compute the slope ($\alpha_\nu$) and offset ($\omega_\nu$) of the 
fit minimizing the $\chi^2$
\begin{equation}
\chi^2 = \sum_{i=1}^{N} [T_\nu (i) - \alpha_\nu \, I_{\rm HI}(i) -  \omega_\nu]^2, \label{eq:1}
\end{equation}
where  $T_\nu$ and 
$I_{\rm HI}$ are the data and template values from maps at a common resolution.
The sum is computed over N pixels within 
sky patches centred on the positions at which the
correlation is performed. 
The minimization yields the following expressions for $\alpha_\nu$ and $\omega_\nu$
\begin{equation}
\alpha_\nu =  \frac{\sum_{i=1}^{N}\vnT (i) \, . \, \vtHI (i) }{\sum_{i=1}^{N}\, \vtHI (i)^2 }  \label{eq:2}
\end{equation}

\begin{equation}
%\omega_\nu = \langle T_\nu\rangle -  \alpha_\nu \times \langle I_{\rm HI}\rangle, \label{eq:2}
\omega_\nu =  \frac{1}{N}\sum_{i=1}^{N} (T_\nu (i) - \alpha_\nu \, I_{\rm HI} (i) ), \label{eq:3}
\end{equation}
where $\vnT$ and $\vtHI$ are the data  and \hi\ 
template vectors with mean values, computed over the N pixels, subtracted. 
The slope of the linear regression $\alpha_\nu, $ hereafter referred to as the correlation measure, is  used to compute the dust emission at frequency $\nu$ 
per unit  \NH. The offset of the linear regression $\omega_\nu$
is used in building a model of the dust emission that is
correlated with the \hi\ template in Appendix~\ref{appendix:model}.

We write the sky emission as the sum of five contributions
\begin{equation}
T_\nu = {T_{\rm D}}(\nu) +  {T_{\rm C}}  + {T_{\rm CIB}}(\nu) + {T_{\rm G}}(\nu) + {T_{\rm N}}(\nu), \label{eq:4}
\end{equation}
where ${T_{\rm D}}(\nu$) is the map of dust emission associated with
the Galactic \hi\ emission, ${T_{\rm C}}$ and ${T_{\rm CIB}}(\nu)$ are the cosmic
microwave and infrared backgrounds,  ${T_{\rm G}}(\nu)$ represents  Galactic emission components 
unrelated to \hi\ emission (dust associated with H$_2$ and \hii\ gas, synchrotron
emission, and free-free), and ${T_{\rm N}}(\nu)$ is the data noise.  These five terms are expressed in units of thermodynamic CMB temperature. 

Combining Eqs.~(\ref{eq:2}) and (\ref{eq:4}), we write the cross-correlation
measure as the sum of five contributions
\begin{align}
\alpha_\nu =   \left(\frac{1}{\sum_{i=1}^{N}\, \vtHI (i)^2 }\right) \, \sum_{i=1}^{N} & [\hat{T}_{\rm D} (\nu,i)  + \hat{T}_{\rm C} (i) + \hat{T}_{\rm CIB} (\nu,i)  \nonumber \\
 &  + \hat{T}_{\rm G} (\nu,i) + \hat{T}_{\rm N}(\nu,i)]. \, \vtHI (i) \label{eq:5}
\end{align}
\begin{equation}
 \alpha_\nu = \alpha_\nu(D_{\rm HI}) + \alpha (C_{\rm HI}) + \alpha_\nu({\rm CIB}_{\rm HI}) + \alpha_\nu(G_{\rm HI}) + \alpha_\nu(N),  \label{eq:6}
 %\alpha_\nu(R_{\rm HI}),  \label{eq:4} 
\end{equation}
where the subscript HI refers to the \hi\ template used in this
paper.
The first term $\alpha_\nu(D_{\rm HI})$ is the dust emission at frequency
$\nu$ per unit \NH, hereafter referred to as the dust emissivity
$\epsilon_{\rm H}(\nu)$.  The second term $\alpha (C_{\rm HI})$ is the chance
correlation between the CMB and the \hi\ template. It is independent
of the frequency $\nu$ because Eqs.~(\ref{eq:4}) and (\ref{eq:5}) are
written in units of thermodynamic CMB temperature.  
The last terms in Eq.~(\ref{eq:6})
represent the cross-correlation of the \hi\ map with the  CIB, the Galactic emission components unrelated with \hi\ emission, and the data noise.  
We take these terms as uncertainties on $\epsilon_{\rm H}(\nu)$.
In Appendix~\ref{appendix:cmb}, we detail how we estimate $ \alpha
(C_{\rm HI})$ to get $\epsilon_{\rm H}(\nu)$ from $\alpha_\nu$.  For part of our
analysis, we circumvent the calculation of $ \alpha (C_{\rm HI})$ by
computing the difference $\alpha^{100}_\nu = \alpha_\nu- \alpha_{\rm 100\,GHz}$.  

We write the standard deviation on the dust emissivity $\epsilon_{\rm H}(\nu)$ as
\begin{equation}
\sigma (\epsilon_{\rm H}(\nu))  = (\sigma_{\rm CIB}^2 +\sigma_{\rm G}^2 + \sigma_{\rm N}^2  + (\delta_{\rm C} \times \alpha (C_{\rm HI}))^2)^{0.5},  \label{eq:7}
\end{equation}
where the first three terms represent the
contributions from CIB anisotropies, the Galactic residuals, and the data
noise.  Here and subsequently, Galactic residuals refer to the
difference between the dust emission and the model derived from the correlation analysis (Appendix~\ref{appendix:model}). They arise
from Galactic emission unrelated with \hi\ ($T_{\rm G}(\nu)$ in
Eq.~(\ref{eq:4})), and also from variations of the dust emissivity on
angular scales smaller than the size of the sky patch used in
computing the correlation measure.  The last term in Eq.~(\ref{eq:7})
is the uncertainty associated with the subtraction of the CMB,
quantified by an uncertainty factor $\delta_{\rm CMB}$ that we estimate in
Appendix~\ref{appendix:cmb} to be 3\,\%. For $\alpha^{100}_\nu$ and a given experiment, the CMB
subtraction is limited only by the relative uncertainty of the
photometric calibration, which is 0.2--0.3\,\% at microwave frequencies
for both \planck\ and \wmap\ \citep{planck2013-p01,Bennett12}.

\subsection{Implementation}
\label{subsec:implementation}

We perform the cross-correlation analysis at two angular
resolutions.  First, we correlate the \hi\ template with the seven \planck\
maps at frequencies of 70\,GHz and greater and the 94\,GHz  channel of \wmap, all smoothed to the $16\arcmin$ resolution of the \hi\ map, i.e. $N_{\rm side} = 512$, with $6\parcm9$ pixels.  The
map smoothing uses a Gaussian approximation for the \planck\ beams.
The cross-correlation with the \DIRBE\ maps is done at a
single $50\arcmin$ resolution.
Second, to extend our analysis to frequencies lower than $70\,$GHz, we
also perform the data analysis using all of the \planck\ and \wmap\ maps smoothed to a
common $60\arcmin$ Gaussian beam \citep{planck2013-p03} at a \healpix\
resolution $N_{\rm side} = 128$ ($27\parcm5$ pixels), combined with a
smoothed and reprojected \hi\ template.  
At frequencies $\nu \le 353\,$GHz, we also perform a simultaneous linear correlation 
of the \Planck\ and \wmap\ maps with two templates, the GASS \hi\ map and the 408\,MHz map
of \citet{Haslam82}.
This corrects the results of the dust-\hi\ correlation for any chance
correlation of the \hi\ spatial template with synchrotron emission.
\citet{Peel12} have shown that, at high Galactic latitudes, 
the level of the dust-correlated emission in the \wmap\ bands does not depend significantly on the frequency of the synchrotron template. 

We perform the cross-correlation over circular sky patches 
$15^\circ$ in diameter centred on \healpix\ pixels.  The
analysis of sky simulations presented in
Appendix~\ref{appendix:uncertainties} shows that the size of the sky
patches is not critical.  
We require the number of unmasked pixels used to compute the correlation measure
and the offset to be higher than one third of the total number of
pixels within a sky patch.  For input maps at $16\arcmin$ angular
resolution projected on \healpix\ grid with $N_{\rm side} = 512$, this
corresponds to a threshold of 4500 pixels.

We compute the correlation measure $\alpha_\nu $ and offset
$\omega_\nu $ at positions corresponding to pixel centres on 
\healpix\ grids with $N_{\rm side} = 32$ and 8 (pixel size
$1\pdeg8$ and $7\pdeg3$, respectively).  The higher resolution grid,
which more finely samples variations of the dust emissivity on the
sky, is used to produce images for display, 
for example the dust emissivity at 353\,GHz presented in Fig.~\ref{fig:IR_HI_correlation}, and the dust model in Appendix~\ref{appendix:model}.  For statistical studies, we use the
lower resolution grid, for which we obtain a correlation measure for
135 sky patches. Because of the sampling of the $15^\circ$ patches at $N_{\rm side} = 8$, each pixel in the input data is part of three sky patches, and these correlation measures
are not independent.

\begin{figure}[!h]
\centering
\includegraphics[width=0.46\textwidth]{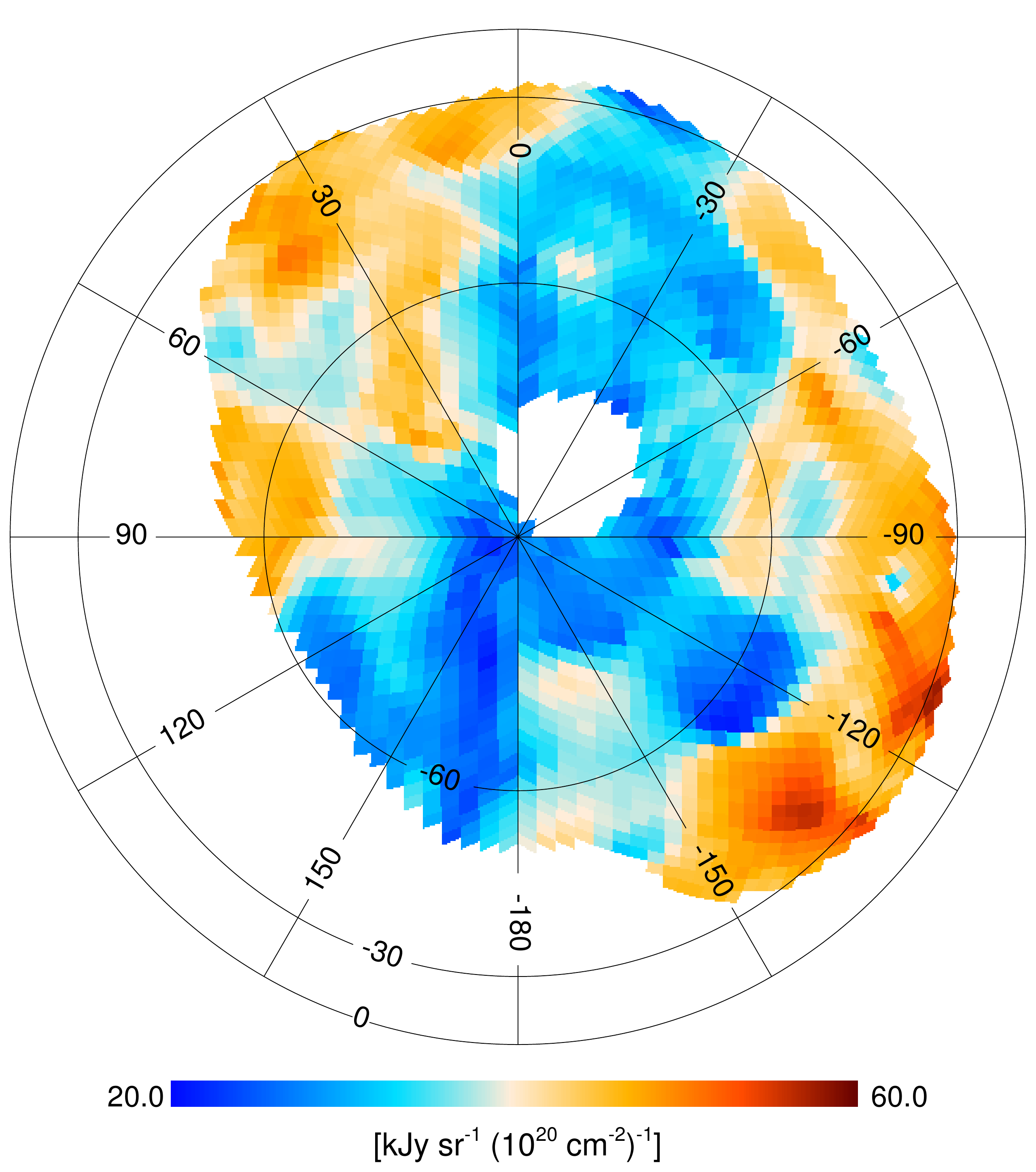}
\caption{Map of the dust emissivity at $353\,$GHz, i.e. the correlation measure $\alpha_{353}$ with 
the CMB contribution $\alpha (C_{\rm HI})$ subtracted [see Eq.~(\ref{eq:6})],  % by subtracting the correlation measure at 100\,GHz ($\Delta \alpha (353\,$GHz), 
The correlation measure  is computed in each pixel correlating the \planck\ map with the \hi\ template over 
a  sky patch with $15^\circ$ diameter centred on it.}
\label{fig:IR_HI_correlation}
\end{figure}

We detail how we quantify the various contributions to the uncertainty
of the dust emissivity in Appendix~\ref{appendix:uncertainties}, including
those associated with the separation of the \hi\ emission between its Galactic and 
MS contributions (Sect.~\ref{subsec:GASS_MS}), 
which is the main source of uncertainty on the \hi\ template used as independent variable in the correlation analysis.
As in \citet{planck2011-7.12}, we do not include any noise weighting in
Eq.~(\ref{eq:1}) because data noise is not the main source of
uncertainty.  For most HFI frequencies, the noise is much lower
than either CIB anisotropies or the differences between the dust emission and the model we fit.

\subsection{Sky masking}
\label{subsec:masking}
% masking

In applying Eqs.~(\ref{eq:2}) and (~\ref{eq:3}), we use a sky mask
that defines the overall part of the sky where we characterize the
correlation of \hi\ and dust, and within this large area the pixels
that are used to compute the correlation measures.  We describe in
this section how we make this mask.

We focus our analysis on low column density gas around the southern
Galactic pole, specifically, \hi\ column densities $\NH \le 6 \times
10^{20}$\,cm$^{-2}$ at Galactic latitudes
$b \le -25^\circ $.  Within this sky area we mask a
$20^\circ$-diameter circle centred at Magellanic longitude and
latitude $l_{\rm MS} =-50^\circ $ and $b_{\rm MS} =0^\circ$, where the radial
velocity of gas in the MS merges with Galactic velocities so that a
Galactic \hi\ template cannot be separated.

To characterize the dust signal associated with the \hi\ gas, we also
need to mask sky pixels where the dust and \hi\ emission are not
correlated.  As in \citet{planck2011-7.12}, we need to identify the sky
pixels where there is significant dust emission from H$_2$ gas.  This
is relatively easy to do at high Galactic latitudes where the gas
column density is the lowest, and the surface filling factor of H$_2$
gas is small.  UV observations \citep{Savage77,Gillmon06} and the
early \planck\ study \citep{planck2011-7.12} show that the fraction of H$_2$
gas can become significant for some sight lines where  \NH\ exceeds $3
\times 10^{20}\,$cm$^{-2}$ or so.
We also need to mask pixels where there is Galactic \hi\ gas with
little or no far infrared counterpart, and bright extragalactic sources.

Following \citet{planck2011-7.12}, we build our mask by iterating the
correlation analysis.  At each step, we build a model of the dust
emission associated with the Galactic \hi\ gas from the results of the
IR-\hi\ correlation (Appendix~\ref{appendix:model}).  We obtain a map
of residuals by subtracting this model from the
input data.  At each iteration, we then compute the standard
deviation of the Gaussian core of the residuals over unmasked
pixels. The mask for the next iteration is set by masking all pixels
where the absolute value of the residual is higher than $3\sigma$.  
The choice of this threshold is not critical. 
For a $5\sigma$ cut, we obtain a mean dust emissivity at $857\,$GHz higher by only  
1\% than the value for a $3\sigma$ cut. 
The standard deviation of the fractional differences between the two sets of dust emissivities computed patch by patch is 3\%.
We use the highest \planck\ frequency, 857\,GHz,
to identify bright far infrared sources and pixels
where the dust emission departs from the model emission estimated from
the \hi\ map.  The iteration rapidly converges to a stable mask.  Once we have
converged for the 857\,GHz frequency channel, we look for outliers
at other frequencies. This is necessary to mask a few infrared galaxies at $100\,\mu$m and bright radio
sources at microwave frequencies. We perform this procedure with the
maps at $16\arcmin$, $50\arcmin$, and $60\arcmin$ resolution, 
obtaining a separate mask for each resolution.

\begin{figure}[!h]
\centering
\includegraphics[width=0.49\textwidth]{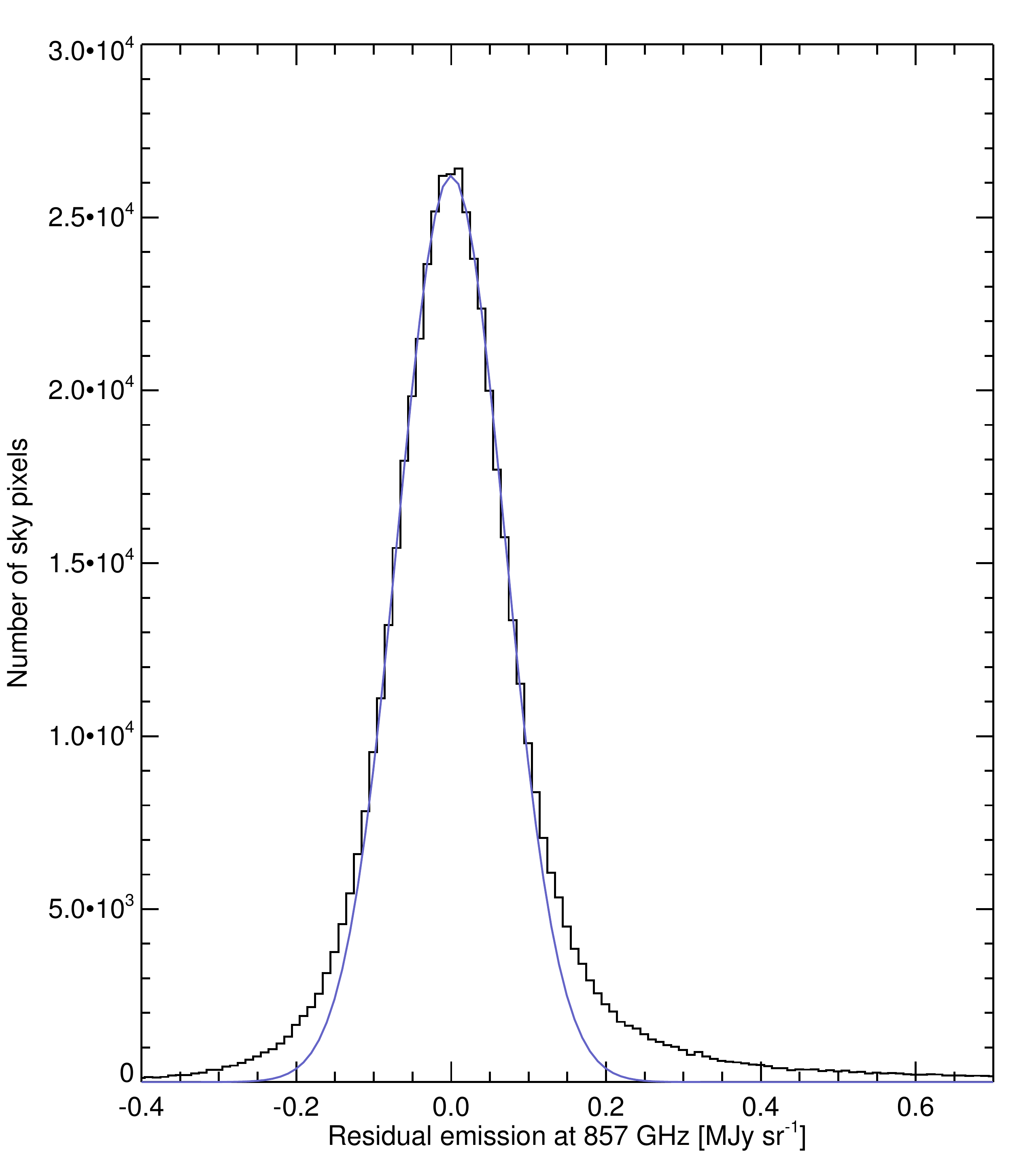}
\caption{Histogram of residual emission at 857\,GHz after
  subtraction of the dust emission associated with HI gas. The blue
  solid line is a Gaussian fit to the core of the histogram, with
  dispersion $\sigma = 0.07$\,MJy\,sr$^{-1}$.  We mask pixels where the
  absolute value of the residual emission is higher than $3\sigma$.
  The positve (negative) wing of the histogram beyond this threshold
  represents 7\,\% ( 2\,\%) of the data. }
\label{fig:histo_residuals}
\end{figure}

Figure~\ref{fig:histo_residuals} presents the histogram of the
residual map at 857\,GHz with $16\arcmin$ resolution.  The mask
discards the positive and negative tails that depart
from the Gaussian fit of the central core of the histogram. These
tails amount to 9\,\% of the total area of the residual map.

\begin{figure}[!h]
\centering
\includegraphics[width=0.49\textwidth]{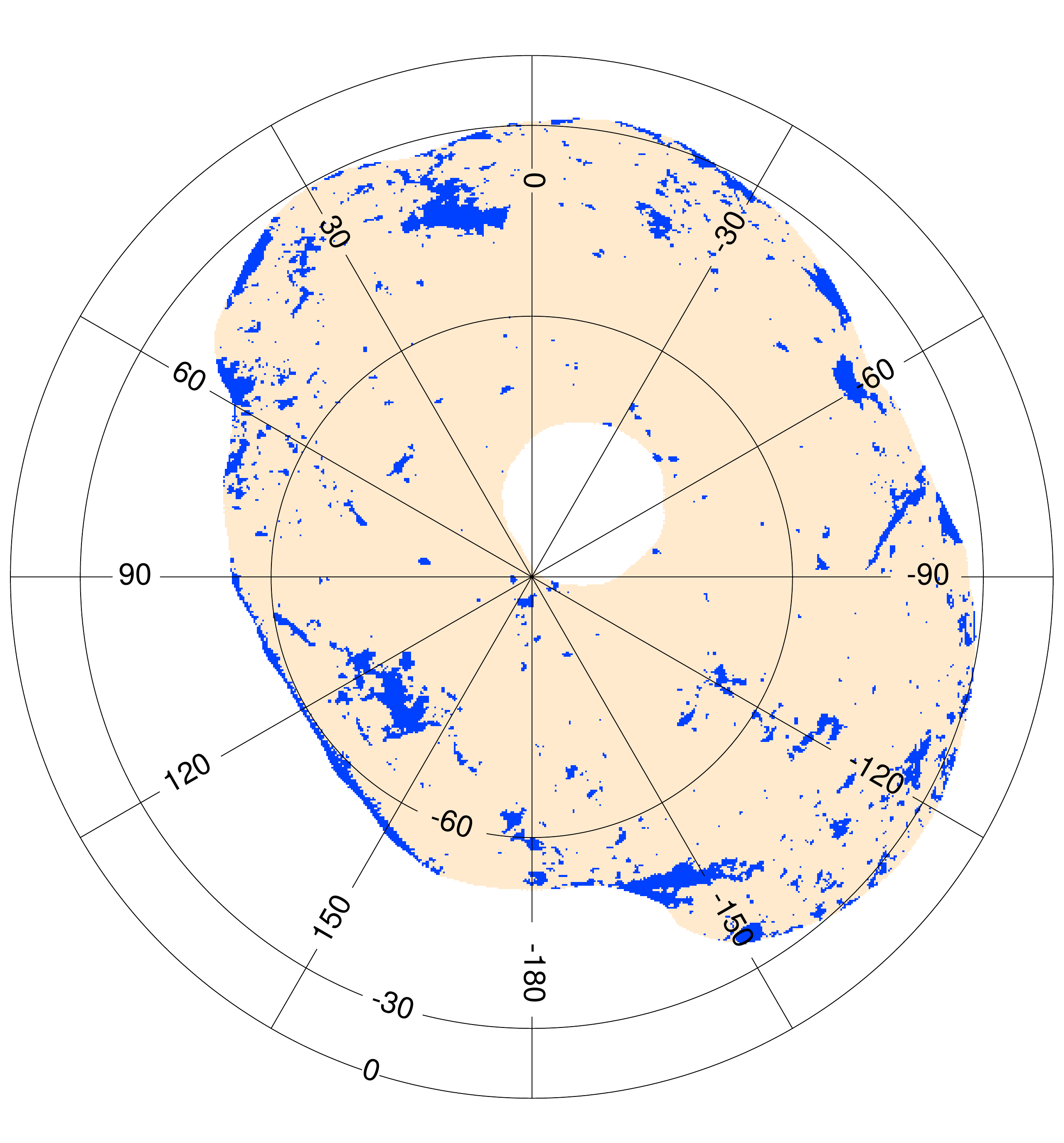}
\caption{Mask for our analysis of the {\planck}-\hi\ correlation. The
coloured area that is not blue defines the data used to compute the correlation measures. 
Within this area, the median \NH\ is $2.1\times
10^{20}\,{\rm H\, cm^{-2}}$, and $N_{\rm HI} < 3 \times 10^{20}\,{\rm H\,
cm^{-2}}$ for 74\,\% of the pixels.
The blue patches correspond to regions where the absolute value of the
residual emission is higher than $3\sigma$ at 857~GHz
(Fig.~\ref{fig:histo_residuals}). The circular hole near the Southern
Galactic pole corresponds to the area where \hi\ gas in the Galaxy
cannot be well separated because the mean radial velocity of the gas
in the MS is within the Galactic range of velocities. }
\label{fig:IR_HI_mask}
\end{figure}

A sky image of the mask used in the analysis of HFI maps at
$16\arcmin$ resolution is shown in Fig.~\ref{fig:IR_HI_mask}.  The
total area not masked is 7500\,deg$^2$ (18\,\% of the
sky).
The median \NH\ is $2.1\times 10^{20}\,{\rm H\, cm^{-2}}$, and 
$\NH < 3 \times 10^{20}$\,H\,cm$^{-2}$ for 74\,\% of the unmasked pixels.

\subsection{Galactic residuals with respect to the dust-\hi\ correlation}
 \label{subsec:GalRes}

 In this section, we describe the Galactic residuals with respect to
 the dust-\hi\ correlation. A power spectrum analysis of the CIB
 anisotropies over the cleanest part of the southern Galactic cap is
 presented in \citet{planck2013-pip56}.

\begin{figure*}[!h]
\centering
\includegraphics[width=0.46\textwidth]{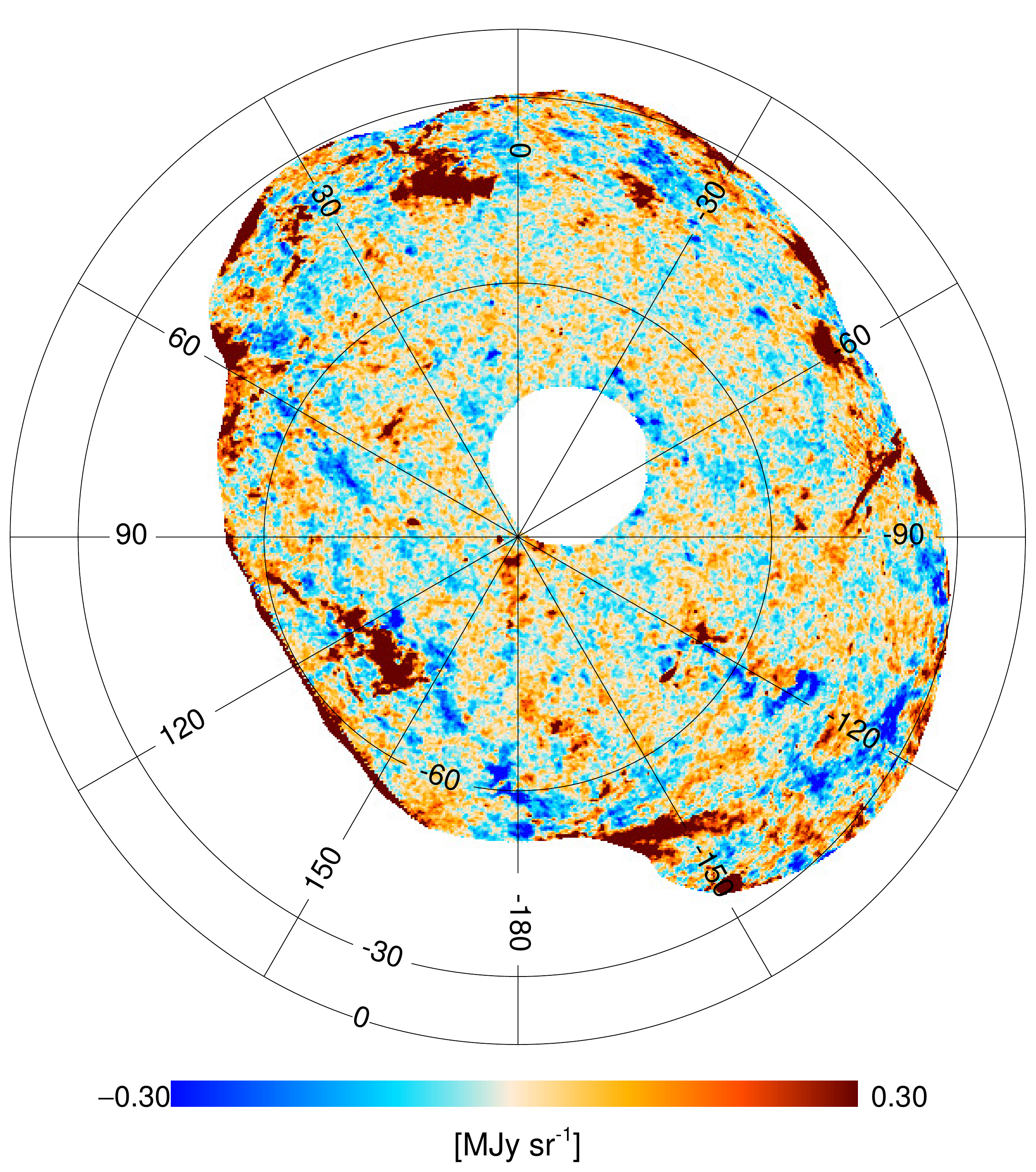}
\includegraphics[width=0.46\textwidth]{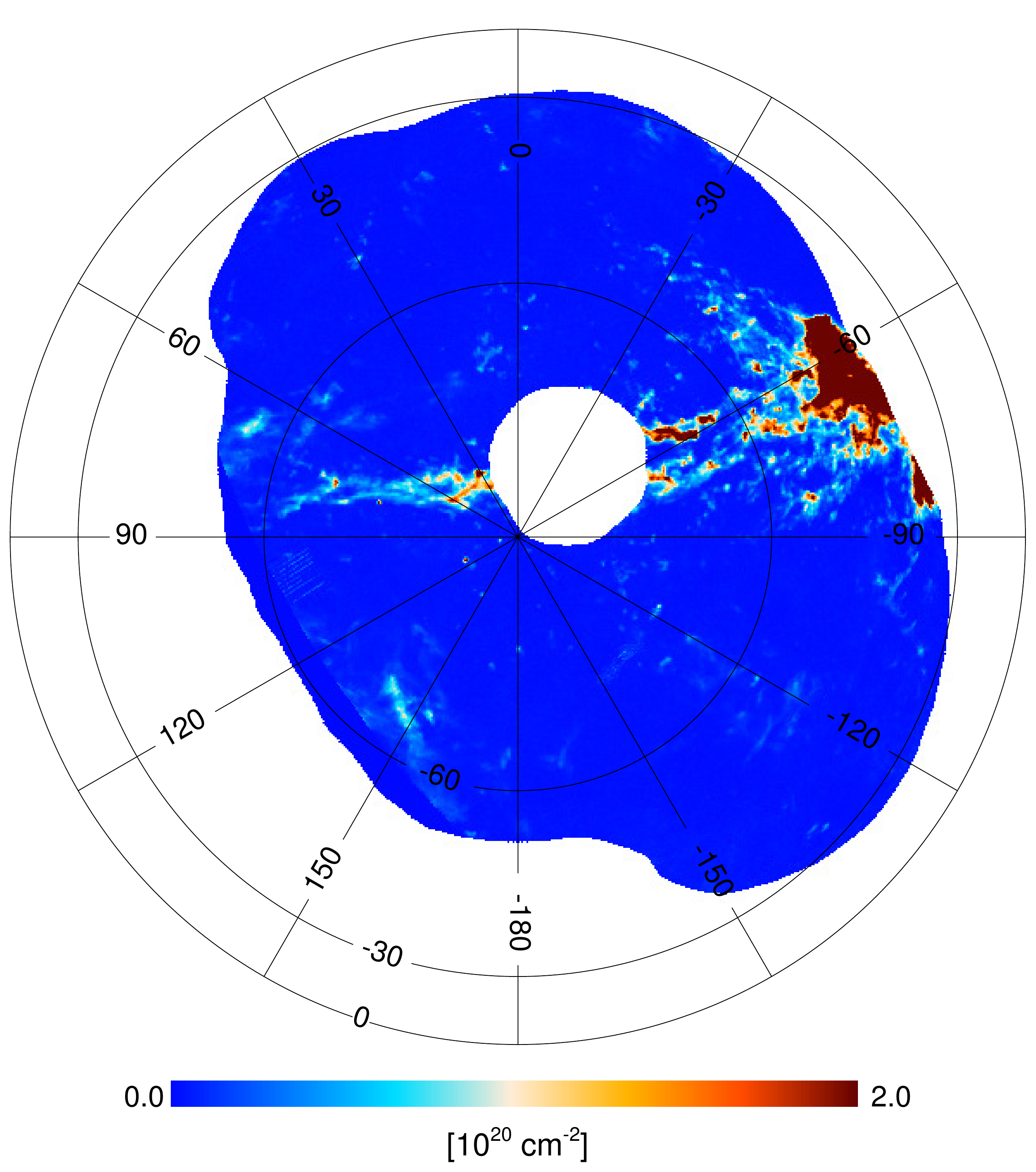}
\caption{{\it Left\/}: Image of the residual emission at 857\,GHz obtained by
subtracting the {\hi}-based model of the dust emission from the input
\planck\ map.  
  {\it Right\/}: Image of \NH\ from the Magellanic Stream (see
  Sect.~\ref{subsec:GASS_MS}), the sum of the IVC and HVC maps in
  Fig.~\ref{fig:IVC_HVC}.  %The data unit is $\rm 10^{20} \, H\, cm^{-2}$.
  }
\label{fig:Residu857}
\end{figure*}

Figure~\ref{fig:Residu857} shows the map of residual emission at
857\,GHz together with the map of \hi\ emission in the MS. The first
striking result from Fig.~\ref{fig:Residu857} is that the residual map
shows no evidence of dust emission from the MS. This result indicates
that the MS is dust poor; it will be detailed in a dedicated
paper. % in Appendix~\ref{appendix:MS}.

The residual map shows localized regions, both positive and negative,
that produce the non-Gaussian wings of the histogram in
Fig.~\ref{fig:histo_residuals}.   The
positive residuals are likely to trace dust emission associated with
molecular gas \citep{Desert88,Reach98,planck2011-7.12}.  
In addition, some positive residuals may be from dust emission associated with
Galactic IVC gas not in the Galactic \hi\ template.

The non-Gaussian tail toward negative residuals was not significant in
the earlier higher resolution \Planck\ study that analysed a much
smaller sky area at low \hi\ column densities.
%??
However, that analysis deduced emissivities for low velocity gas
and IVC gas independently, and did find many examples of IVCs with less than
half the typical emissivity.  If such gas were included in the
Galactic \hi\ template for $|v_{\rm dev}| \le 45$\,km\,s\mo, then
negative residuals could arise.
Another interesting possible interpretation, which needs to be
tested, is that negative residuals correspond to \hi\ gas at Galactic
velocities with no or deficient dust emission, akin to the MS, or to
typical HVC gas \citep{Peek09,planck2011-7.12}.
We do not discuss further these regions that are masked in our data
analysis.  Instead, we focus our analysis on the fainter residuals of
Galactic emission that together with CIB anisotropies make the Gaussian core
of the histogram in Fig.~\ref{fig:histo_residuals}.

\begin{figure}[!h]
\centering
\includegraphics[width=0.49\textwidth]{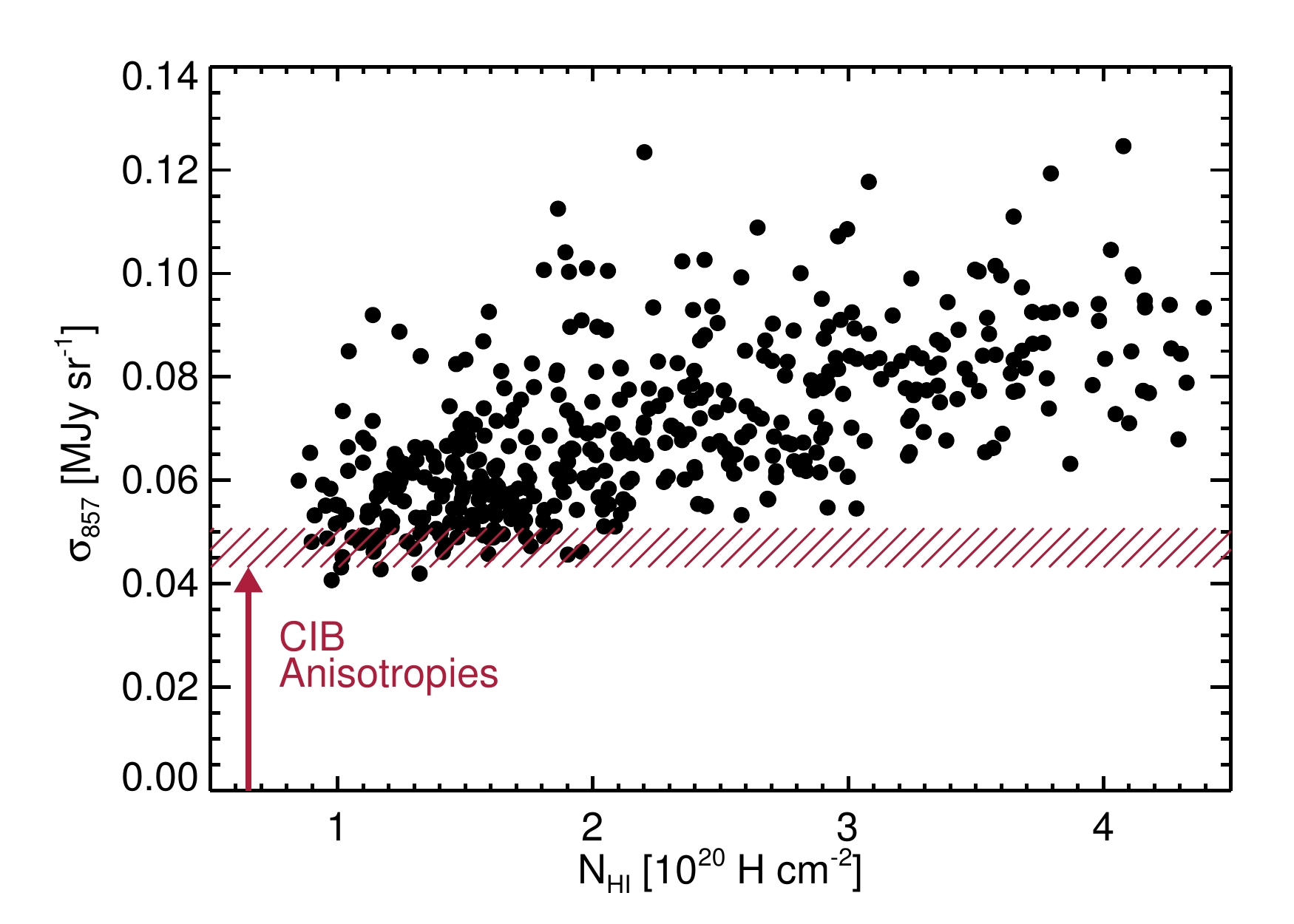}
 \caption{Standard deviation $\sigma_{857}$ of the residuals with
respect to the {\Planck}-\hi\ correlation at 857\,GHz versus the mean
\NH, both computed within circular sky patches with $5^\circ$ diameter
and over unmasked pixels.
The red hatched strip marks the contribution of CIB anisotropies to the residuals
at $16\arcmin$ resolution, computed from the CIB model in
\citet{planck2013-pip56}. The width of the strip represents the expected scatter
($\pm 1\, \sigma$) of this contribution.  Both the scattered
distribution of data points above CIB anisotropies strip and the increase in
the mean $\sigma_{857}$ with \NH\ arise from residuals with a
Galactic origin (Appendix~\ref{appendix:simus_GalRes}).}
\label{fig:sigma_residuals}
\end{figure}

To characterize the Gaussian component of the residuals with respect
to the dust-\hi\ correlation, we compute the standard deviation  $\sigma_{857}$ of the
residual map at 857\,GHz within circular apertures of 
%$2\pdeg5$ radius -- be consistent
$5^\circ$ diameter centred on $N_{\rm side}= 16$ pixels.
We choose this aperture size to be smaller than the sky patches used
to compute the dust emissivity so as to sample more finely $\sigma_{857}$.  Within each $5^\circ$
aperture, we compute the standard deviation  of the
residual 857\,GHz map and the mean \NH\ over unmasked pixels,
requiring at least 1000 of the maximum 1500 pixels available at
$N_{\rm side} =512$.
In Fig.~\ref{fig:sigma_residuals}, $\sigma_{857}$ is plotted versus
the mean \NH.  The hatched strip in the figure indicates the
contribution to $\sigma_{857}$ from CIB anisotropies at $16\arcmin$
resolution, as computed using the model power spectrum in
\citet{planck2013-pip56}. Most values of $\sigma_{857}$ are above the
strip. Since the contribution of noise to $\sigma_{857}$ is
negligible, there is a significant contribution to $\sigma_{857}$ from
residuals with a Galactic origin.  
The statistical properties of $\sigma_{857}$ -- the mean trend with
increasing \NH\ and the large scatter around this trend in
Fig.~\ref{fig:sigma_residuals} -- can be accounted for by a simple
model where the Galactic residuals arise from variations of the dust
emissivity on scales lower than the $15^\circ$ diameter of the
patches in our correlation analysis. In
Appendix~\ref{appendix:simus_GalRes}, we quantify this interpretation
with simulations.

The ratio of the dispersions from
Galactic residuals and from CIB anisotropies increases towards higher
frequencies, but it decreases with decreasing patch size used in the
underlying correlation analysis and with better angular resolution of
the \hi\ template map (Appendix~\ref{appendix:uncertainties}). Thereby 
an obvious Galactic contribution in the faintest fields was not
noticed in the earlier study with the GBT of \citet{planck2011-7.12},
but they did find an increase in the standard deviation of the
residuals with the mean column density (see their Fig.~12). 

Unlike the localized features that make the non-Gaussian part of the
histogram in Fig.~\ref{fig:histo_residuals}, the Gaussian contribution
cannot be masked out.  As discussed in \citet{planck2013-pip56}, it 
significantly biases the power spectrum of CIB anisotropies at $\ell < 100$, depending on
the range of \NH\ within the part of the sky used for the analysis.

\begin{figure*}[!h]
\centering
\includegraphics[width=0.46\textwidth]{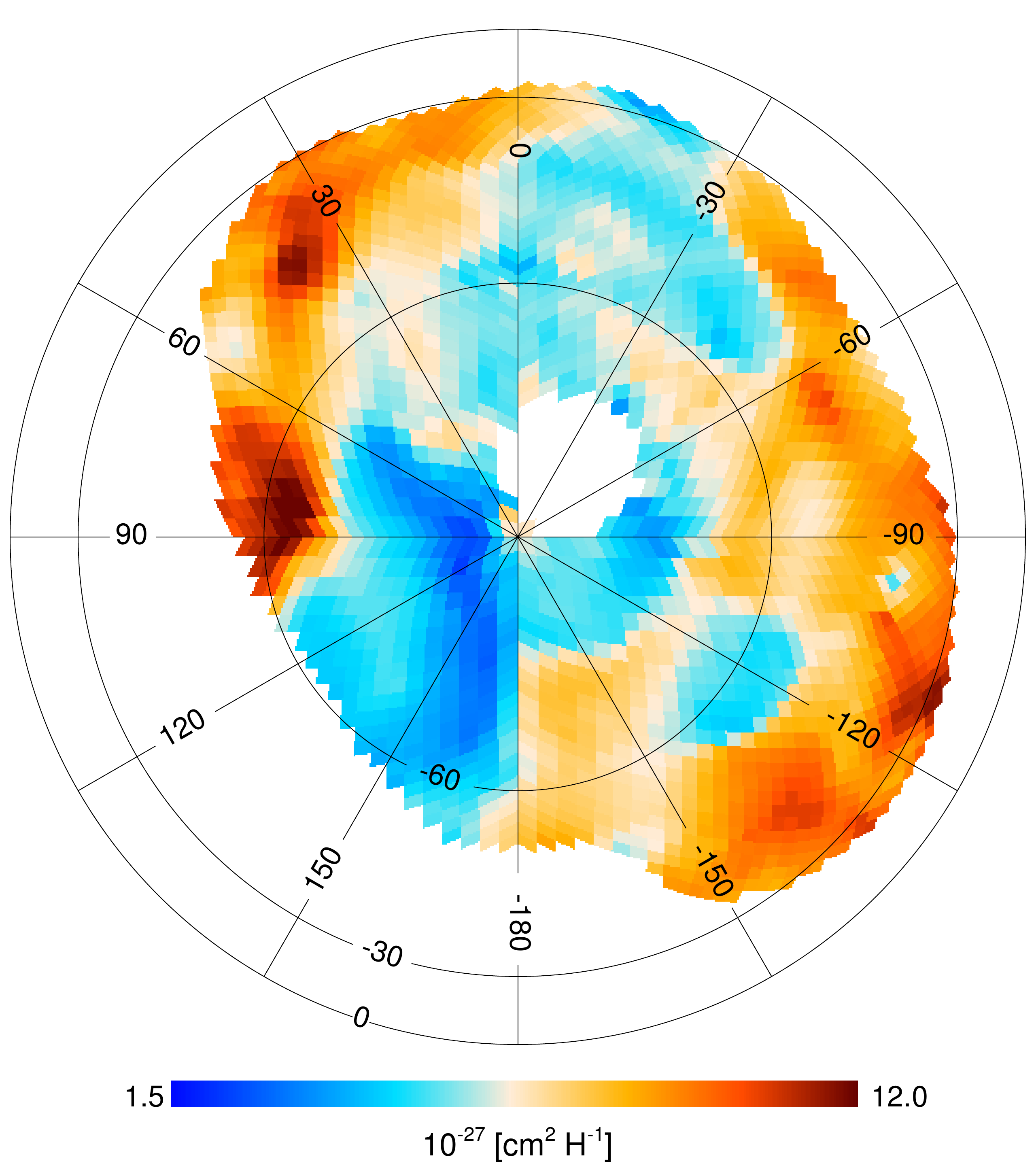}
\includegraphics[width=0.46\textwidth]{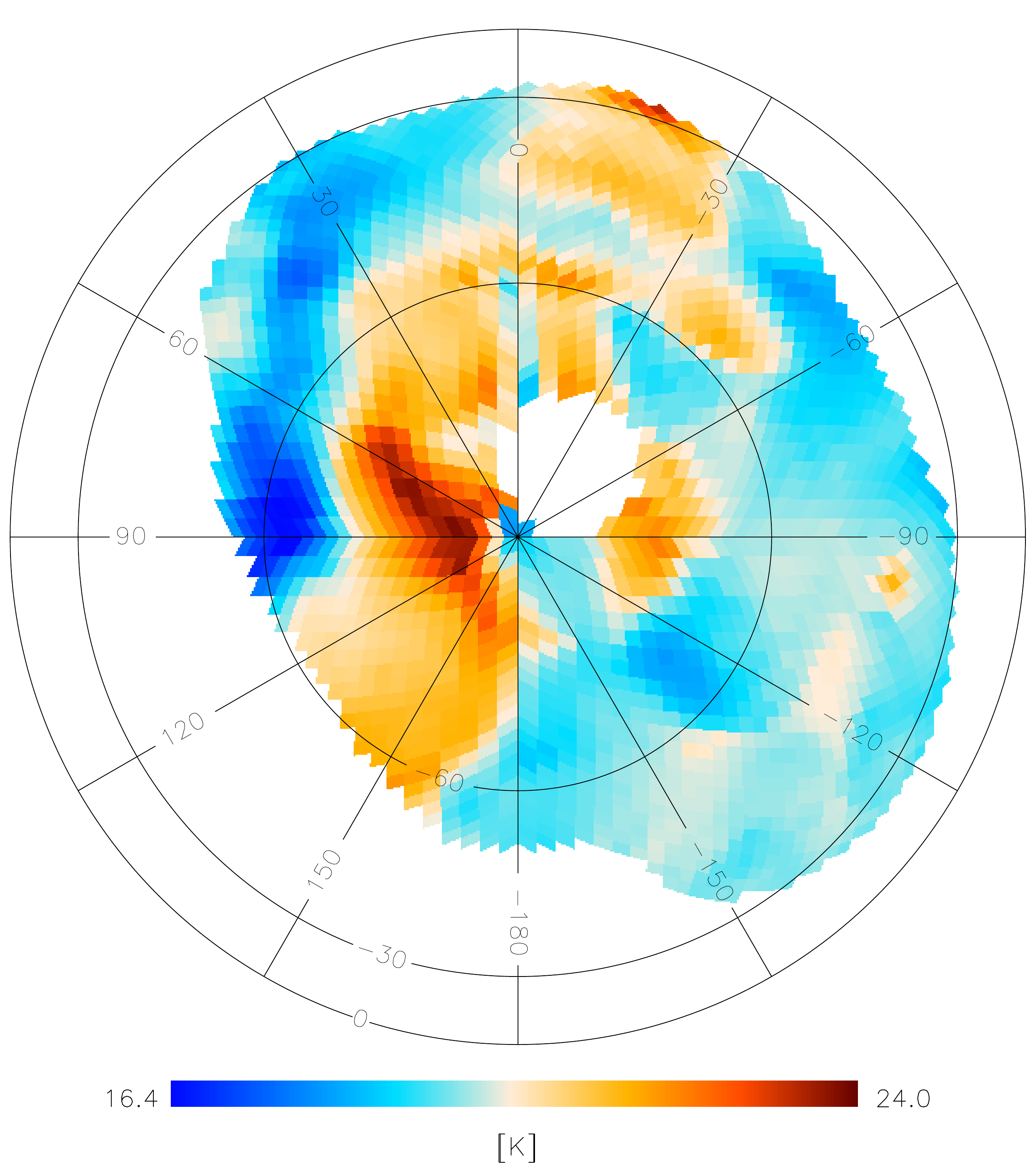}
\caption{{\it Left\/}: Map of the dust opacity  $\sigma_{\rm H}(353\,$GHz) in Eq.~(\ref{eq:dust_opacity}).
{\it Right\/}: Colour temperature map inferred from the ratio between the dust
emissivities at $100\,\mu$m from \DIRBE\ and 857\,GHz from \planck,
with a spectral index of the dust emissivity $\beta_{\rm FIR} =1.65$.
This figure reveals that the temperature and submillimetre opacity of dust are anti-correlated. }
\label{fig:col_temp}
\end{figure*}

\begin{figure}[!h]
\centering
\includegraphics[width=0.5\textwidth]{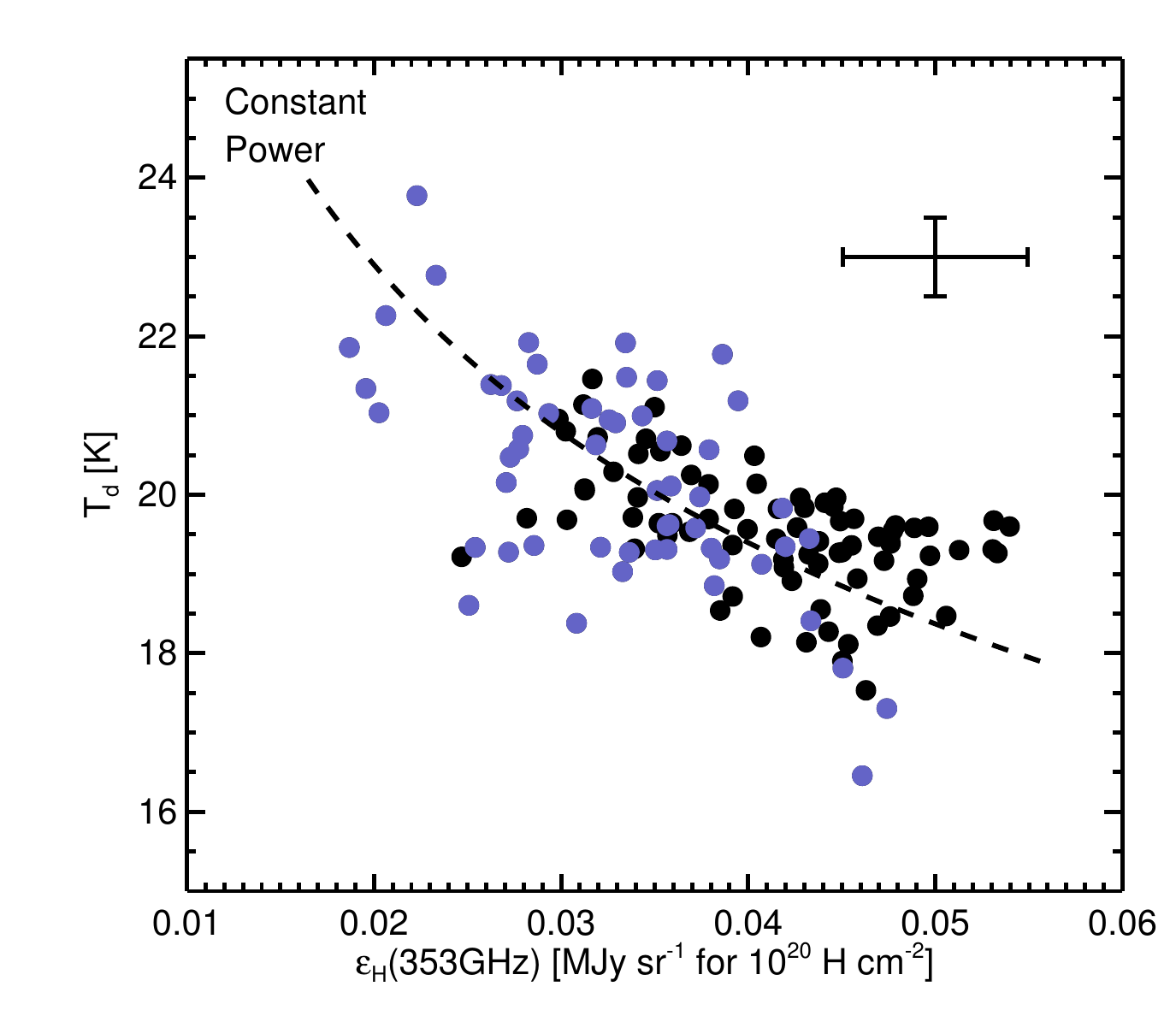}
\includegraphics[width=0.5\textwidth]{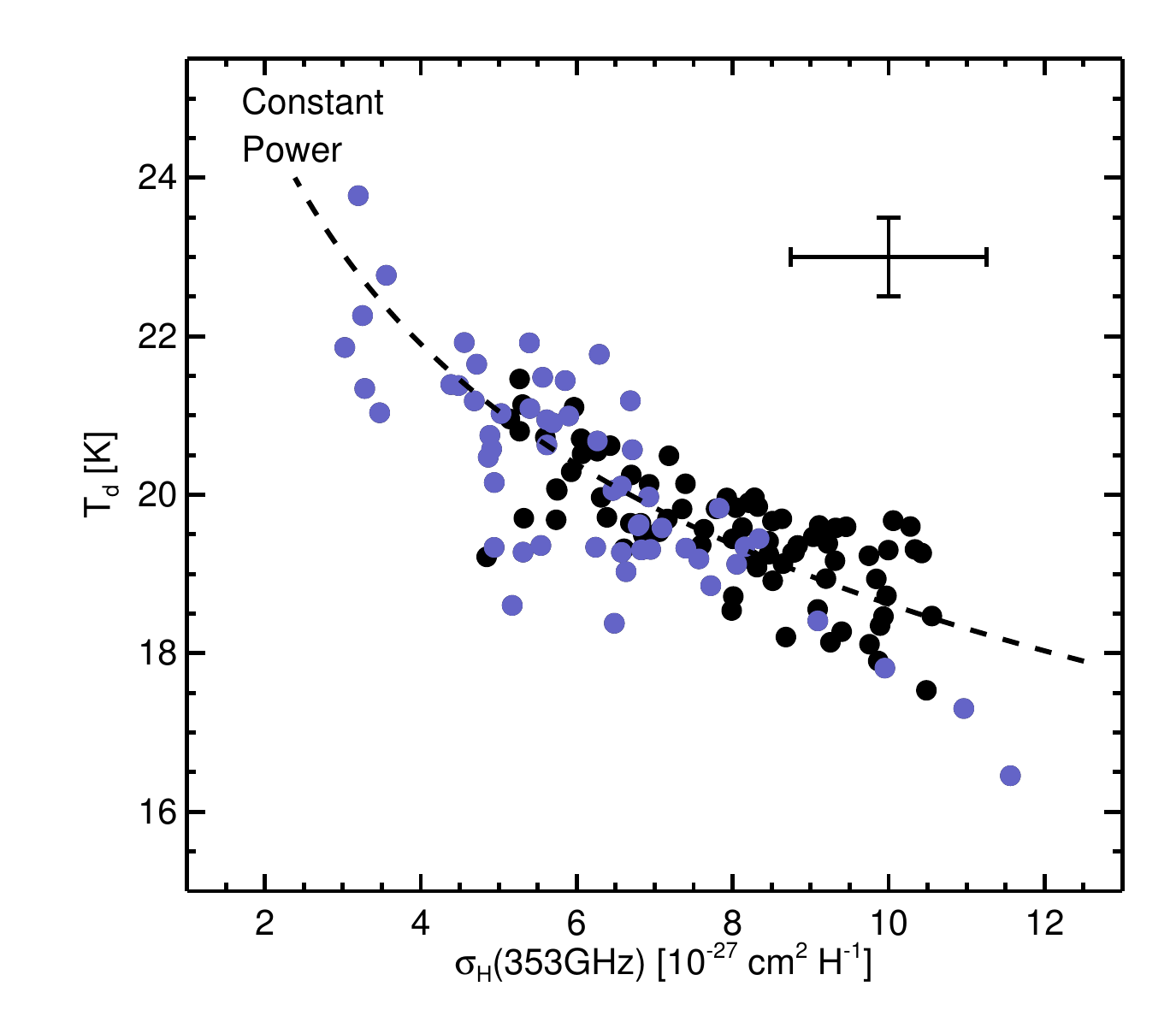}
\caption{{\it Top\/}: Dust colour temperature $T_{\rm d}$ versus dust
emissivity at $353\,$GHz, two independent observables (Fig.~\ref{fig:IR_HI_correlation}), with typical error bars at the top right.
% the top X label needs to read "per" not "for"
The dashed line represents the expected dependency of $T_{\rm d}$ on the dust emissivity for a fixed emitted power of $3.4\times10^{-31}$\,W\,H\mo.  
The blue dots identify data for sky patches centred at Galactic latitudes $b \le -60^\circ$.
{\it Bottom\/}: $T_{\rm d}$ versus dust opacity at $353\,$GHz,
re-expressing the same data in the form plotted by \citet{planck2011-7.12}
and \citet{Martin12}.  }
\label{fig:tau353_Tdust}
\end{figure}

\begin{figure}[!h]
\centering
\includegraphics[width=0.5\textwidth]{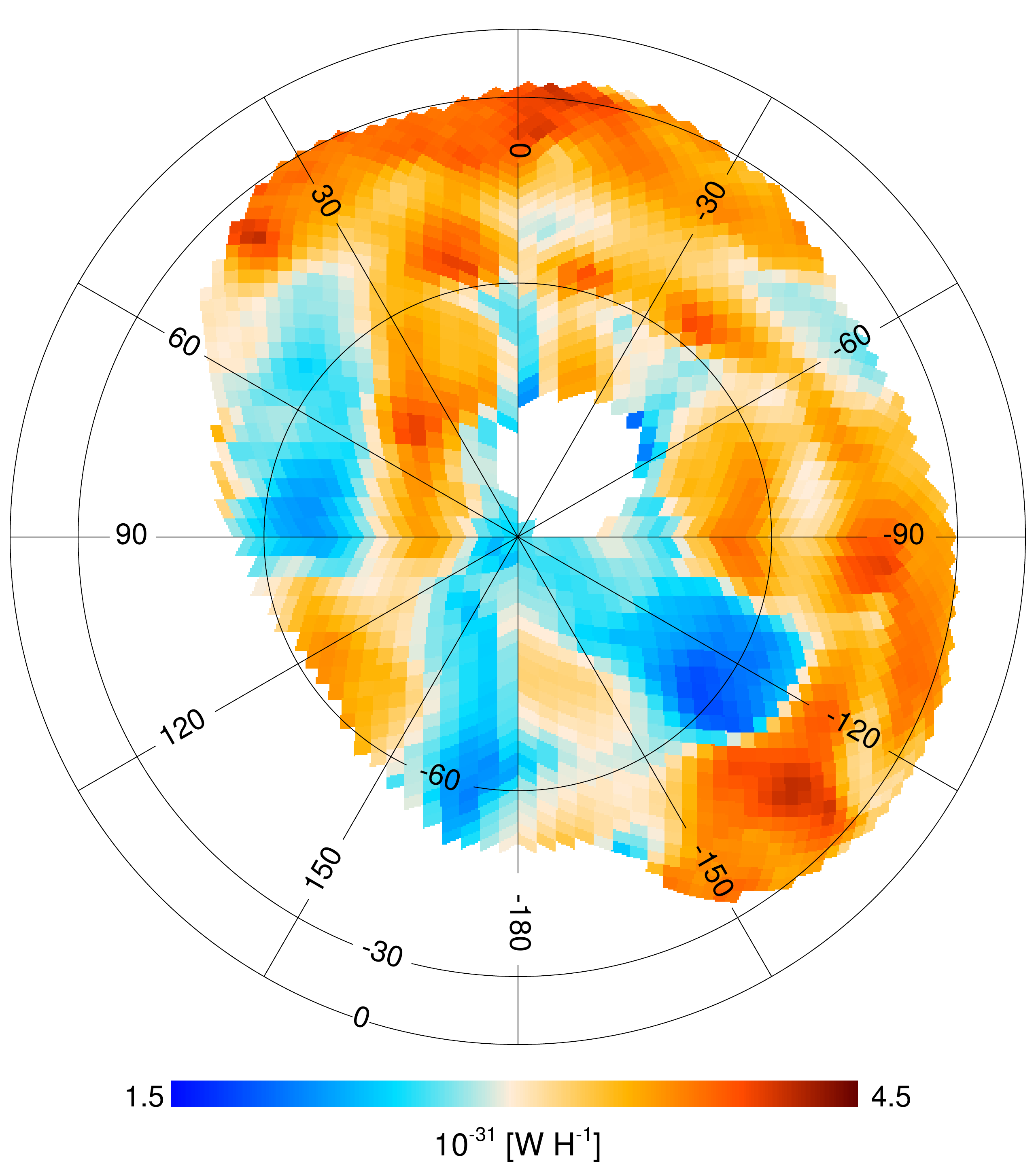}
\caption{Map of the specific power radiated by dust at far infrared wavelengths per H. % in units of $10^{-31} \, {\rm W\, H^{-1}}$.  
This figure displays spatial variations of the specific dust power,  which may be decomposed as the sum of two parts correlated 
with the opacity and temperature maps (see Fig.~\ref{fig:col_temp}), respectively. }
\label{fig:lum_H}
\end{figure}

\section{Dust emission properties across the southern Galactic cap}
\label{sec:southern_Cap}

In this section, we use the results from our analysis of the dust-\hi\
correlation to describe how dust emission properties vary across the
southern Galactic cap.

\subsection{Dust temperature and opacity}
\label{subsec:dust_emissivity}

At frequencies higher than $353$\,GHz, our analysis extends that of \citet{planck2011-7.12} to
a wider area.  The dust emissivities are consistent with earlier values, once we correct them 
for the change in calibration of the 857 and $545\,$GHz data that occurred after the
publication of the \planck\ Early Papers \citep{planck2013-p03f}.
The dust emissivity is observed to vary over the sky in a correlated
way between contiguous frequencies.\footnote{\citet{planck2011-7.12} reported
  a systematic difference between the dust emissivities measured for
  local velocity gas and IVCs.  This is difficult to
  confirm in our field where much of the gas in the IVC velocity range is low
  metallicity gas that belongs to the MS.}
In units of MJy\,sr$^{-1}$ per $\rm 10^{20}\, H \, cm^{-2}$, the dust
emissivity at 857\,GHz ranges from 0.20 to 0.57 with a mean 0.43. 
\footnote{This range is much higher than the fractional uncertainty of 13\,\% on the emissivity. See
 Appendix~\ref{appendix:uncertainties}.}  The emissivity also varies
by nearly a factor of three at 353\,GHz (see Fig.~\ref{fig:IR_HI_correlation}), and by a factor of four at 100\microns.  
The fact that we work on a large contiguous sky area allows us to map
these variations over the sky and assess their nature.  

Figure~\ref{fig:col_temp} displays maps of the dust temperature and submillimetre opacity. 
The map of colour temperature $T_{\rm d}$ is derived 
from the ratio between the dust emissivities at $100\,\mu$m
from \DIRBE\ and at 857\,GHz from \planck, $R(3000,857)$.  
We do not use the dust emissivities from the 140 and $240\,\mu$m \DIRBE\ bands because these maps are noiser (see
Fig.~\ref{fig:IR_HI_uncertainties}).
The colour ratio is converted into a colour temperature assuming a
greybody spectrum
\begin{equation}
\label{eq:greybody}
I_\nu  = {\rm cc}(T_{\rm d},\beta) \tau_{\nu_0} \, (\nu/\nu_0)^{\beta} \,  B_\nu(T_{\rm d}),
\end{equation}
where $cc$ is the colour-correction \citep{planck2013-p03d}, $B_\nu$ is the Planck
function, $T_{\rm d}$ is the dust temperature, and $\beta$ is the dust spectral
index.  In the far infrared, we adopt $\beta_{\rm FIR}= 1.65$, the value found fitting a greybody to the mean dust SED
at $\nu \ge 353\,$GHz.
The reference frequency $\nu_0$ and the optical depth there
$\tau_{\nu_0}$, divide out in the colour ratio.
The mean colour temperature is $19.8\,$K, in good agreement with what
is reported for the same part of the sky in \citet{planck2013-p06b} for the same
$\beta_{\rm FIR}$.
The dust opacity is computed from the dust emissivity and
colour temperature:
\begin{equation}
\label{eq:dust_opacity}
\sigma_{\rm H}(\nu)  = \epsilon_{\rm H}(\nu) / B_\nu(T_{\rm d}),
\end{equation}
the equivalent of the optical depth divided by \NH.

The two  maps in Fig.~\ref{fig:col_temp}
illustrate an anti-correlation between the dust opacity and the
colour temperature, first reported in \citet{planck2011-7.12}. Our analysis
confirms their result over a wider sky area.  The anti-correlation
is at odds with the expected increase in the dust emissivity with dust
temperature.  It suggests that the temperature is a response to 
variations in dust emission properties and not in the heating rate of
dust.
To support this interpretation, in Fig.~\ref{fig:tau353_Tdust} we plot
the dust temperature versus the dust emissivity and opacity at
$353\,$GHz.  
As in earlier studies where different data sets and sky regions have
been analysed \citep{planck2011-7.12,Martin12,Roy13}, we find that the dust
temperature is anti-correlated with the dust emissivity and opacity in
such a way that the far infrared specific dust power (i.e. the thermal
emission integrated over the far infrared SED, per H) is constant.  The
dashed line in each panel corresponds to the mean value of the far infrared
power, $3.4\times10^{-31} {\rm W\, H^{-1}}$, as also found by
\citet{planck2013-p06b} for high latitude dust. 

To check that the anti-correlation does not depend on our assumption of  a fixed $\beta_{\rm FIR}$ used to compute the colour temperatures, 
we repeat our analysis with dust temperatures and opacities derived from
a greybody  fit to the dust emissivities  at 100\microns\ and the \Planck\ 353, 545 and $857\,$GHz frequencies, for each sky patch.   
The dust temperatures from these fits are closely correlated to the colour temperatures determined from the 100\microns\ and $857\,$GHz  colour ratio. 
The mean temperature is $19.8\,$K for both sets of dust temperatures because 
the $\beta_{\rm FIR}$,  $1.65$,  used in the  calculation of colour temperatures is the mean of the values derived from the greybody fits.
We find that variations of the dust spectral index do not change the anti-correlation  
between dust opacity and temperature, but they increase the scatter of the data points by about 20\%. 

The far infrared power emitted by dust
equals that absorbed from the interstellar radiation field (ISRF) and
so, as discussed by \citet{planck2011-7.12} and \citet{Martin12}, the fact
that the power is quite constant has two implications.
(1) Increases (decreases) in the equilibrium value of $T_{\rm d}$ are a
response to decreases (increases) in the dust far infrared opacity (the
ability of the dust to \emph{emit} and thus cool).\quad
(2) The optical/UV \emph{absorption} opacity of dust must be
relatively unchanged, given that variations in the strength of the
ISRF are probably small within the local ISM.  
Thus, an observational constraint to be understood in grain modeling
is that the ratio of far infrared to optical/UV opacity changes within the
diffuse ISM. 

The anti-correlation between $T_{\rm d}$ and $\sigma_{\rm H}(353$GHz) at constant power does not fully characterize 
the spatial variations of the dust emission properties.   The scatter of the data points in 
Fig.~\ref{fig:tau353_Tdust} around the line of constant power is not noise. 
Figure~\ref{fig:lum_H} displays 
variations over the southern polar cap of the specific power radiated by dust at far-IR wavelengths per H  (Fig.~\ref{fig:col_temp}).
They could result from variations in the dust-to-gas ratio, the dust absorption cross section per H of star light, and/or the ISRF intensity.  
The dust-to-H mass ratio is inferred from spectroscopic measurements of elements depletions 
to vary in the local interstellar medium from 0.4\,\% in warm gas to 1\,\% in cold neutral medium \citep{Jenkins09}.

\subsection{Dust evolution within the diffuse ISM}
\label{subsec:dust_evol}

Our analysis provides evidence of a varying ratio between the dust
opacity at far infrared and visible/UV wavelengths, strengthening the early
results from \citet{planck2011-7.12}.  These two \planck\ papers extend to
the diffuse atomic ISM results reported in many studies for the
translucent sections of molecular clouds
\citep{Cambresy01,Stepnik03,planck2011-7.13,Martin12,Roy13} .
Evidence of dust evolution in the diffuse ISM from far-IR observations of large dust grains  was first reported by \citet{Bot09}.

The observations of dust evolution in molecular clouds are often
related to grain growth associated with mantle formation or grain
coagulation/aggregation.  Model calculations do indeed show that the
variations in the far infrared dust opacity per unit $A_v$ may be accounted for by
grain coagulation \citep{Kohler12}.  The fact that such variations are
now observed in \hi\ gas, where densities are not high enough for
coagulation to occur, challenges this interpretation.  It would be
more satisfactory to propose an interpretation that would account for
opacity variations in both the diffuse ISM and
molecular clouds.  \citet{Jones12} and \citet{Jones13}
take steps in this direction by introducing evolution of carbon dust
composition and properties into their dust model.  A quantitative
modeling of the data has yet to be done within this new framework, but
the results presented by \citet{Jones13} are encouraging. The
variations in the far infrared opacity and temperature of dust could trace
the degree of processing by UV photons of hydrocarbon dust formed
within the ISM.

Alternatively, the variations of the far infrared dust opacity could result
from changes in the composition and structure of silicate dust.  At
the temperature of interstellar dust grains in the diffuse
ISM, low energy transitions, associated with disorder
in the structure of amorphous solids on atomic scales, contribute to
the far infrared dust opacity. This contribution depends on the dust
temperature and on the composition and structure of the grains
\citep{Meny07}.  The dust opacity of silicates is 
observed in laboratory experiments \citep{Coupeaud11} to depend on
parameters describing the amorphous structure of the grains, which may
evolve in interstellar space through, for example, exposure to cosmic
rays.

A different perspective is considered in \citet{Martin12}. Dust
evolution might not be ongoing now within the diffuse ISM.  Instead,
the observations might reflect the varying composition of interstellar
dust after evolution both within molecular clouds and while recyling
back to the diffuse ISM, reaching different end points.

\begin{figure}[!h]
\centering
\includegraphics[width=0.46\textwidth]{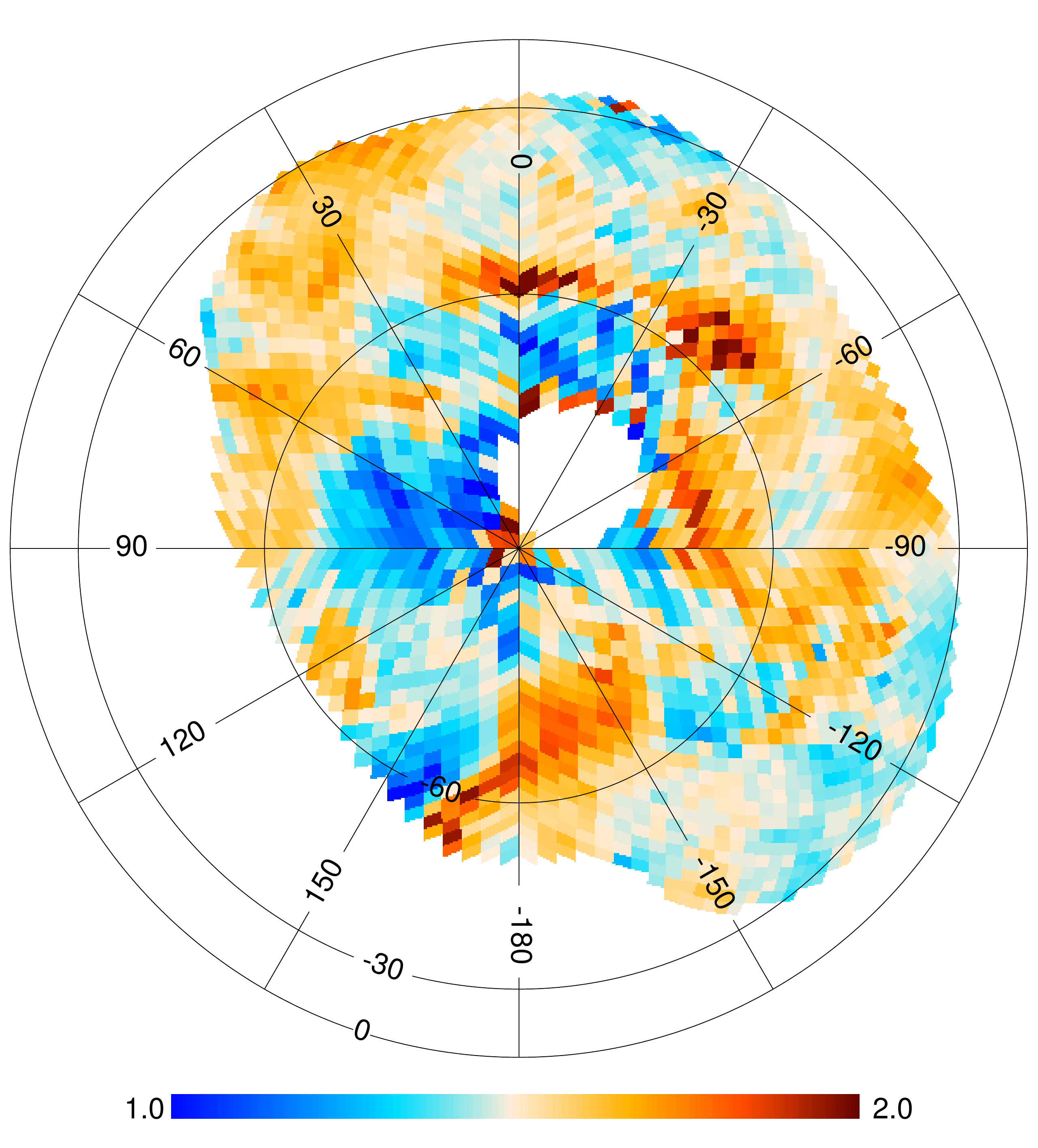}
\caption{Spectral index $\beta_{\rm mm}$ of the dust emission derived from
the ratio between correlation measures at 353 and $217\,$GHz (both
corrected for the CMB contribution by subtracting the correlation
measure at 100\,GHz) and the colour temperature map in
Fig.~\ref{fig:col_temp}.
}
\label{fig:beta_map}
\end{figure}

\begin{figure}[!h]
\centering
\includegraphics[width=0.49\textwidth]{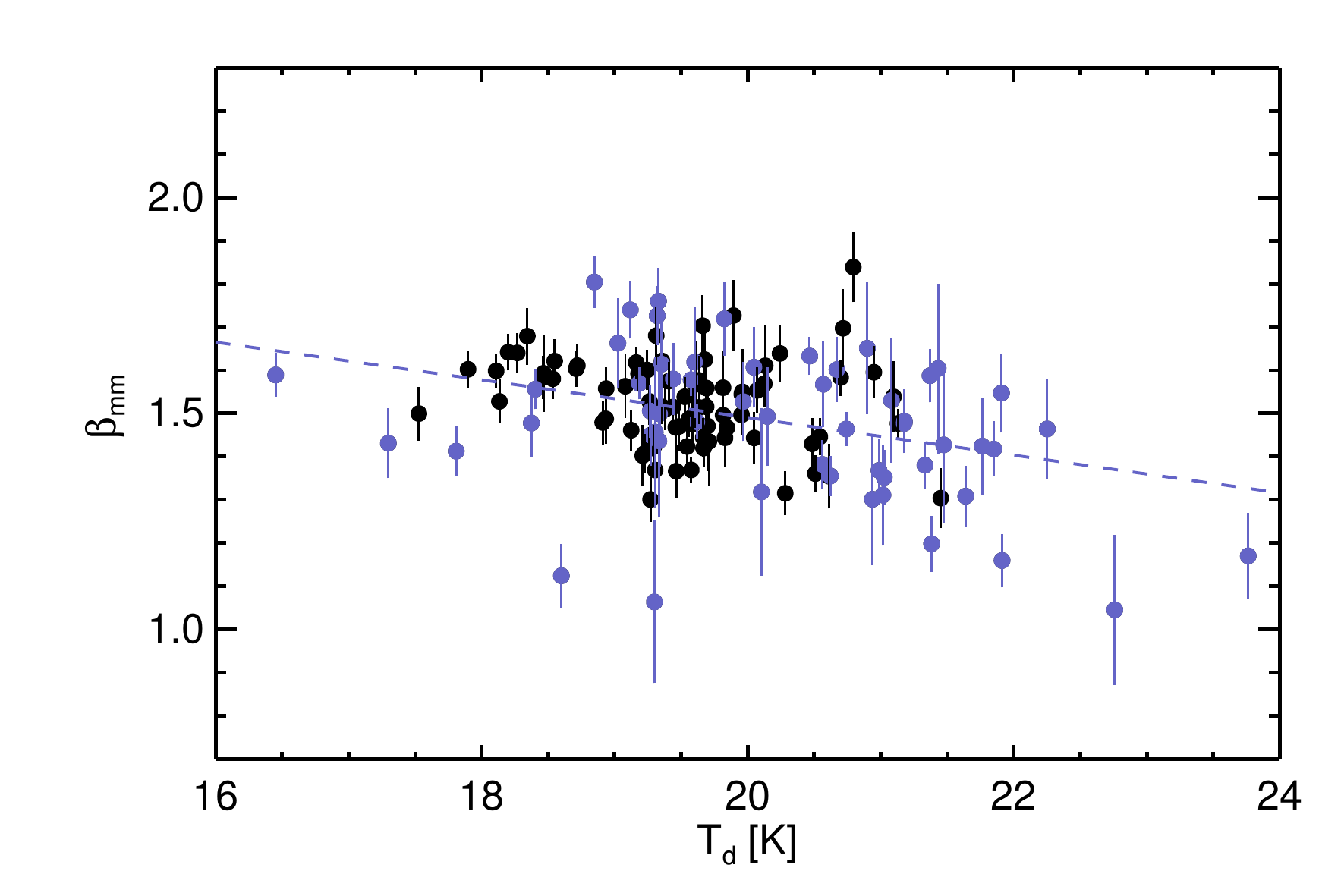}
\caption{Spectral index $\beta_{\rm mm}$ versus $T_{\rm d}$ for the 135 sky
  patches. The blue dots distinguish patches centred at Galactic
  latitude $b\le -60^\circ$.  The uncertainties are derived from
  simulations.  The dashed line is a linear regression of $\beta_{\rm mm}$ on
  $T_{\rm d}$,  slope $(-0.043 \pm 0.009)$\,K$^{-1}$.
}
 \label{fig:beta_plot}
\end{figure}

\section{The dust spectral index from submillimetre to millimetre wavelengths}
\label{sec:microwave_index}

Our analysis of the \planck\ data allows us to measure the spectral
index of the thermal dust emission from submillimetre to millimetre wavelengths
$\beta_{\rm mm}$.  This complements measurements of the spectral index at
far infrared wavelengths $ \beta_{\rm FIR}$ in \citet{planck2013-p06b} and many earlier
studies \citep[e.g.][]{Dupac03}.

\subsection{Measuring the spectral index}

For each circular sky patch, we compute the colour ratio $R_{100}(353,217) =
\alpha^{100}_{\rm 353GHz}/\alpha^{100}_{\rm 217GHz}$, where
$\alpha^{100}_{\nu}$ is the correlation measure at frequency $\nu$
corrected for the CMB contribution by subtracting the correlation
measure at $100\,$GHz (Sect.~\ref{subsec:cc}).  The colour ratio is
converted into a spectral index using a greybody spectrum (Eq.~\ref{eq:greybody}).
We compute $R_{100}(353,217)$ for a grid of values of $\beta_{\rm mm}$ and
$T_{\rm d}$.  For each sky patch, adopting the colour temperature determined
above independently from the $R(3000,857)$ colour ratio, we find the
value of $\beta_{\rm mm}$ that gives a match with the observed
$R_{100}(353,217) $.  We obtain the $\beta_{\rm mm}$ map presented in
Fig.~\ref{fig:beta_map}.

The mean value and standard deviation (dispersion) of $\beta_{\rm mm}$ are
1.51 and 0.13 for \Planck\ maps without subtraction of the model of
zodiacal emission, and 1.51 and 0.16 for maps with the model
subtracted. The standard deviation of the patch by patch difference
between these two $\beta_{\rm mm}$ values is 0.10, only slightly lower
than the dispersion of each.
The mean $\beta_{\rm mm}$ is in good agreement with the value of 1.53 estimated for the more
diffuse atomic regions of the Galactic disk by \citet{planck2013-XIV}, but
it is lower than values close to 2 derived from
the analysis of \COBE\ data at higher frequencies
\citep{Boulanger96,Finkbeiner99}. For comparison, we computed a value of 
$\beta_{\rm FIR}$ for each sky patch by fitting a greybody 
to the dust emissivities at the high frequency \planck\ channels ($\nu \ge 353\,$GHz) and 
at $100\,\mu$m. The difference $\beta_{\rm FIR}-\beta_{\rm mm}$ has a median value of  0.15, and shows 
no systematic dependence   on the colour temperature $T_{\rm d}$.

For the derivation of $\beta_{\rm mm}$, we have assumed that the dust
emission at 100\,GHz is well approximated by a greybody
extrapolation from 353 to 100\,GHz. To check that
this assumption does not introduce a bias, we repeat the data analysis
on \Planck\ maps in which the CMB anisotropies have been subtracted
using the CMB map obtained with {\tt SMICA} 
\citep{planck2013-p06}.  This allows us to compute the spectral index
$\beta_{\rm mm}(\tt SMICA)$ directly from the ratio between the 353 and
$217\,$GHz correlation measures.  The mean value of the differences
$\beta_{\rm mm} - \beta_{\rm mm}(\tt SMICA) $ is negligible, i.e. there is no
bias.

\subsection{Variations with dust temperature}

Many studies, starting with the early work of \citet{Dupac03}, have
reported an anti-correlation between $\beta_{\rm FIR}$ and dust
temperature.
Laboratory data on amorphous silicates indicate that, at the
temperature of dust grains in the diffuse ISM, it is at millimetre wavelengths
that the variations of the spectral index may be the largest
\citep{Coupeaud11}.  These laboratory results and astronomical data,
have been interpreted within a model where variations in the dust
spectral index stem from the contribution of low energy transitions,
associated with disorder in the structure of amorphous solids on
atomic scales, to the dust opacity
\citep{Meny07,Paradis11}. Variations of $\beta_{\rm mm}$ are also
predicted to be possible signatures of the evolution of carbon dust
\citep{Jones13}.
 
Our analysis allows us to look for such variations over a frequency
range where the determination of the spectral index is to a large
extent decoupled from that of the dust temperature.  We determine the
dust colour temperature $T_{\rm d}$ and the spectral index $\beta_{\rm mm}$ from
two independent colour ratios, whereas in far infrared studies the spectral
index $\beta_{\rm FIR}$ and temperature $T_{\rm d}$ are determined
simultaneously from a spectral fit of the SED \citep{Schnee09,planck2013-p06b}.  Althought $T_{\rm d}$ is used in the
conversion of $R_{100}(353,217) $ into $\beta_{\rm mm}$, the uncertainty of $T_{\rm d}$
has a marginal impact.  Furthermore, the photometric uncertainty of far infrared
data is higher than that at $\nu \le 353\,$GHz, where the data
calibration is done on the CMB dipole.

We start quantifying the uncertainties of $\beta_{\rm mm}$ using the
numerical simulations presented in the companion \Planck\ paper
\citep{planck2014-XXII} that extends this work to dust polarization.  These simulations include \hi-correlated dust
emission with a fixed %emission temperature $19.4\,$K and a
spectral index 1.5, dust emission uncorrelated with \hi\ with a
spectral index of 2, noise, CIB anisotropies, and free-free emission.  We
analyse 800 realizations of simulated maps at 100, 143, 217, and
$353\,$GHz with the same procedure as used on the \Planck\ data.  For
each sky patch, we obtain 800 values of $\beta_{\rm mm}$. The additional
components do not bias the estimate of $\beta_{\rm mm}$, but introduce
scatter around the mean input value of 1.5.  We use the standard
deviation of the extracted $\beta_{\rm mm}$ values as a noise estimate
$\sigma_\beta$ for each sky patch. 

The noise on $\beta_{\rm mm}$ shows a
systematic increase towards low \NH, something that we also observe
for the \Planck\ analysis.  We also measure the standard deviation of
$\beta_{\rm mm}$ over sky patches for each simulation. We find a value of
$0.079\pm 0.01$, lower than the dispersion 0.13 measured on the
\Planck\ data. If the simulations provide a good estimate of the
uncertainties, the higher dispersion for the data shows that
$\beta_{\rm mm}$ has some variance.
This can be appreciated in Fig.~\ref{fig:beta_plot}, where the values
of $\beta_{\rm mm}$ with their uncertainties are plotted versus the dust
temperature $T_{\rm d}$.  The plot also displays the result of a linear
regression, which has a slope of
%$(-0.043 \pm 0.0085)$\,K$^{-1}$), 
$(-0.043 \pm 0.009)$\,K$^{-1}$.
Using the set of temperatures obtained from the greybody fits increases the spread of  the data points in Fig. ~\ref{fig:beta_plot}. 
The slope is changed to $(-0.053 \pm 0.007)$\,K$^{-1}$.
The non-zero slope implies some variation of $\beta_{\rm mm}$, and also suggests that
$\beta_{\rm mm}$ and $T_{\rm d}$ are anti-correlated.
This would extend to the millimetre range a result that has been
reported in many studies for $\beta_{\rm FIR}$ versus $T_{\rm d}$, but the
variations here are small and perhaps only marginally significant. The constancy of $\beta_{\rm mm}$  is an observational constraint on the nature
of the  process at the origin of variations of the far-IR dust opacity (Sect.~\ref{subsec:dust_evol}).  We
note that \citet{planck2013-XIV} do not find evidence of an anti-correlation
in their analysis of \Planck\ observations of the diffuse emission in
the Galactic disk.

\section{The spectral energy distribution of Galactic dust in the diffuse ISM}
\label{sec:Dust_SED}

At the \Planck-LFI
and \wmap\ frequencies, the signal-to-noise ratio on the dust
emissivity for a given sky patch is very low because the signal is
very faint compared to CMB anisotropies and noise.  However, by
averaging the emissivities over sky patches, we obtain an SED of dust
emission spanning the full spectral range and computed consistently at
all frequencies (Sect.~\ref{subsec:mean_SED}).  We present greybody
fits of the thermal emission of dust at $\nu \ge 100\,$GHz in
Sect.~\ref{subsec:greybody}.  The SED is compared with existing models
in Sect.~\ref{subsec:dust_models}.

%%%%%%%%%%%%%%%%%%%%%
% Table 1 Mean Dust SED
%%%%%%%%%%%%%%%%%%%%%%
\begin{table*}[tmb]
\begingroup
\newdimen\tblskip \tblskip=5pt
\caption{\label{tab:mean_sed16} Mean SED of dust emissivity from \hi\ correlation}
\nointerlineskip
\vskip -3mm
\tiny
\setbox\tablebox=\vbox{
   \newdimen\digitwidth 
   \setbox0=\hbox{\rm 0} 
   \digitwidth=\wd0 
   \catcode`*=\active 
   \def*{\kern\digitwidth}
   \newdimen\signwidth 
   \setbox0=\hbox{+} 
   \signwidth=\wd0 
   \catcode`!=\active 
   \def!{\kern\signwidth}
\halign{
\hbox to 4.4 cm{#\leaderfil}\tabskip 1.7em&
\hfil #\hfil\tabskip 0.8em&
\hfil #\hfil&
\hfil #\hfil&
\hfil #\hfil&
\hfil #\hfil&
\hfil #\hfil&
\hfil #\hfil&
\hfil #\hfil&
\hfil #\hfil&
\hfil #\hfil&
\hfil #\hfil\tabskip=0pt\cr
\noalign{\doubleline}
\omit&\multispan{11}\hfil F{\sc requency} [GHz]\hfil\cr
\omit&\multispan{11}\hfil E{\sc xperiment}\hfil\cr
\noalign{\vskip -3pt}
\omit&\multispan{11}\hrulefill\cr
\noalign{\vskip 2pt}
\omit&  70&  94& 100& 143& 217& 353& 545& 857&1249&2143&2997\cr
\omit\hfil Q{\sc uantity}\hfil& LFI& WMAP& HFI& HFI& HFI& HFI& HFI& HFI& DIRBE& DIRBE& DIRBE\cr
\noalign{\vskip 4pt\hrule\vskip 3pt}
$\epsilon_{\rm H}(\nu)$ [\MJH]&0.00027&0.00045&0.00067& 0.0020& 0.0086&  0.039&   0.14&   0.43&   0.84&    1.1&   0.63\cr
$\sigma_{\rm stat}$ [\MJH]&$2.8\times 10^{-5}$&$8.9\times 10^{-5}$&$2.8\times 10^{-5}$&$7.9\times 10^{-5}$& $3.0\times 10^{-4}$& 0.0013& 0.0045&  0.013&  0.027&  0.048&  0.022\cr
${\rm phot}_{unc}$ [\,\%]& 0.5& 0.2&  0.5&  0.5&  0.5&  1.2& 10.0& 10.0& 11.6& 10.6& 13.6\cr
$\sigma_{\rm tot}$ [\MJH]&$2.8\times 10^{-5}$&$8.9\times 10^{-5}$&$2.8\times 10^{-5}$&$7.9\times 10^{-5}$&$3.0\times 10^{-4}$& 0.0014&  0.015&  0.045&   0.10&   0.13&  0.088\cr
${\rm cc}$&0.96& 0.98& 1.09& 1.02& 1.12& 1.11& 1.10& 1.02& 1.00& 0.94& 0.92\cr
${\rm uc}$&  7.54&   4.63&   4.10& 2.69&   2.07&  3.48&\dots&\dots&\dots&\dots&\dots\cr 
\noalign{\vskip 5pt\hrule\vskip 3pt}}}
\endPlancktablewide
\  \  \  $\epsilon_{\rm H}(\nu)\equiv$ \vtop{\hsize=170mm\noindent\strut Mean dust emissivity $\epsilon_{\rm H}(\nu)$ expressed as monochromatic brightness  at the reference frequencies, derived from correlation of the maps with the Galactic \hi\ template.  Not colour corrected.\strut }\par
\  \  \   $\sigma_{\rm stat}\equiv$ Statistical uncertainty ($1\sigma$) of the mean emissivities.\par
\  \  \   ${\rm phot}_{\rm unc}~(\,\%)\equiv$ Uncertainties of the absolute calibration [\,\%] from \citet{planck2013-p01}, \citet{Bennett12}, and \citet{Hauser98}.\par
 \  \  \   $\sigma_{\rm tot}\equiv$ Total uncertainty combining statistical and photometric uncertainties  [MJy\,sr$^{-1}$ per $10^{20}\,{\rm H\,cm}^{-2}$].\par
\  \  \   ${\rm cc}\equiv$ Colour-correction factors in Eq.~(\ref{eq:greybody}) computed with the greybody parameters listed in
Table~\ref{tab:greybody_fits}.\par
\  \  \   ${\rm uc}\equiv$ Unit conversion factors from MJy\,sr$^{-1}$ to thermodynamic (CMB) temperatures in mK.\par

\endgroup
\end{table*}

%%%%%%%%%%%%%%%%%%%%%
% Table 2  Mean microwave SED
%%%%%%%%%%%%%%%%%%%%%%

\begin{table*}[tmb]
\begingroup
\newdimen\tblskip \tblskip=5pt
\caption{\label{tab:mean_sed60} Mean microwave SED from \hi\ correlation}
\nointerlineskip
\vskip -3mm
%\footnotesize
\tiny
\setbox\tablebox=\vbox{
   \newdimen\digitwidth 
   \setbox0=\hbox{\rm 0} 
   \digitwidth=\wd0 
   \catcode`*=\active 
   \def*{\kern\digitwidth}
   \newdimen\signwidth 
   \setbox0=\hbox{+} 
   \signwidth=\wd0 
   \catcode`!=\active 
   \def!{\kern\signwidth}
\halign{
% Template
\hbox to 6.7 cm{#\leaderfil}\tabskip 1.7em&
\hfil #\hfil\tabskip 0.8em&
\hfil #\hfil&
\hfil #\hfil&
\hfil #\hfil&
\hfil #\hfil&
\hfil #\hfil&
\hfil #\hfil&
\hfil #\hfil&
\hfil #\hfil&
\hfil #\hfil&
\hfil #\hfil&
\hfil #\hfil\tabskip=0pt\cr
\noalign{\doubleline}
\omit&\multispan{12}\hfil F{\sc requency} [GHz]\hfil\cr
\omit&\multispan{12}\hfil E{\sc xperiment}\hfil\cr
\noalign{\vskip -3pt}
\omit&\multispan{12}\hrulefill\cr
\noalign{\vskip 2pt}
\omit&  23&  28.4&  33&  41&  44.1&  61&  70.4&  94& 100& 143& 217& 353\cr
\omit\hfil Q{\sc uantity}\hfil& WMAP& LFI& WMAP& WMAP& LFI& WMAP& LFI& WMAP& HFI& HFI& HFI& HFI\cr
\noalign{\vskip 4pt\hrule\vskip 3pt}
$\epsilon_{\rm H}(\nu)$  [\microKH]&    17.&    9.6&    6.7&    3.7&    3.0&    2.0&    1.7&    1.8&    2.1&    3.2&    6.0&    10.4\cr
$\sigma_{\rm stat}$  [\microKH]&    1.4&   0.92&   0.60&   0.38&   0.31&   0.23&   0.17&   0.26&  0.087&   0.12&   0.19&   0.31\cr
$\epsilon^\prime_{\rm H}(\nu)$ [\microKH]&     14.&    7.8&    5.4&    3.1&    2.5&    1.9&    1.6&    1.6&    2.2&    3.2&    6.0&    10.3\cr
$\sigma^\prime_{\rm stat}$  [\microKH]&    1.2&   0.72&   0.64&   0.42&   0.34&   0.27&   0.20&   0.27&   0.11&   0.12&   0.19&   0.31\cr
$uc_K$&   1.01&  0.92&   1.03&   1.04&   1.06&   1.10&   1.15&   1.26&   1.26&   1.69&   2.99&   13.3\cr
\noalign{\vskip 5pt\hrule\vskip 3pt}}}
\endPlancktablewide
\  \  \   $\epsilon_{\rm H}$ and $\epsilon^\prime_{\rm H}\equiv$ \vtop{\hsize=165mm\noindent\strut Mean dust emissivity expressed as
monochromatic brightness at the reference frequencies 
from the correlation of the maps with the Galactic \hi\ template alone, and with both the Galactic \hi\ template
and the $408\,$MHz map, respectively.  Not colour corrected.\strut}\par
\  \  \   $\sigma_{\rm stat}$ and $\sigma^\prime_{\rm stat}\equiv$ Statistical uncertainty ($1\sigma$) of the brightness
temperatures $T_{\rm b}$ and $T^\prime_{\rm b}$.\par
\  \  \   ${\rm uc}_K\equiv$ \vtop{\hsize=165mm\noindent\strut  Unit conversion factors  from brightness (Rayleigh-Jeans)
to thermodynamic (CMB) temperature. For \wmap\, the conversion factors are computed at the reference frequency, while for 
\Planck\,  they are computed assuming a constant $\nu\,I_\nu$ within the spectral band. \strut}\par
\endgroup
\end{table*}

%%%%%%%%%%%%%%%%%%%%%%%%%%%%%%%

\subsection{The SED of the mean dust emissivity}
\label{subsec:mean_SED}

We produce a mean SED of dust in the diffuse ISM by averaging the
correlation measures, after correction for the CMB contribution as
described in Appendix~\ref{appendix:cmb}, over the 135 sky patches on
our lower resolution grid (Sect.~\ref{subsec:implementation}).  This
SED characterizes the mean emission properties of dust in atomic gas
in the local ISM. 
The statistical uncertainty of  the mean SED is computed from
the standard deviation of individual measurements divided by the
square root of the number of independent sky patches (135/3) used.  
On average, each pixel of the images is part of 3 sky patches. 
This is why we consider that the number of independent sky patches  is the total number divided by 3. 
This standard estimate is appropriate for the noisier low frequency
data.  For the emissivities at higher frequencies, we observe large
variations over the sky (Sect.~\ref{subsec:dust_emissivity}).
However, analysis of our simulations
(Appendix~\ref{appendix:uncertainties}) shows that the uncertainties,
including the variations of the emission properties over the sky, 
average out when we compute the mean dust emissivity over sky patches.
Mean emissivities with statistical and
photometric uncertainties are listed in Table~\ref{tab:mean_sed16} for the
$16\arcmin$ resolution maps at $\nu \ge 70\,$GHz.

\begin{figure}[!h]
\centering
\includegraphics[width=88mm]{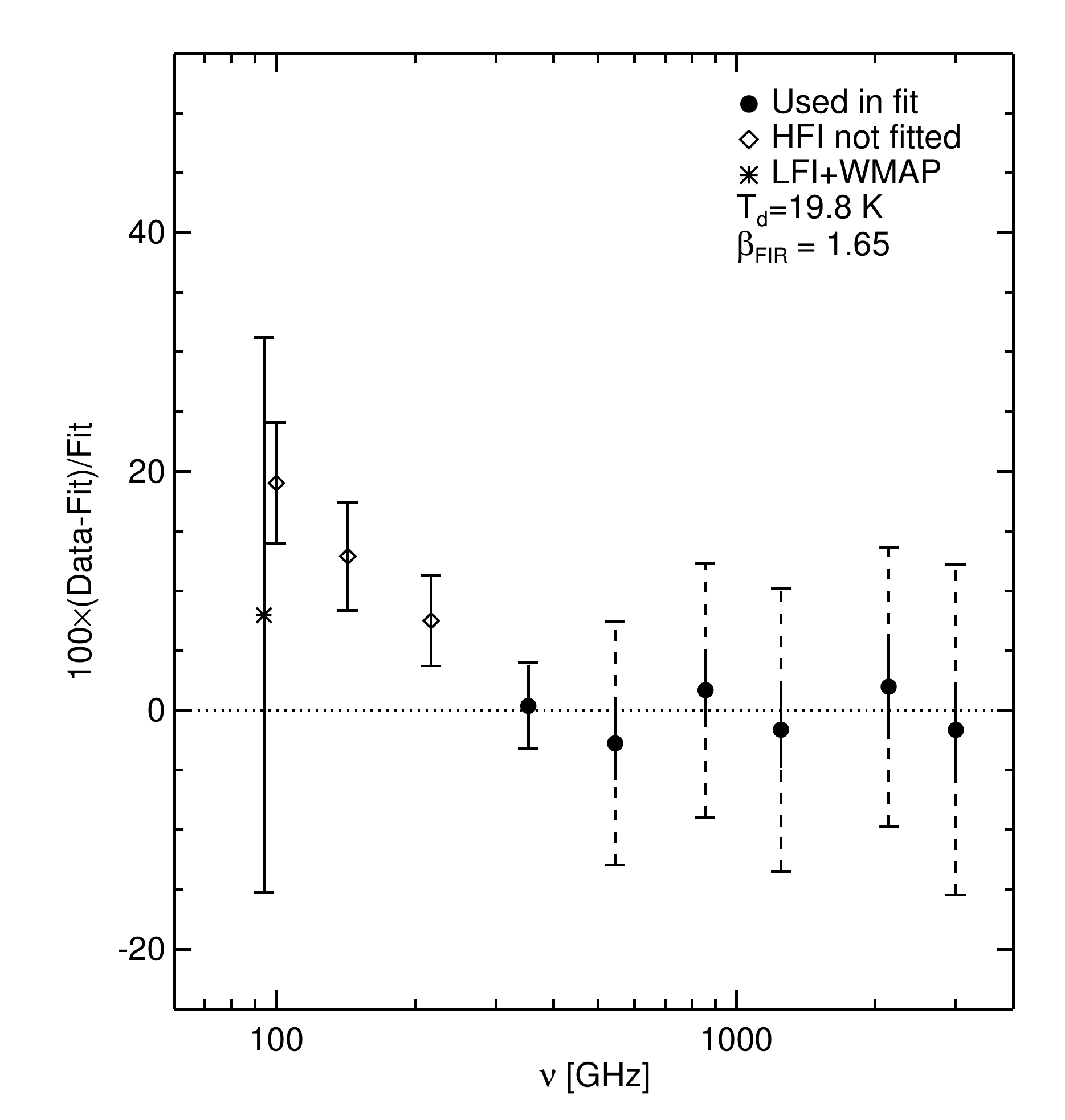}
\includegraphics[width=88mm]{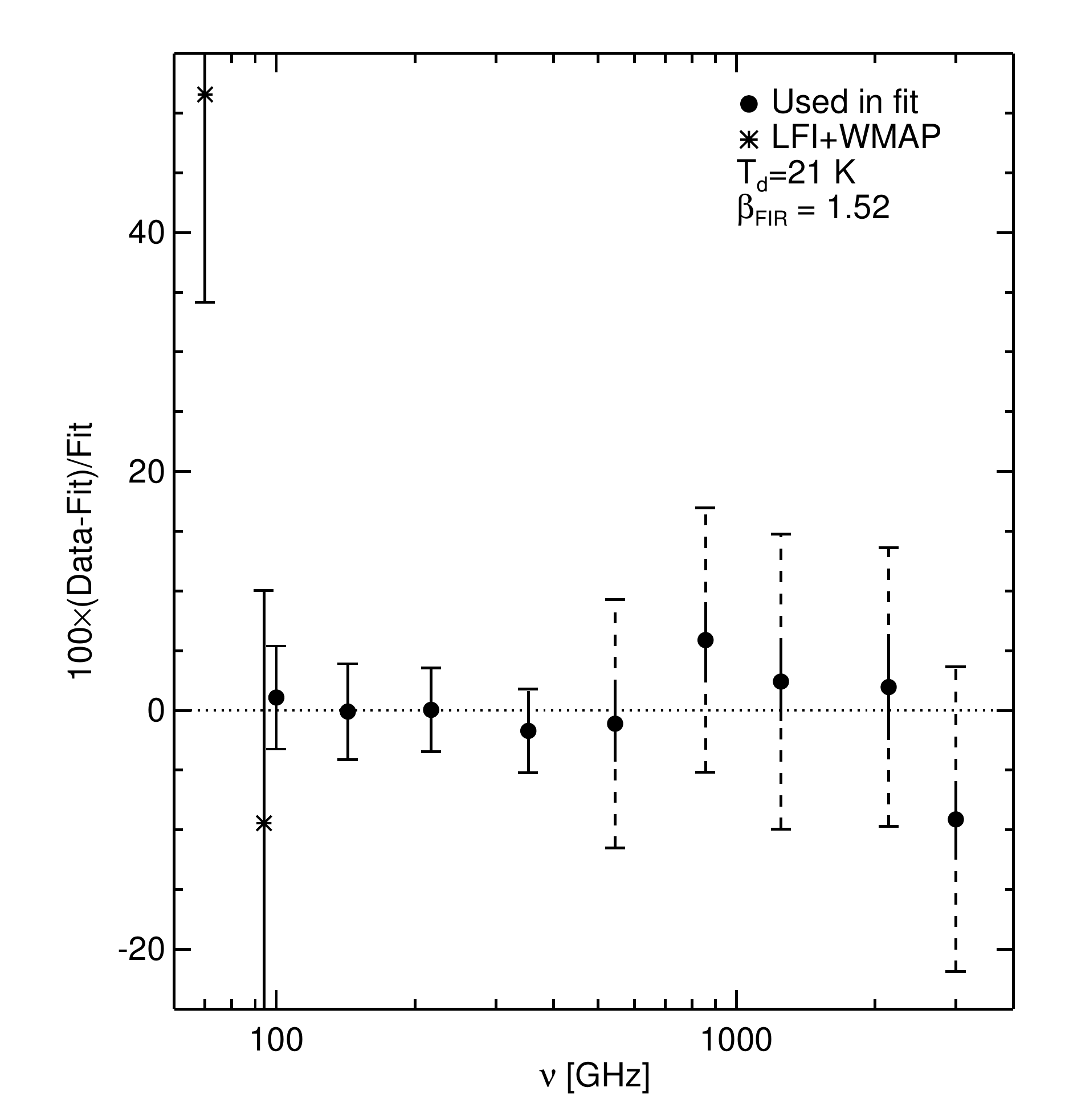}
\caption{{\it Top\/}: Residuals from a greybody fit of the mean dust
  SED at $\nu \ge 353$\,GHz, using one spectral index.  Dashed error
  bars are the quadratic sum of the statistical error
  (solid) and the photometric uncertainty. The photometric uncertainty
  is dominant at $\nu \ge 545\,$GHz and negligible for the lower
  frequencies.
  {\it Bottom\/}: Residuals from a greybody fit to all data points down
  to 100\,GHz, again using a single spectral index. }
\label{fig:sed_greybody_fit}
\end{figure}

%%%%%%%%%%%%%%%%%%%%%
% Table 3  greybody fits
%%%%%%%%%%%%%%%%%%%%%%
\begin{table*}[tmb]
\begingroup
\newdimen\tblskip \tblskip=5pt
\caption{\label{tab:greybody_fits} Parameters from greybody fits of
  the mean dust SED}
\nointerlineskip
\vskip -3mm
%\footnotesize
\tiny
\setbox\tablebox=\vbox{
   \newdimen\digitwidth 
   \setbox0=\hbox{\rm 0} 
   \digitwidth=\wd0 
   \catcode`*=\active 
   \def*{\kern\digitwidth}
   \newdimen\signwidth 
   \setbox0=\hbox{+} 
   \signwidth=\wd0 
   \catcode`!=\active 
   \def!{\kern\signwidth}
\halign{
% Template
\hbox to 4.9 cm{#\leaderfil}\tabskip 1.7em&
\hfil #\hfil\tabskip 1.5em&
\hfil #\hfil&
\hfil #\hfil&
\hfil #\hfil&
\hfil #\hfil\tabskip=0pt\cr
\noalign{\doubleline}
\omit&\multispan{5}\hfil M{\sc odel} P{\sc arameters} \hfil\cr
\noalign{\vskip -3pt}
\omit&\multispan{5}\hrulefill\cr
\noalign{\vskip 2pt}
\omit&  $\sigma_{\rm H}(353\,$GHz)& $T_{\rm d}$& $\beta_{\rm FIR}$& $\beta_{\rm mm}$& $\chi^2/{\rm DOF}$\cr
\omit\hfil M{\sc odel}\hfil&   $\rm [cm^2 \, H^{-1}]$& [K]\cr
\noalign{\vskip 4pt\hrule\vskip 3pt}
Without subtraction of zodiacal emission\hfil\cr
\hglue 2em$\nu \ge 353$\,GHz&               $(7.3 \pm 0.65) \times 10^{-27}$&  $19.8 \pm 1.0$&   $1.65 \pm 0.10$& \dots& 0.05*\cr
\hglue 2em$\nu \ge 100$\,GHz&               $(6.9 \pm 0.5*) \times 10^{-27}$&    $21.0 \pm 0.7$&   $1.52 \pm 0.03$&  \dots& 0.22*\cr
\hglue 2em$\nu \ge 100$\,GHz with 2 $\beta$&$(7.3 \pm 0.6*) \times 10^{-27}$&   $19.8 \pm 1.0$& $1.65 \pm 0.10$& $1.52 \pm 0.03$&  0.041\cr
\noalign{\vskip 3pt}
With subtraction of zodiacal emission\hfil\cr
\hglue 2em$\nu \ge 353$\,GHz&               $(7.1 \pm 0.65) \times 10^{-27}$&  $19.9 \pm 1.0$&   $1.65 \pm 0.10$& \dots&  0.07*\cr
\hglue 2em$\nu \ge 100$\,GHz&               $(6.8 \pm 0.5*) \times 10^{-27}$&   $21.0 \pm 0.7$&    $1.53 \pm 0.03$& \dots&  0.19*\cr
\hglue 2em$\nu \ge 100$\,GHz with 2 $\beta$&$(7.2 \pm 0.6*) \times 10^{-27}$&  $19.9 \pm 1.0$&  $1.65 \pm 0.10$&  $1.54 \pm 0.03$&  0.060\cr
\noalign{\vskip 5pt\hrule\vskip 3pt}}}
\endPlancktablewide
\  \  \   \hskip 43pt $\sigma_{\rm H}(353\,{\rm GHz})\equiv$ Dust opacity at $353\,$GHz from greybody fit.\par
\  \  \   \hskip 43pt $T_{\rm d}\equiv$ Dust temperature from greybody fit. \par 
\  \  \   \hskip 43pt $\beta_{\rm FIR}\equiv$ Spectral index for $\nu \geq 353\,$GHz for models 1 and 3,
and for $\nu \geq 100\,$GHz for model 2.\par
\  \  \   \hskip 43pt $\beta_{\rm mm}\equiv$Spectral index for $\nu \leq 353\,$GHz for  model 3. \par
\  \  \   \hskip 43pt $\chi^2/DOF\equiv$ $\chi^2 $ of the fit per degree of freedom. \par
\endgroup
\end{table*}

\subsection{Greybody fits}
\label{subsec:greybody}

We characterize  the dust SED with greybody fits. 
The mean emissivities are weighted using uncertainties that are the
quadratic combination of the statistical and photometric
uncertainties.  We map the $\chi^2$ for greybody spectra over the parameter space to determine the
best fit parameters listed in Table~\ref{tab:greybody_fits}. 
We report parameters from data without and with subtraction of the
zodiacal emission model \citep{planck2013-pip88}.  The differences in
fit parameters are within the uncertainties.
This is to be expected because the zodiacal emission is a slowly
varying function uncorrelated with the spatial fluctuations of the \hi\ template within the
$15^\circ$ patches.

 All of the best fits have a
$\chi^2$ per degree of freedom much lower than 1, because the statistical 
and photometric uncertainties are correlated across
frequencies.  To test our fits and to estimate error bars on the parameters, we run a Monte-Carlo simulation that takes these correlations into account.  
We assume that the photometric uncertainties are correlated for the three DIRBE frequencies, for the two highest HFI frequencies calibrated on planets, and for
the four lowest HFI frequencies calibrated on the CMB dipole. 
For the statistical errors, we use the frequency-dependent decomposition into Galactic, CMB, CIB, and noise contributions inferred from the sky simulations
in Appendix~\ref{appendix:uncertainties}. The sky simulations ignore the decorrelation from far infrared to microwave frequencies of CIB anisotropies \citep{planck2013-pip56} and of Galactic residuals due to variations in dust temperature. 
These two shortcomings are not an issue, because they mainly impact the modeling of the statistical uncertainties at far infrared frequencies where the  photometric uncertainties are dominant.
We apply our fits to a greybody spectrum with $\beta_{\rm FIR}=\beta_{\rm mm}=1.55$ and $T_{\rm d}=19.8\,$K, combined with 1000 realizations of the statistical and photometric uncertainties. 
For each realization,  we obtain a set of values for the parameters of the fit. For each of the three fits in Table~\ref{tab:greybody_fits}, 
we compute the average and standard deviation of the parameters. The average values match the input values, showing
that correlated uncertainties do not bias the fit.  We list the standard deviations from the Monte Carlo simulation as error bars for the fit parameters in Table~\ref{tab:greybody_fits}. 
We are confident about this estimate of the errors because the $\chi^2$ values obtained for the data fits are in the core of the $\chi^2$ distribution for the Monte Carlo simulation.   In other words, the simulation accounts for the low values of the $\chi^2$ per degree of freedom in Table~\ref{tab:greybody_fits}.

The first fit is for frequencies $\nu \ge 353\,$GHz. It is directly
comparable to the fits presented in the all-sky analysis of
\citet{planck2013-p06b}.  The spectral index that we find, $\beta =1.65 \pm
0.10$, agrees with the mean value used in Sect.~\ref{sec:southern_Cap}
to compute colour temperatures, but it is greater than the values of $\beta_{\rm mm} =1.51 \pm 0.13 $ derived
from the $R_{100}(353,217) $ ratio in Sect.~\ref{sec:microwave_index}.
The second fit extends the greybody fit with a single spectral index
down to $100\,$GHz.  This fit yields a spectral index of $1.52\pm
0.03$ in agreement with the mean value inferred from the above
$R_{100}(353,217) $ ratio.
For the latter, the dispersion about the mean is higher than the
uncertainty from the fit, which is more like an uncertainty of the
mean.

The third fit, again from 100 to 3000\,GHz, uses separate 
spectral indices for frequencies higher and lower than $353\,$GHz.
With this extra parameter, a significantly lower $\chi^2$ per degree
of freedom is achieved, and systematic departures from the fit
(Fig.~\ref{fig:sed_greybody_fit}) are removed.
The best fit is obtained for a higher spectral index at high
frequency.  The difference between the two spectral indices, $\beta_{\rm FIR}-\beta_{\rm mm}$, is 0.13 for the data not corrected for zodiacal emission.
We use our Monte Carlo simulations to test whether the reduction of the $\chi^2$ per degree of freedom 
between the fits with one and two spectral indices (factors 3.7 and 5.4 for the SEDs with and without subtraction of the zodiacal light model) 
is statistically significant. We obtain a reduction of the $\chi^2$ by a factor greater than 3.5 
for less than 5\,\% of the realizations.  Based on this test, we consider that the variation of the spectral index between far infrared and millimetre wavelengths, 
quantified by the third fit is statistically significant. \citet{planck2013-XIV} reach the same conclusion for the diffuse dust emission in the inner Galactic plane. 

The values of the opacity $\sigma_{\rm H} (353\,$GHz) for all fits listed in
Table~\ref{tab:greybody_fits} are consistent with a mean value of $ (7.1 \pm 0.6)
\times 10^{-27} \,{\rm cm^2\, H}^{-1} $, as obtained for the first fit
using data with the zodiacal emission subtracted.  This mean value
agrees with that of \citet{planck2013-p06b} for low column density.
For an dust-to-H mass ratio of 1\,\% \citep{Jenkins09}, the specific absorption coefficient per unit dust
mass is $\kappa_\nu = 0.43 \pm 0.04 \,{\rm cm^2\, g^{-1}}$ at $850\,\mu$m.

Residuals of the first two greybody fits are plotted in
Fig.~\ref{fig:sed_greybody_fit}. The top panel shows that the
extrapolation to $\nu < 353\,$GHz of the first fit departs
progressively from the data points toward lower frequencies.  The
bottom panel shows the residuals of the second fit of the SED from 100
to $3000\,$GHz with a single spectral index. The 3000 and 857\,GHz
data points depart from the fit by more than the statistical
uncertainties. The differences are within the photometric
uncertainties listed in Table~\ref{tab:greybody_fits}, but in opposite
directions for the \DIRBE\ 100\microns\ and the \planck\ 857\,GHz
emissivities.  The residuals do not show the $\sim 10\%$ excess emission at $500\,\mu$m  with respect to greybody fits that 
has been reported for the Large Magellanic Cloud \citep{Gordon10}.
We also point out that the residuals to the fits do not show any excess emission in the
100 and $217\,$GHz spectral bands, which could be coming from the CO(1-0) and CO(2-1) lines \citep{planck2013-p03a}.

\begin{figure}[!h]
\centering
\includegraphics[width=88mm]{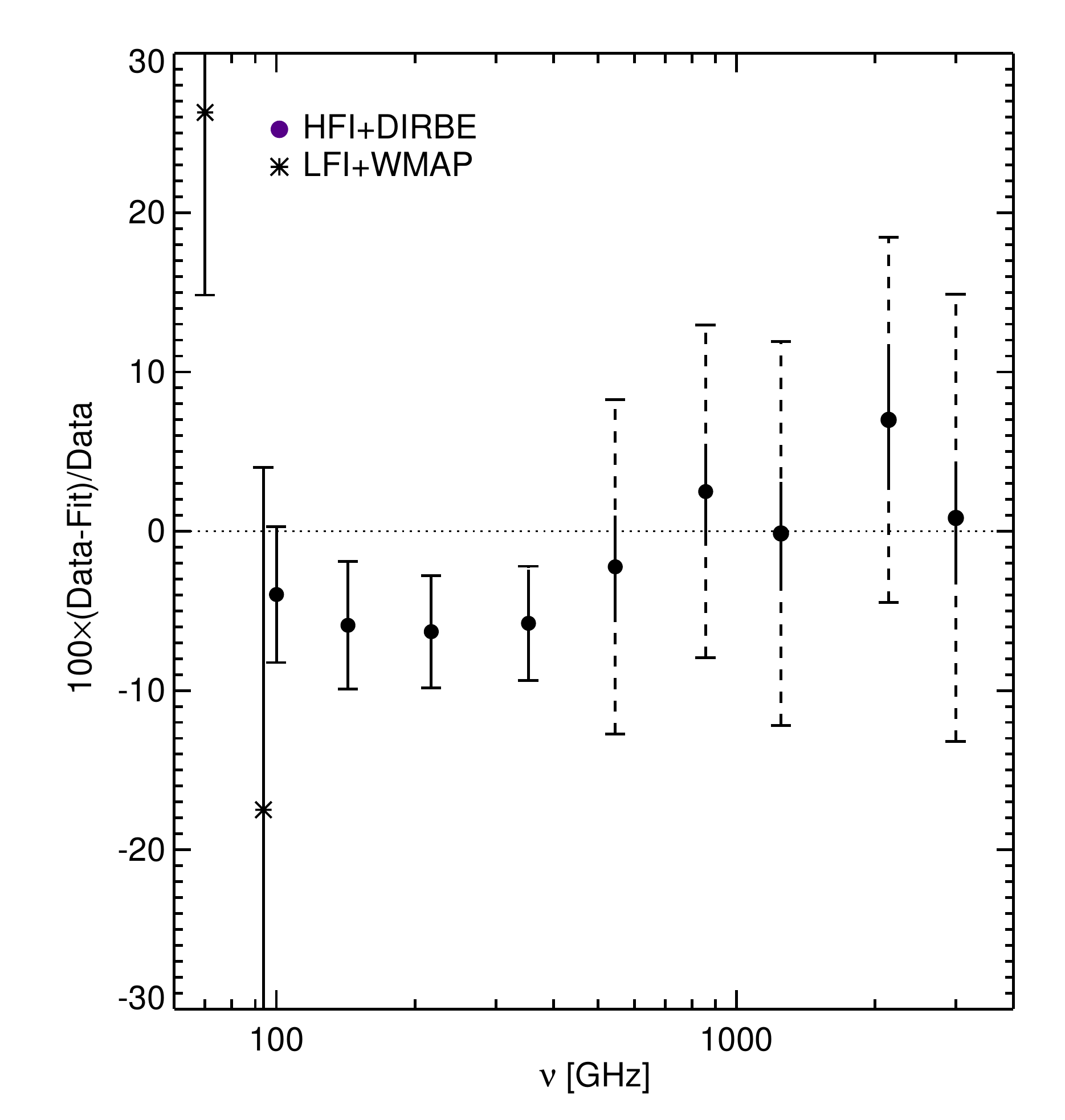}
\includegraphics[width=88mm]{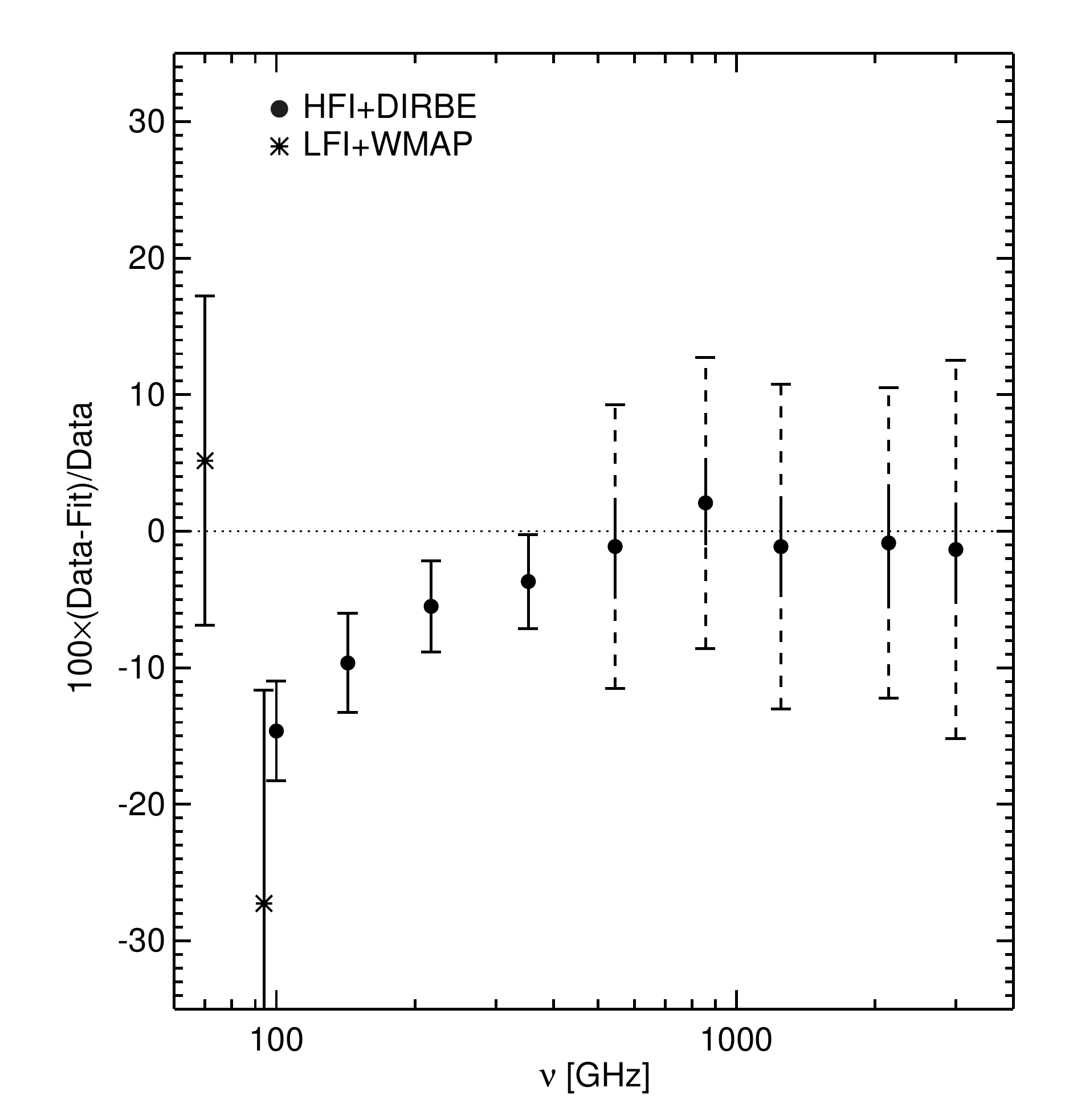}
\caption{Same as Fig.~\ref{fig:sed_greybody_fit}, but for residuals
from fits of the mean dust SED with the DUSTEM (top panel) and DL07
(bottom panel) dust models.
}
\label{fig:dustem_DL07_fit}
\end{figure}

\subsection{Comparison with dust models}
\label{subsec:dust_models}

%align sentence order with Fig. 12
In this section, we compare the mean SED from \planck\ with two
models of the thermal dust emission.  We fit the mean SED in
Table~\ref{tab:mean_sed16} with the dust models presented in
\citet{Compiegne11} and \citet{DraineLi07}, hereafter the DUSTEM and
DL07 models.  For both models, we fit the scaling factor $G_0$ of the mean interstellar  radiation field in the Solar
Neighbourhood from \citet{Mathis83}, and another scaling parameter,
$f_{\rm SED}$, that allows for differences in the normalization of the
dust emission per unit gas mass.  
The two parameters of the fit are quite independent.  The value of
$f_{\rm SED}$ is constrained by the submillimetre data points, while $G_0$ is
constrained by the peak of the SED.
For the DUSTEM model, the best fit is obtained for $G_0=1.0$ and
$f_{\rm SED}=1.05$, whereas for the DL07 model we find $G_0=0.7$ and
$f_{\rm SED}=1.45$.
The residuals from these two fits are shown in
Fig.~\ref{fig:dustem_DL07_fit}. Both models fit the data within 5\,\% at
$\nu \ge 353$\,GHz. They depart from the data at lower
frequencies by 5 to 15\,\%.
We note that both models use the same optical properties for silicates 
from \citet{LiDraine01},  who introduced a flattening of the emissivity law at $\lambda \ge 250\,\mu$m  to match 
the SED of \citet{Finkbeiner99}. They differ in their modeling of carbon dust. 

This comparison shows that none of the models provides a fully
satisfactory fit of the \Planck\ SED. For the DL07 model, it also
shows that there is a significant difference in the dust emission per
unit gas mass, which
is higher than what may be accounted for by dust within the diffuse
ionized gas \citep{Gaensler08}, even in the most favourable hypothesis
where its spatial distribution is highly correlated with \hi\
emission.

\section{Microwave dust emission}
\label{sec:microwave}

We extend our analysis of the thermal dust emission by analyzing the microwave SED of dust that combines the \planck\ and \wmap\ spectral channels.  We present the 
SED and discuss several spectral decompositions.

\subsection{Microwave SED of dust emission}
\label{sec:microwave_sed}

The microwave  SED of dust emission in the diffuse ISM at $23 \le \nu \le 353\,$GHz, obtained 
by averaging the correlation measures for the $60\arcmin$ resolution maps
over the 135 sky patches on our lower resolution grid (Sect.~\ref{subsec:implementation}),
is listed in Table~\ref{tab:mean_sed60}.  
The statistical uncertainty of  the mean SED is computed from
the standard deviation of individual measurements,  after
correction for the CMB contribution as
described in Appendix~\ref{appendix:cmb}, divided by the
square root of the number of independent sky patches (135/3) used.  
These error-bars include variations of the dust SED across the southern polar cap and 
uncertainties in the CMB subtraction. The mean difference between the two independent estimates  
of the CMB presented in Appendix~\ref{appendix:cmb} is 
one order of magnitude lower than the minimum of the dust SED at 60--70\,GHz.

Table~\ref{tab:mean_sed60} lists two SEDs. 
In this section, we use the SED, $\epsilon^\prime_{\rm H}(\nu)$, computed from
emissivities corrected for the chance correlation of the \hi\ template
with synchrotron emission by fitting the \planck\ and \wmap\ data
simultaneously with two templates (Sect.~\ref{subsec:implementation}).
%, the \hi\ GASS map and the 408\,MHz maps of \citet{Haslam82}, using
%the method presented in \citet{Davies06}.  
The synchrotron template impacts the dust SED only at the lowest frequencies.

\begin{figure}[!h]
\centering
\includegraphics[width=0.49\textwidth]{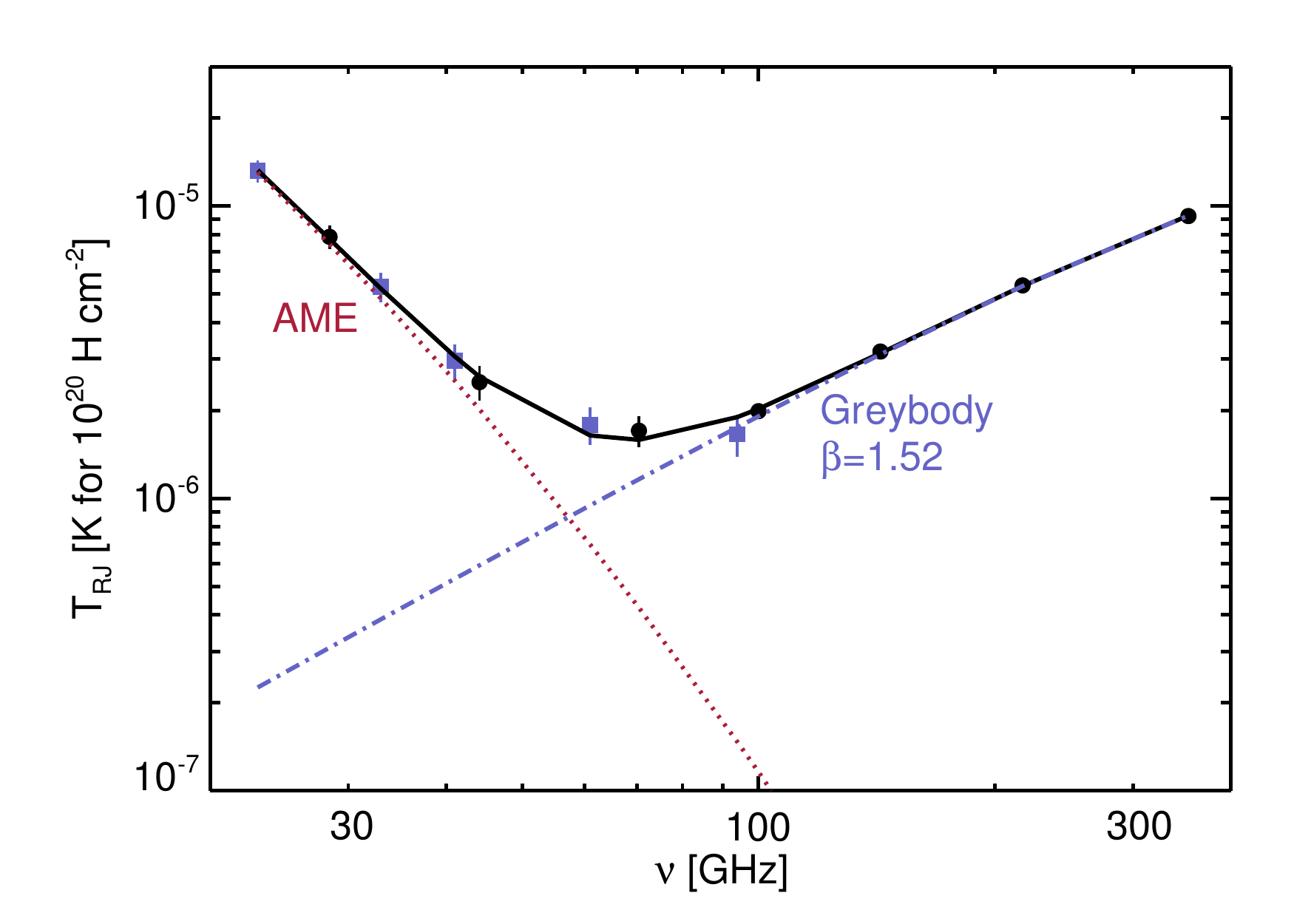}
\includegraphics[width=0.49\textwidth]{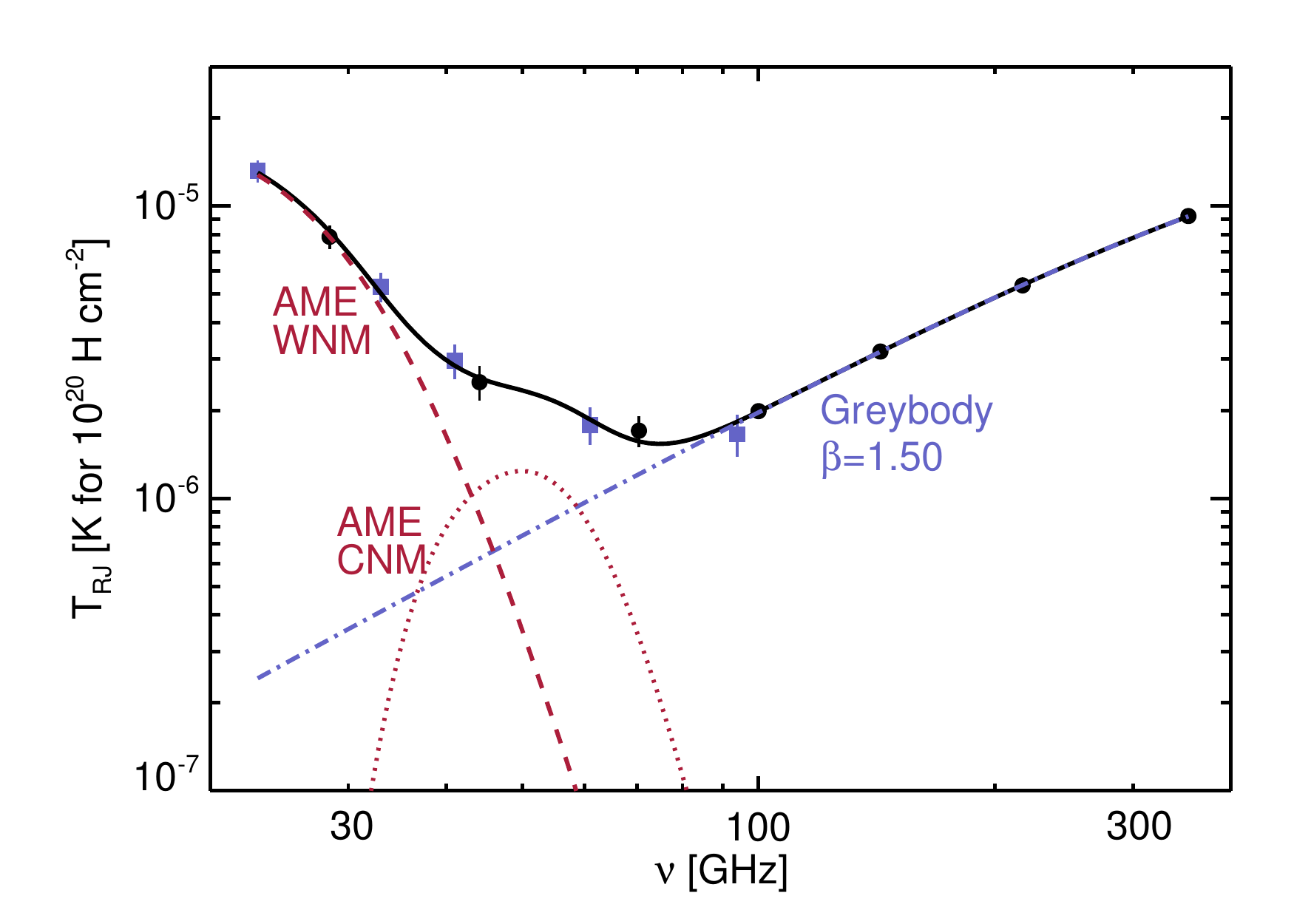}
\caption{Mean microwave dust SED obtained by cross-correlating the \planck\ and \wmap\ data with the \hi\ and 408\,MHz
templates (blacks dots for \planck\ and blue squares for \wmap). 
{\it Top\/}: Model~1 in Table~\ref{tab:microwave_fits}, with two emission components combining AME
and greybody thermal dust emission with $\beta =1.52$.  The AME
is fitted with the analytical model in Eq.~(\ref{eq:Bonaldi}).
{\it Bottom\/}: Spectral fit for model~3, where the AME is fitted with two {\tt SPDUST} spectra peaking at different frequencies.
}
\label{fig:microwave_sed}
\end{figure}

The microwave SED is displayed in Fig.~\ref{fig:microwave_sed}.
We check in two ways that this SED
is not contaminated by  free-free emission correlated with the
\hi\ map.  First, we find that the $70\,$GHz emission is not reduced
if we compute the mean dust SED after masking the southern extension
of the Orion-Eridanus super-bubble to high Galactic latitudes, the
area of brightest H$_\alpha$ emission at $b < -30^\circ$.  
Second, we check that the correlation between the H$_\alpha$ emission and the \hi\
column density has a negligible impact on the dust SED by doing a three 
template fit,  over the part of the
southern Galactic cap covered by the survey of WHAM (Wisconsin H-Alpha Mapper) survey \citep{Haffner03}. 
The photometry of diffuse $H_\alpha$ emission in the all-sky map of \citet{Dickinson03} is not reliable on degrees scale outside of this area. 

\begin{figure}[!h]
\centering
\includegraphics[width=0.49\textwidth]{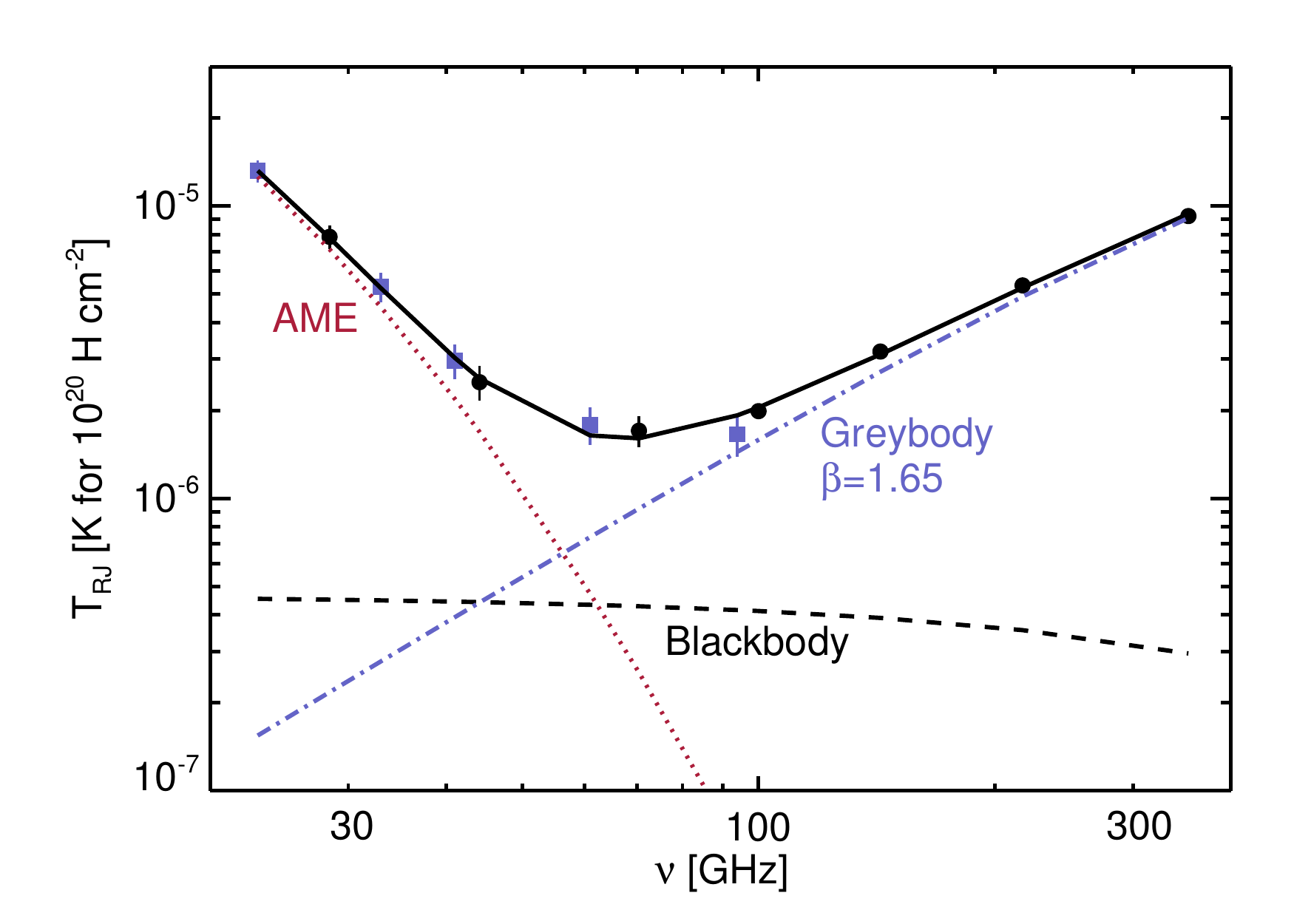}
\includegraphics[width=0.49\textwidth]{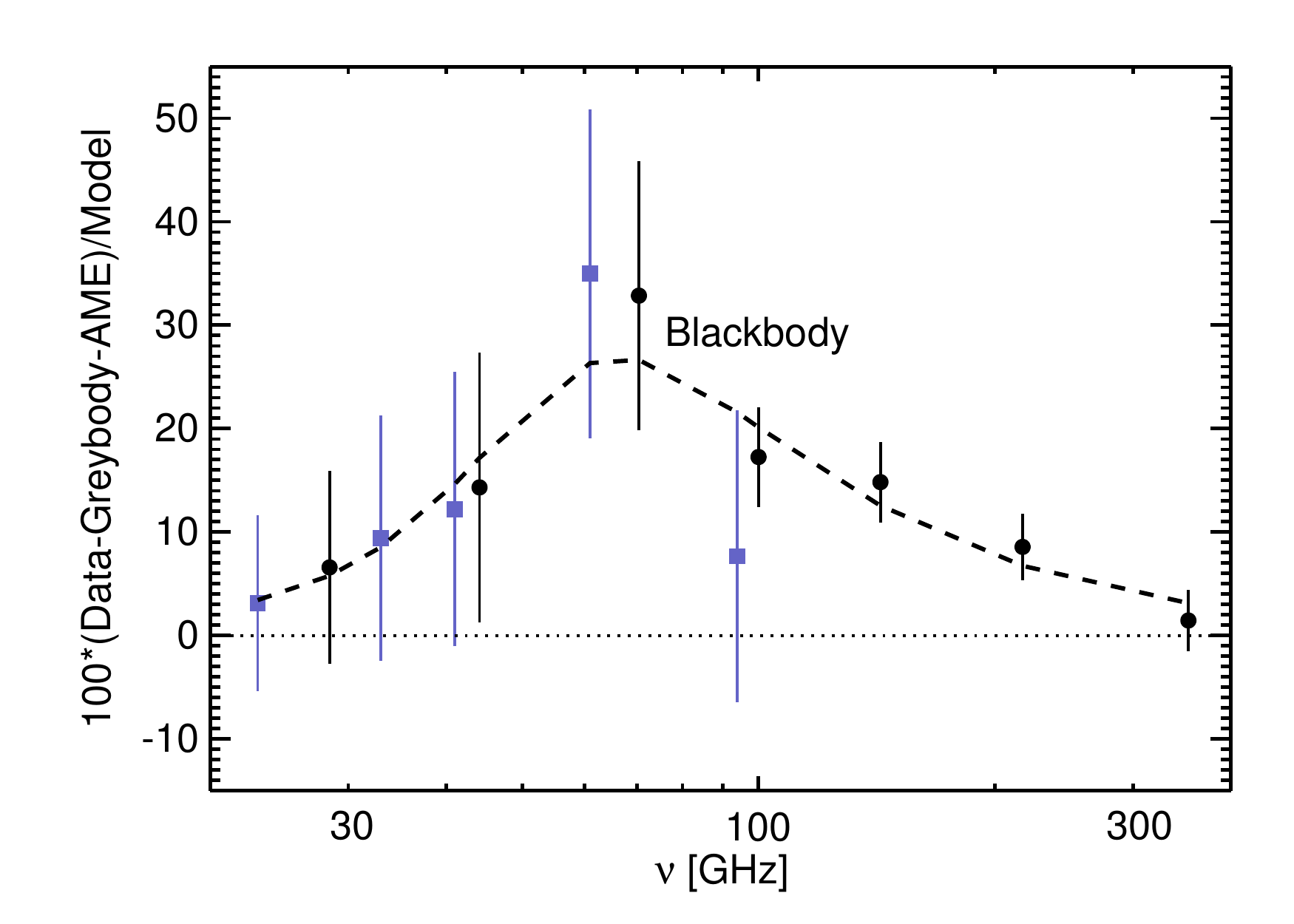}
\caption{{\it Top\/}: Same microwave dust SED as in Fig.~\ref{fig:microwave_sed}
with the spectral fit for model~2 in Table~\ref{tab:microwave_fits}.
The spectral index of the greybody is fixed to the value $\beta_{\rm FIR} = 1.65$ inferred from the fit
of the dust SED at $\nu \ge 353\,$GHz (Table~\ref{tab:greybody_fits}).
The AME is fitted with the analytical model in Eq.~(\ref{eq:Bonaldi}).
This fit includes a third component represented by a 
blackbody spectrum at the same temperature (19.8\,K) as that of the greybody. 
{\it Bottom\/}: Blackbody component in model~2 as a fractional residual after
subtraction of the AME and greybody emission from the total,
compared to the data residuals.}
\label{fig:spdust_fits}
\end{figure}

\subsection{Separation of the thermal emission of dust from AME}

The SED in Fig.~\ref{fig:microwave_sed}  is dominated by thermal dust
emission at the high frequencies and AME at low
frequencies. 
We perform several spectral fits to separate the two emission components. 
The model parameters are listed in Table~\ref{tab:microwave_fits}.
In this section we present the fits with models~1 and 3 displayed in Fig.~\ref{fig:microwave_sed}.
Both models use a greybody spectrum at a fixed temperature of 19.8\,K 
for the dust thermal emission, but they differ in the way the AME is fitted.

In model~1, we fit the AME with the analytical model introduced by
\citet{Bonaldi07}, which in the log(Brightness)-log($\nu$) plane is a
parabola parametrized by peak frequency $\nu_{\rm p}$\footnote{The spectrum peaks at frequency $\nu_{\rm p}$ in flux units} 
and slope $-m_{60}$ at
$60\,$GHz.  Thus
\begin{equation}
\log \left (\frac{T_{\rm b}(\nu)}{T_{\rm b}(\nu_{\rm p})} \right )= -2 \,
\log(\nu/\nu_{\rm p}) +m_{60}\, \frac{\left[\log(\nu/\nu_{\rm p})\right]^2}{2\,
  \log(\nu_{\rm p}/60\,{\rm GHz})}, \label{eq:Bonaldi}
\end{equation} 
where $T_{\rm b}$ is the AME brightness (Rayleigh-Jeans) temperature and
$\nu$ is the frequency in gigahertz.  \citet{planck2013-XII} show that this
analytical function provides a good fit to the AME spectra derived
from their analysis of the \planck\ and \wmap\ maps along a section of
the Gould Belt at intermediate Galactic latitudes. In model 3, we fit the AME combining two spectra labeled WNM and CNM,
which were computed with the physical {\tt SPDUST}  model \citep{AliHaimoud09,Silsbee11} using standard parameters for
the warm and cold neutral medium from Table~1 in \citet{Draine99}. This model allows us to check
whether our determination of the microwave emission from dust depends on the spectral template used for the AME. We do not aim 
at proposing and discussing a physical fit of the AME.

In model~1, we fit the 12~data points of the SED from 23 to 353\,GHz with five
free parameters: the specific opacity $\sigma_{\rm H}(353\,$GHz); the spectral index  $\beta_{\rm mm}$ for 
the greybody; $\nu_{\rm p}$; $m_{60}$; and the AME brightness temperature $T_{\rm b}(23\,{\rm GHz})$.
%In model 2, we fix the spectral index to the far infrared value $\beta_{\rm FIR}= 1.65$ in Table~\ref{tab:greybody_fits}.
%To account for the flattening of the SED with decreasing frequency we introduce an additional component
In model~3, we also fit five free parameters. The AME parameters are the amplitudes of the two AME spectra,
$A_{\rm WNM}(23\,{\rm GHz})$ and  $A_{\rm CNM}(41\,{\rm GHz})$, plus a frequency shift $\nu_{\rm shift}$
of the CNM {\tt SPDUST} spectrum. This shift is an empirical means to account for the dependency of the peak frequency of the 
AME emission on physical parameters such as the gas density and the minimum grain size \citep{Ysard11,Thiem11}.
\cite{Thiem11} present a fit of the AME SED determined with WMAP data by \cite{Miville08} with two AME spectra that
have clearly distinct peak frequencies. The peak frequencies of the WNM and CNM {\tt SPDUST} spectra we use are 24.3 and $30\,$GHz in flux units. 
We find that we need to introduce a positive shift of $25\,$GHz of the CNM spectrum to obtain a good fit. This shift moves the peak 
of the CNM {\tt SPDUST} spectrum to $55 \,$GHz in flux units ($51 \,$GHz in brightness temperature, Fig.~\ref{fig:microwave_sed}). 

The two models provide a very good fit of all data points. % with $\chi^2$ per degree of freedom lower than one. 
They yield similar results for the greybody parameters that characterize the dust thermal emission. These parameters 
match the corresponding ones derived from the fit of the data at $\nu \ge 70\,$GHz in Sect.~\ref{subsec:greybody}. They do not depend
on the way the AME is modelled. The $\chi^2$ per degree of freedom of all fits is lower than unity.
As for the greybody fits in Sect.~\ref{subsec:greybody}, this results from the significant correlation of uncertainties across frequencies.
To take this correlation into account, we run a Monte-Carlo simulation of each fit. We use each of the models in Table~\ref{tab:microwave_fits} as the input SED. We compute 1000 realizations of the 
data uncertainties using the results of a Principal Component Analysis
of the 135 SEDs measured on the individual sky patches to parametrize the correlation across frequencies. We perform the spectral fits on each realization. 
The simulations show that the fit results are not biased, and provide the errors-bars in Table~\ref{tab:microwave_fits}.
We also find that the large errors-bars on the  AME parameters for model 1 are highly correlated.

%%%%%%%%%%%%%%%%%%%%%
% Table 4  Microwave fits
%%%%%%%%%%%%%%%%%%%%%%
\begin{table*}[tmb]
\begingroup
\newdimen\tblskip \tblskip=5pt
\caption{\label{tab:microwave_fits} Spectral fits of  the mean microwave dust SED}
\nointerlineskip
\vskip -3mm
\tiny
\setbox\tablebox=\vbox{
   \newdimen\digitwidth 
   \setbox0=\hbox{\rm 0} 
   \digitwidth=\wd0 
   \catcode`*=\active 
   \def*{\kern\digitwidth}
   \newdimen\signwidth 
   \setbox0=\hbox{+} 
   \signwidth=\wd0 
   \catcode`!=\active 
   \def!{\kern\signwidth}
\halign{
\hbox to 2.7 cm{#\leaderfil}\tabskip 2.0em&
\hfil #\hfil\tabskip 1.0em&
\hfil #\hfil&
\hfil #\hfil\tabskip 2.5em&
\hfil #\hfil&
\hfil #\hfil\tabskip 1.0em&
\hfil #\hfil&
\hfil #\hfil\tabskip=0pt\cr
\noalign{\doubleline}
\omit& \multispan{6}\hfil M{\sc odel} P{\sc arameters} \hfil\cr
\noalign{\vskip -3pt}
\omit&\multispan{6}\hrulefill\cr
\noalign{\vskip 2pt}
\omit&  \multispan{3}\hfil AME\hfil\cr
\omit&  \multispan{3}\hfil Analytical model\hfil&&  \multispan{2}\hfil Greybody\hfil\cr
\noalign{\vskip -3pt}
\omit&\multispan{3}\hrulefill&  BB& \multispan{2}\hrulefill\cr
\noalign{\vskip 2pt}
\omit\hfil Model\hfil&  $T_{\rm b}{(23\,{\rm GHz})}$& $\nu_{\rm p}$& $-m_{60}$& $\tau_{\rm BB}$& $\sigma_{\rm H}(353\,$GHz)& $\beta_{\rm mm}$&  $\chi^2/{\rm DOF}$\cr
\noalign{\vskip 4pt\hrule\vskip 3pt}
%\omit&   $\rm [cm^2 \, H^{-1}]$&& $\rm [\mu K \,cm^2\,H^{-1}]$& [GHz]&&$\rm [cm^2 \, H^{-1}]$&\cr
1&         $13.0 \pm 1.1 \, 10^{-20}$& $11 \pm 7 $& $1.4 \pm 0.7$&  \dots&  $7.4 \pm 0.23 \, 10^{-27}$&  $1.52\pm 0.03$& 0.27\cr
2&         $12.6 \pm 1.2 \, 10^{-20}$& $19 \pm 6 $& $2.2 \pm 1.0$&  $2.4 \pm 0.51 \, 10^{-28}$&  $7.3 \pm 0.24 \, 10^{-27}$&  $1.65$& 0.42\cr
\noalign{\vskip 4pt\hrule\vskip 3pt}
\noalign{\vskip 3pt}
\omit&  \multispan{3}\hfil {\tt SPDUST} spectra\hfil&&&&\cr
\noalign{\vskip -3pt}
\omit&\multispan{3}\hrulefill&\cr
\noalign{\vskip 2pt}
\omit&  $A_{\rm WNM}(23\,{\rm GHz})$& $A_{\rm CNM}(41\,{\rm GHz})$& $\nu_{\rm shift}$&&&&\cr
\noalign{\vskip 4pt\hrule\vskip 3pt}
3&          $12.8 \pm 1.3 \, 10^{-20}$& $0.88 \pm 0.26 \,  10^{-20}$& $25 \pm 3$&  \dots&  $7.4 \pm 0.26 \, 10^{-27}$&  $1.50\pm 0.04$& 0.21\cr
4&         $12.2 \pm 1.2 \, 10^{-20}$& $0.71 \pm 0.25 \,  10^{-20}$& $24.5 \pm 3$&  $2.4 \pm 0.54 \, 10^{-28}$&  $7.3 \pm 0.26 \, 10^{-27}$&  $1.65$& 0.34\cr
\noalign{\vskip 5pt\hrule\vskip 3pt}}}
\endPlancktablewide
\  \  \   \hskip 20pt $T_{\rm b}(23\,{\rm GHz})\equiv$ Brightness temperature, in $\mu$K\,cm$^2$\,H$^{-1}$, of AME at $23\,$GHz for models 1 and 2.\par
\  \  \   \hskip 20pt $\nu_{\rm p}$ and $-m_{60}\equiv$ Peak frequency in gigahertz and slope at $60\,$GHz of AME spectrum in Eq.~(\ref{eq:Bonaldi}) for models 1 and 2.\par
\  \  \   \hskip 20pt $A_{\rm WNM}$ and $A_{\rm CNM}\equiv$ Maximum brightness temperature of WNM and CNM {\tt SPDUST}  spectra, in $\rm \mu K \,cm^2\,H^{-1}$, for models 3 and 4. \par
\  \  \   \hskip 20pt $\nu_{\rm shift}\equiv$ Frequency shift in gigahertz of the CNM {\tt SPDUST}  spectrum for models 3 and 4. \par
\  \  \   \hskip 20pt $\tau_{\rm BB}\equiv$ Specific opacity of the blackbody component, in cm$^2\,$H$^{-1}$, for models 2 and 4. \par
\  \  \   \hskip 20pt $\sigma_{\rm H}(353\,{\rm GHz})\equiv$ Specific dust opacity at $353\,$GHz of greybody in cm$^2\,$H$^{-1}$.\par
\  \  \   \hskip 20pt $\beta_{\rm mm}\equiv$ Spectral index of the greybody component. The spectral index is fixed to 1.65
for models 2 and 4. \par
\  \  \   \hskip 20pt The temperature is 19.8\,K for the greybody and blackbody components for all models.\par
\  \  \   \hskip 20pt $\chi^2/{\rm DOF}\equiv$ $\chi^2 $ of the fit per degree of freedom. \par
\endgroup
\end{table*}

\subsection{Spectral fit with an additional emission component}
\label{subsec:flattening}

In this section, we discuss models~2 and 4 in Table~\ref{tab:microwave_fits},  
where we fix the spectral index of the greybody to the value $\beta_{\rm FIR} = 1.65$
inferred from the fit of the SED at $\nu \ge 353\,$GHz.
To account for the flattening of the dust SED at lower frequencies,
we add a third emission
component to the AME and the greybody.  This additional component is
assumed to have a blackbody spectrum with the same temperature
19.8\,K as that of the thermal dust emission.  We refer to this as
the blackbody (BB) component.  For the frequency range over which this component is significant, the
blackbody spectrum is a good approximation of magnetic dipole emission
from ferro-magnetic particles or magnetic inclusions in dust grains,
as modelled by \citet{Draine13}. 
Model~2 uses the same analytical model for the AME as
model 1; model~4 uses the same two {\tt SPDUST}  spectra as model~3.
As in models~1 and 3, we fit for five parameters since the amplitude of the BB component replaces
the spectral index  of the greybody as a free parameter.

The three components model provides a good fit to all 12~data
points (top panel in Fig.\ref{fig:microwave_sed}).  
In particular, when added to the
greybody component, the blackbody component accounts for the
flattening of the spectral index of the thermal dust emission towards
microwave frequencies.
The specific opacity we find for the BB component is the same for both models. 
At 100\,GHz, the blackbody component amounts to $(26\pm6)\,\%$ of the
greybody dust emission. This fraction is within the range of plausible
values for dipolar magnetic emission within the model of
\citet{Draine13}, and somewhat lower than the value
reported by \citet{Draine12} to fit the SED of  dust emission from the Small Magellanic Cloud \citep{Bot10,Israel10,planck2011-6.4b}.

Magnetic dipole emission is not a unique way to account for the 
flattening of the dust SED at $\nu \le 353\,$GHz. 
We cannot exclude alternative interpretations.  First, the
blackbody component may be a phenomenological way to introduce the
progressive flattening of the thermal dust emission at long
wavelengths observed in laboratory experiments on amorphous silicate
particles \citep{Coupeaud11}. Within this interpretation it would
represent the contribution from low energy transitions to the opacity
of interstellar silicates \citep{Meny07}.  Second, the flattening of the dust SED could be due to an 
increasing contribution of carbon dust towards millimetre wavelengths. 
In the dust model of \citet{Jones13},
the emission from amorphous carbon grains becomes dominant at $\lambda > 1\,$mm for 
a spectral index at microwave frequencies in agreement with that measured on the data. 

The physical interpretation of the additional emission component that
would account for the flattening of the dust SED at microwave
frequencies is further discussed in \citet{planck2014-XXII}, where the SED of
the polarized dust emission is presented. 
The three interpretations proposed here make
different predictions for the dust polarization SED. 
Dipole magnetic emission from iron inclusions would decrease the polarization of the
thermal dust emission from silicate grains because the two polarization angles are $90^\circ$ apart \citep{Draine13}. 
Polarization may also allow us to distinguish between the carbon and silicate contributions to the SED 
flattening, if only the emission from silicates is polarized.

\section{Summary}
\label{sec:summary}

In a $7500\,$deg$^2$ cap around the southern Galactic pole, we
characterize the correlation between far infrared and microwave \planck\
emisison and \NH\ from the \hi\ GASS survey. This study covers the 
part of the southern sky best suited to study the structure of the CMB
and CIB. We characterize the correlation between dust and gas and the
SED of the dust emission.  The data analysis yields four main
scientific results.

\smallskip
\noindent
(1) The \hi\ correlation analysis allows us to separate the dust
emission from the CIB and CMB anisotropies, and to map the emission
properties of dust at high Galactic latitudes.  We map the dust
temperature, and its submillimetre emissivity and opacity.  The
variations of the dust emissivity at 353\,GHz are surprisingly
large, ranging over a factor close to three.  The dust temperature is
observed to be anti-correlated with the dust opacity.  We interpret
these results as evidence of dust evolution within the diffuse ISM, 
and discuss them within the context of existing models of dust.  The
mean dust opacity is measured to be $(7.1 \pm 0.6) \times 10^{-27}
\,{\rm cm^2\, H}^{-1} \times (\nu/{\rm 353\,GHz})^{1.53\pm 0.03}$, for $100 \le
\nu \le  353\,$GHz.  This is a reference value to estimate hydrogen
masses from dust emission at submillimetre and millimetre wavelengths.

\noindent
(2) Using a colour ratio between 353 and 217\,GHz that is free from
CMB, we determine the spectral index $\beta_{\rm mm}$ of the dust
emission.  We find a mean value of 1.51 that is
remarkably constant over the field of our investigation; the standard
deviation is 0.13. Variations of $\beta_{\rm mm}$ show no clear trend with
the $353\,$GHz dust emissivity, nor with the dust temperature.
We compare $\beta_{\rm mm}$ with the spectral index $\beta_{\rm FIR}$ 
derived from greybody fits at $\nu \ge 353\,$GHz.
We find a systematic difference of $\beta_{\rm mm}-\beta_{\rm FIR} =0.15$.

\noindent
(3) We fit the SED of the microwave emission correlated with \hi\ from
23 to 353\,GHz with two components, a parametric model or {\tt SPDUST}  spectra for AME, and 
a greybody for the thermal dust emission.
We show that the flattening of the dust SED at $\nu \le 353\,$GHz  
can be accounted for with an additional blackbody component.  
This additional component, which accounts for $(26\pm6)\,\%$
of the dust emission at $100\,$GHz, could represent magnetic dipole
emission.  Alternatively, it could represent the contribution from low
energy transitions in amorphous solids to the opacity of interstellar silicates, 
or an increasing contribution from carbon dust. These interpretations make different predictions
for the dust polarization SED measured by \Planck.

\noindent
(4) We analyse the residuals with respect to the dust-\hi\
correlation.  We identify a Galactic contribution to these residuals,
which we model with variations of the dust emissivity on angular
scales smaller than the $15^\circ$ patches of our correlation analysis.  This
model of the residuals is used to quantify uncertainties of the CIB
power spectrum in a companion \Planck\ paper \citep{planck2013-pip56}.

\smallskip

These results are important for defining
future models of dust emission. Such models will need to include the
evolution/variation of dust properties within the diffuse ISM. They
are also valuable inputs to CIB and CMB studies.  In \citet{planck2013-pip56},
our analysis is used to determine the power spectrum of CIB anisotropies over
a field more than an order of magnitude higher than in earlier
studies. The spectral characterization of the dust emission is being
combined with the all-sky analysis in \citet{planck2013-p06b} to prepare a
model of dust emission at microwave frequencies for CMB studies.

This paper opens the way to additional studies of the dust-\hi\
correlation. The methodology introduced in this paper is
of general use to studies of the dust-\hi\ correlation with diverse
science objectives. We have focused our scientific analysis on the emission
properties of Galactic dust, leaving for further studies several
aspects of the dust-gas correlation.  In a future paper we will use
the same data and method to quantify an upper limit on the dust-to-gas mass ratio  in 
the MS gas, and to characterize \hi\ gas at Galactic
velocities with no or only a faint counterpart in the \Planck\ maps.
The clouds of excess dust emission with respect to the \hi\ model also
deserve further attention, to investigate where H$_2$ forms
within the diffuse ISM.

\begin{acknowledgements}
The development of \planck\ has been supported by: ESA; CNES and
CNRS/INSU-IN2P3-INP (France); ASI, CNR, and INAF (Italy); NASA and DoE
(USA); STFC and UKSA (UK); CSIC, MICINN, JA and RES (Spain); Tekes,
AoF and CSC (Finland); DLR and MPG (Germany); CSA (Canada); DTU Space
(Denmark); SER/SSO (Switzerland); RCN (Norway); SFI (Ireland);
FCT/MCTES (Portugal); and PRACE (EU).  A description of the Planck
Collaboration and a list of its members, including the technical or
scientific activities in which they have been involved, can be found
at
\url{http://www.sciops.esa.int/index.php?project=planck&page=Planck_Collaboration}.
The Parkes Radio Telescope is part of the Australia Telescope, which
is funded by the Commonwealth of Australia for operation as a National
Facility managed by CSIRO. The research leading to these results has received funding from the European Research Council 
under the European Union's Seventh Framework Programme (FP7/2007-2013) / ERC grant agreement n¡ 267934.
\end{acknowledgements}

\bibliographystyle{aa}
\bibliography{PIP82_Final.bbl}
%\bibliography{Planck_bib,MagStream_bib}

\appendix

\section{Model of the dust emission}
\label{appendix:model}

We detail how we construct a map of the model of the dust emission
that is spatially correlated with the \hi\ emission.  The model of the
dust emission $M$ is written as
\begin{equation}
\label{eq:dust_model}
M (\nu)  = A_{\rm H} (\nu) \times {\rm HI} + B(\nu), 
\end{equation}
where $A_{\rm H}$ is a map at resolution $N_{side} = 512$  built from the 
correlation measure $\alpha_\nu$ in Eq.~(\ref{eq:2}), $B$
is an offset map built from $\omega_\nu$ in Eq.~(\ref{eq:3}), and $\rm HI$ is the \NH\ template for the \hi\ GASS
data.

The $A_{\rm H}$ and $B$ maps are computed from the results of the dust-\hi\
correlation analysis over $15^\circ$ diameter patches, sampled on
\healpix\ pixels with a resolution $N_{\rm side}=32$.

Specifically, at each frequency, $A_{\rm H}$ and $B$ maps are derived from
the correlation measure and the offset maps (Sect.~\ref{subsec:cc}).
Next we correct the correlation measures and the offsets for the CMB
contributions, following the procedure presented in
Appendix~\ref{appendix:cmb}.  The offset map is also corrected for the
CIB monopole using the values determined in \citet{planck2013-p06b}.
Subsequently, we obtain the desired $A_{\rm H}$ map by interpolating the map of
correlation measures from $N_{\rm side}=32$ to 512 of the original data
using a Gaussian kernel with a standard deviation equal to the
1\pdeg8 pixel size at $N_{\rm side}= 32$.  This final $A_{\rm H}$ map is a
slightly smoothed version of the initial map of the correlation
measures. We follow the same procedure to interpolate the map of
offsets $\omega_\nu$ and get the desired $B$ map.

The CMB anisotropies and the noise increase the uncertainty of
the dust emissivity and dust model for $\nu \le 217\,$GHz. To reduce
these uncertainties at these low frequencies, in \citet{planck2013-pip56} but not in this paper for which
this is not necessary,  we choose to
extrapolate the $353\,$GHz model using the greybody function in
equation~(\ref{eq:greybody}) for the mean temperature of 19.8\,K and
the map of spectral indices from Sect.~\ref{sec:microwave_index}.

\section{CMB contribution to correlation measures}
\label{appendix:cmb}

Here is how we proceed to find the CMB contribution to the correlation
measures, i.e. the $\alpha (C_{\rm HI})$ term in Eq.~\ref{eq:6} in units
of thermodyamic (CMB) temperature.  The correlation measures corrected
for the CMB contributions are used in Sect.~\ref{sec:Dust_SED} to
compute the mean SED averaged over all sky patches, and in
Appendix~\ref{appendix:model} for the dust model.

We assume that the dust SED at $100 \le \nu \le 353\,$GHz is well
approximated by a greybody spectrum with the spectral indices $ \beta_{\rm mm}$
determined in Sect.~\ref{sec:microwave_index} and the mean dust
temperature of $19.8\,$K.  For each sky patch, we perform a linear fit
between the correlation measures at 100, 143, 217, and $353\,$GHz and
the greybody SED normalized to unity at $353\,$GHz, with weights taking
into account the uncertainties of the correlation measures. The slope
of the fit is the dust emissivity at $353\,$GHz, while the offset is
our estimate of $\alpha (C_{\rm HI})$.

\begin{figure}[!h]
\centering
\includegraphics[width=0.49\textwidth]{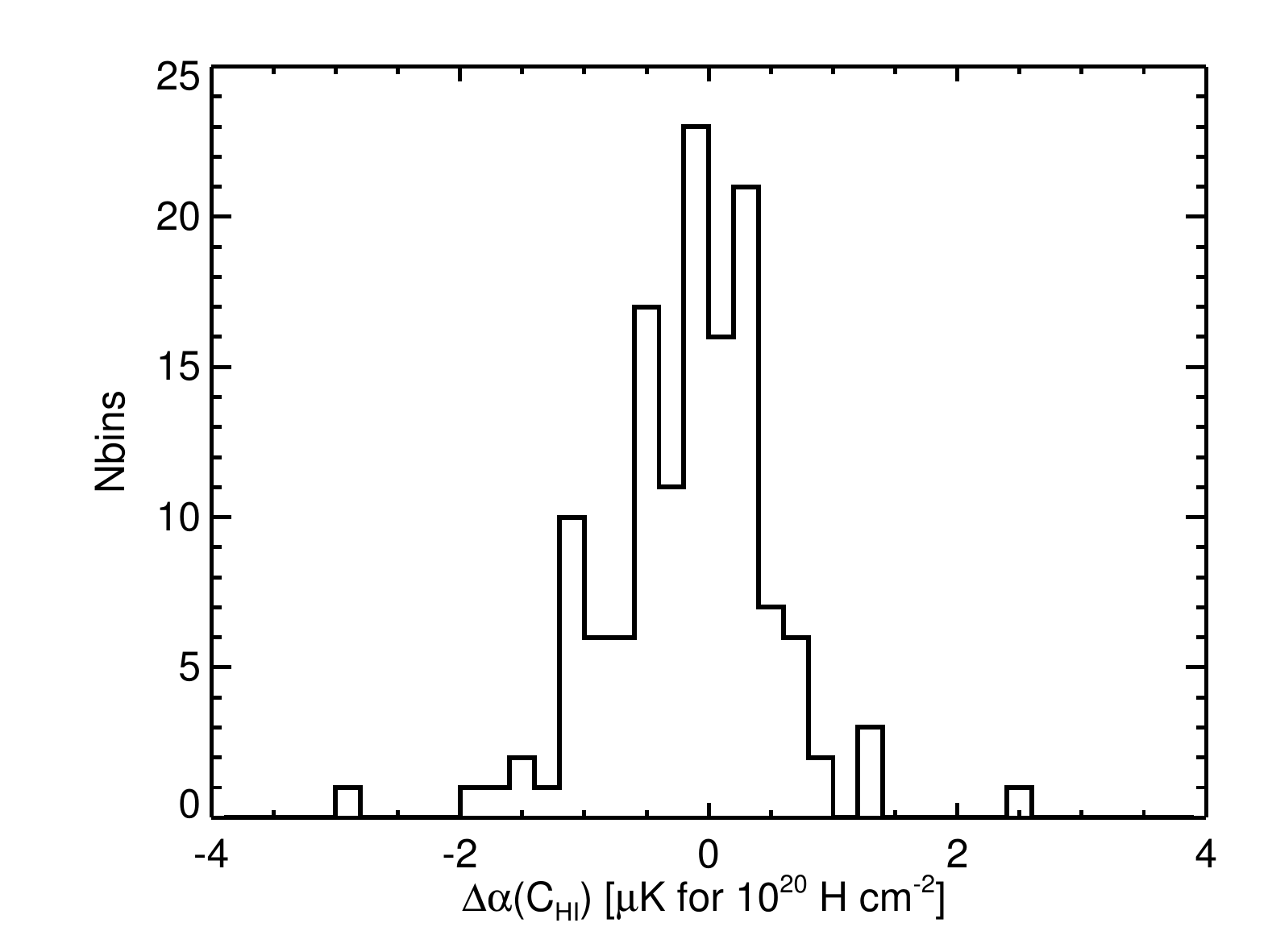}
\caption{Histogram of the difference between two estimates of $\alpha
(C_{\rm HI})$ (the correlation measure between the CMB and the \hi\
template), found assuming a greybody spectrum for the dust emissivity
or calculated with the {\tt SMICA} CMB map.  The standard deviation of
the difference, $0.7\, \mu$K per $10^{20}\,{\rm H\,cm^{-2}}$, is 3\,\%
of the standard deviation of $\alpha (C_{\rm HI})$. }
%?? X axis needs "per" rather than "for"
\label{fig:histo_CMB}
\end{figure}

For comparison, we also quantify the cross-correlation between the CMB
and the \hi\ map using the {\tt SMICA} map presented in the \planck\
component separation paper \citep{planck2013-p06}.  A histogram of the
difference between the two values of $\alpha (C_{\rm HI})$ for the 135 sky
patches at $N_{\rm side}=8$ is presented in Fig.~\ref{fig:histo_CMB}.  
The standard deviation $0.7\, \mu$K per
$10^{20}\,{\rm H\,cm^{-2}}$ represents only 3\,\% of the standard deviation
of the $\alpha (C_{\rm HI})$ values. We consider this percentage as our
uncertainty factor $\delta_{CMB}$ on the CMB correction in
Eq.~(\ref{eq:7}).
The mean difference ($-0.15 \, \mu$K per $10^{20}\,{\rm H\,cm^{-2}}$) is
within the expected statistical error. 

%This  result shows that equation~(\ref{eq:greybody}) with the spectral indices $\beta_{\rm mm}$ and the mean dust temperature
%derived from our correlation analysis provide an estimate of the dust thermal emission from 100 to $353\,$GHz consistent with the CMB map from component separation.

\section{Uncertainty of the dust emissivity}
\label{appendix:uncertainties}

In this appendix, we quantify the uncertainty of the dust
emissivity. In the first subsection, we quantify the uncertainties
from the correlation analysis.  In the second, we assess the
uncertainties associated with the definition of the Galactic \hi\
template that depends on the separation between Galactic and MS
emission (see Sect.~\ref{subsec:GASS}).  Finally, we discuss
uncertainties associated with subtraction of the zodiacal emission.

\subsection{Correlation analysis}

We describe how we estimate each of the contributions to  $\sigma (\epsilon_{\rm H}) $ [Eq.~(\ref{eq:7})], the
uncertainty of the dust emissivity.
% starting with the contribution from the noise.  
%
At each \planck\ frequency, we obtain a noise map by computing and dividing by two
the difference of the two maps made out of the first and second halves
of each stable pointing period \citep{planck2013-p03}.  For the
\DIRBE\ frequencies, we compute one Gaussian realization of the noise
using the maps of data uncertainty. 
The noise maps are cross-correlated with the \hi\ template using the
same mask and over the same sky patches. The standard deviation of the
correlation measures over all the sky patches yields the noise
contribution to $\sigma (\epsilon_{\rm H})$ at each of the \planck\ and
\DIRBE\ frequencies.

To estimate the additional contributions to $\sigma (\epsilon_{\rm H})$, we
use sky simulations of the Galactic emission and CMB and CIB
anisotropies.  For the Galactic maps, we consider only dust
emission. We compute dust maps by multiplying the \hi\ template with a
Gaussian realization of the dust emissivity map as
described in Appendix~\ref{appendix:simus_GalRes}. For the CMB and CIB
anisotropies, we compute Gaussian realizations using the power spectra
of the \planck~best-fit CMB model in \citet{planck2013-p08}, and of
the CIB model at 857\,GHz in \citet{planck2013-pip56}.  We scale CIB anisotropy
simulations at 857\,GHz to the full set of \planck-HFI and \DIRBE\
frequencies using a mean SED of CIB anisotropies.  This SED is a greybody fit
to the $C_{\ell}$ values at $\ell=500$ in \citet{planck2013-pip56}.  The spectral
index is $\beta= 1$ and the temperature $18.3\,$K.  We use 100
realizations of each of the Galactic, CIB and CMB maps. We
cross-correlate each of the simulated maps with the \hi\ template
using the same circular sky patches with $15^\circ$ diameter as for
the data analysis.

\begin{figure}[!h]
\centering
\includegraphics[width=0.49\textwidth]{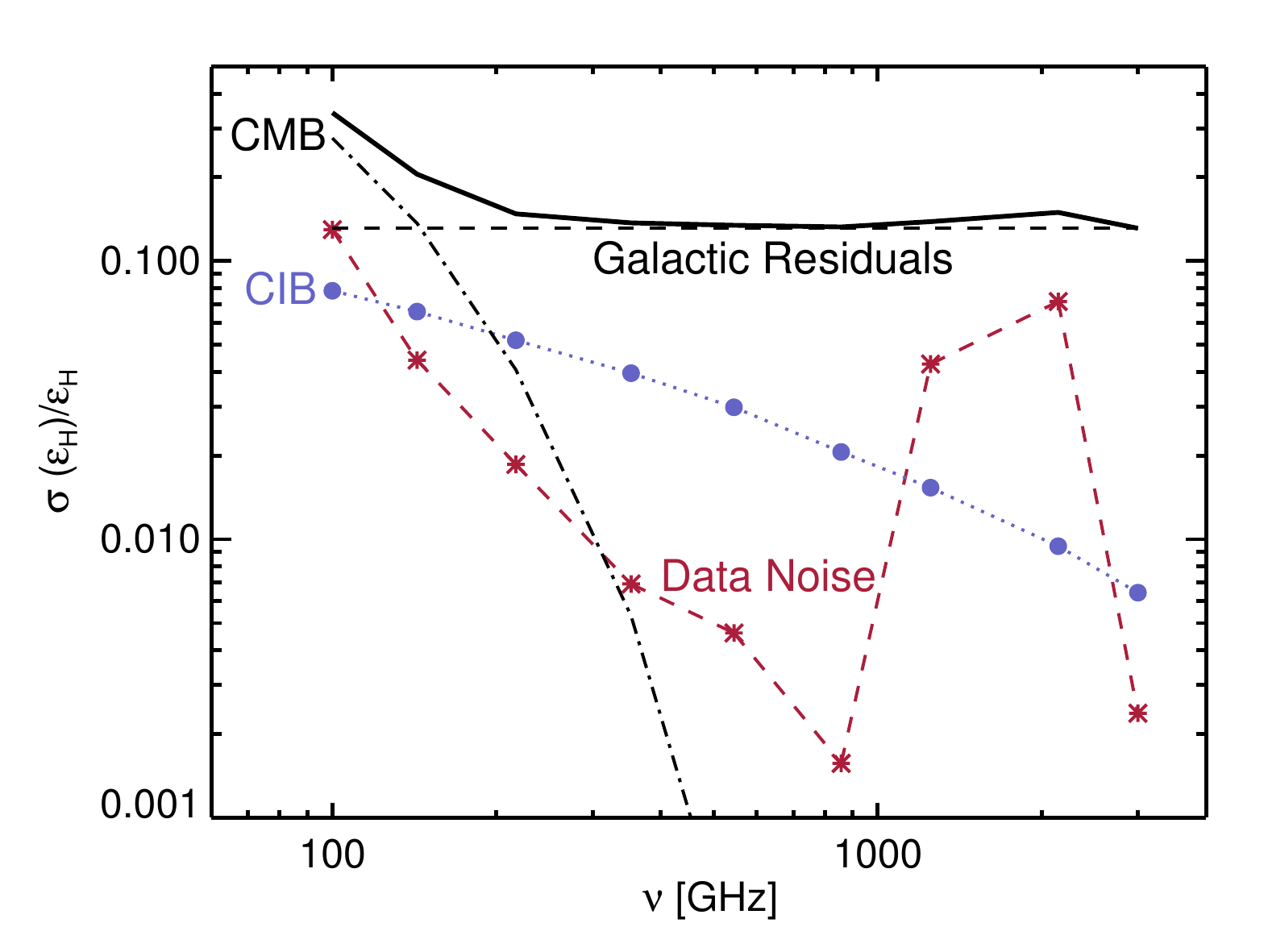}
\caption{ Fractional uncertainty (solid line) of the dust emissivities
  $\epsilon_{\rm H}$, normalized to the mean dust SED in
  Table~\ref{tab:mean_sed16}.  This total consists of contributions
  from Galactic residuals (black dashed line), noise (red dashed
  line with stars), CIB anisotropies (blue dotted line), and the CMB
  correction (black dash-dotted line).  The Galactic residual
  contribution is dominant at $\nu > 217\,$GHz, and the CMB
  contribution is dominant at lower frequencies.  }
\label{fig:IR_HI_uncertainties}
\end{figure}

Each component is analysed separately from the others to estimate its
specific contribution to the error budget.  The uncertainty of the
dust emissivity is quantified by comparing the emissivity derived from
the correlation analysis with the mean value of the input emissivity
map for each sky patch and each realization.  For the CMB contribution,
we use a fractional error $\delta_{CMB}$ of 3\,\% from
Appendix~\ref{appendix:cmb}.
In Fig.~\ref{fig:IR_HI_uncertainties}, the four contributions to the
fractional error $\sigma (\epsilon_{\rm H}) / \epsilon_{\rm H}$ are plotted versus
frequency.  The total uncertainty is the top solid line. We find that
the Galactic residual contribution is dominant at $\nu > 217\,$GHz, and
the CMB contribution is dominant at lower frequencies. The noise
is significant for the 140 and $240\,\mu$m bands and for the lowest
HFI frequencies.

These results depend on the size of the sky patches and on the angular
resolution. To quantify this dependence, we repeat the analysis of the
simulations for sky patches with diameters of $5^\circ$ and
$7\pdeg5$. We find that the contributions from noise and CIB anisotropies
scale with the inverse of the diameter, while the Galactic
contribution remains roughly constant.  The ratio between the CIB and
Galactic contributions also increases when we use a template with
higher angular resolution. These two effects contribute to make the
CIB contribution to the uncertainties more important for the low
column density fields in \citet{planck2011-7.12} than in our study.

The simulations show that the uncertainties do not bias our estimates
of the dust emissivity.  At all frequencies, the mean emissivity
averaged over all sky patches and all simulated maps is equal to the
mean input value within statistical errors.  We also find that the
uncertainty of the mean emissivity is roughly independent of the size
of the sky patches. The diameter that we use is thus not a critical
aspect of our data analysis.

The Galactic and CIB contributions to the uncertainty of the dust
emissivity are correlated between frequencies because variations of the
SED of dust and CIB anisotropies are not taken into account.  This reason is a
simplification, but the data analysis does show that the residual maps,
obtained after subtracting the dust model
(Appendix~\ref{appendix:model}) from the data, are highly correlated
between frequencies.

\subsection{Galactic \hi\ template}

To assess the uncertainties associated with the separation of the \hi\
emission into Galactic and MS components (Sect.~\ref{subsec:GASS_MS}),
we follow \citet{planck2011-7.12} in correlating the \planck\ maps with three
\hi\ maps for the low velocity gas (the original single tempate), and
for the IVC and HVC components (Sect.~\ref{subsec:GASS_IVC}).  We
perform this analysis over the same sky patches, using the same mask, as
in our cross-correlation with a single Galactic \hi\ template
(Sect.~\ref{subsec:masking}). We obtain dust emissivities for each of
the three \hi\ velocity components. The emissivities for the low
velocity component are very close to those reported in the paper for
our analysis with a single template. For example, at 857\,GHz the
fractional difference between the two sets of values (the ratio
between the difference and the mean value computed for each sky patch)
has a $1\sigma$ dispersion of 1.1\,\%, which is small compared to the
main uncertainties in Fig.~\ref{fig:IR_HI_uncertainties}.  The mean
difference between the two sets of values is negligible.
%?? emissivities for the other components discussed in a separate paper

\subsection{Subtraction of the zodiacal emission}

We end this appendix by comparing dust emissivities obtained from the
analysis of \planck\ maps with and without subtraction of the zodiacal
emission.  We find that the differences are minor.  For example, at
857\,GHz, the fractional difference in correlation measures has a
mean of zero and a standard deviation of 1.4\,\%, which is one order of
magnitude lower than the total uncertainty in
Fig.~\ref{fig:IR_HI_uncertainties}.  The differences are 
highest, but still small (up to 5\,\%), in sky patches near the southern
Galactic pole that are close to the zodiacal bands and where the
Galactic emission is faint.

\section{Simulations of Galactic residuals to the dust-\hi\ correlation}
 \label{appendix:simus_GalRes}

A histogram of the residuals with respect to the dust-\hi\ correlation
is shown in Fig.~\ref{fig:histo_residuals}. This appendix describes
how we simulate the Galactic contribution to the Gaussian part of this
histogram.  These simulations are used in
Appendix~\ref{appendix:uncertainties} to estimate the contribution of
Galactic residuals to the uncertainty of the dust emissivities, and in
\citet{planck2013-pip56} to assess the associated contamination of the CIB power
spectra.

It is beyond the scope of this appendix to explore fully the origin
and nature of the Galactic residuals. We briefly discuss and quantify
two possible contributions.\quad  
(1) The residual Galactic emission could trace dust associated with
diffuse ionized gas that is not spatially correlated with the \hi\
template.  The column density of this warm ionized medium is known to
account for $\sim 20\,\%$ of the total gas column density over the high
latitude sky \citep{Gaensler08}.\quad
(2) The Galactic residuals could arise from variations of the dust
emissivity on angular scales smaller than the $15^\circ$ diameter of
the sky patches used in our correlation analysis.  These variations
would be the extension to small scales of the variations mapped by our
correlation analysis (Fig.~\ref{fig:IR_HI_correlation}).  These two
contributions are not mutually exclusive: it is possible that each
contributes.  We do not consider residual emission from molecular gas, however, 
because the molecular fraction of the gas is known from UV
observations to be low at column densities lower than $3\times
10^{20} \,{\rm H\, cm^{-2}}$ \citep{Savage77,Gillmon06}.

We produce sky simulations including each of these hypothetical
contributions to the Galactic residuals and realizations of the CIB
power spectrum. We process these simulated maps through the same
correlation analysis as used on the \Planck\ 857\,GHz map. 
The simulations show that for each hypothesis we can match the
amplitude and scatter of the values of $\sigma_{857}$ in
Fig.~\ref{fig:sigma_residuals}; %, within astrophysical constraints (e.g. the column density of HII gas).
however, it is only when the simulated maps include significant
variations in the dust emissivity that the simulations match the
systematic trend of $\sigma_{857}$ growing with increasing \NH.  
We find that simulations can account for the main statistical
properties of the Galactic residuals at 857\,GHz when the map of
variable dust emissivty is a Gaussian realization of a $k^{-2.8}$ power
spectrum, without needing any contribution from the warm ionized
medium.  The map of the dust emissivity is normalized to reproduce the
mean value and the standard deviation measured from the correlation of
the 857\,GHz map and the \hi\ template.  We make multiple
realizations of this specific model that are used in
Appendix~\ref{appendix:uncertainties} and \citet{planck2013-pip56}.  The
simulated maps at 857\,GHz are scaled to other frequencies using the
mean SED in Table~\ref{tab:mean_sed16}.  The simulations do not take
into account the anti-correlation between the dust temperature and
opacity.

%\end{appendix}
\raggedright
\end{document}